\documentclass{aa}  

\usepackage{graphicx}
\usepackage{txfonts}
\usepackage{lipsum}
\usepackage{subcaption}        
\usepackage{lscape}             
\usepackage{placeins}           
\usepackage{orcidlink}
\usepackage{xcolor}
\usepackage{amsmath}
\newcommand{\spx}{SPHERE\textsc{x}\,}
\usepackage{wrapfig}
\usepackage{multicol}
\usepackage{sidecap}
\usepackage{hyperref}

\begin{document}

   \title{Hidden Monsters with SPHERE\textsc{x}}
   \subtitle{I. A goldmine for heavily reddened quasars at cosmic noon}
   \titlerunning{76 SPHERE\textsc{x}-confirmed heavily reddened quasars at cosmic noon}

   \authorrunning{M. Stepney}

   \author{Matthew Stepney\inst{1}\fnmsep\inst{2}\fnmsep\inst{3}\fnmsep\thanks{Corresponding author: mstepney@cata.cl}
        \and Manda Banerji\inst{4}
        \and Franz E. Bauer \inst{2}
        \and Roberto J. Assef\inst{3} 
        \and Guodong Li\inst{5}
        }

   \institute{Centro de Astrofisica Y Tecnologias Afines, Av. Vicuña Mackenna 4860, San Joaquín, Santiago, Chile
   \and Instituto de Alta Investigación, Universidad de Tarapacá, Casilla 7D,
    Arica, Chile
   \and Instituto de Estudios Astrofísicos, Facultad de Ingeniería y Ciencias, Universidad Diego Portales, Av. Ejército Libertador 441, Santiago, Chile 
   \and School of Physics and Astronomy, University of Southampton, Southampton, SO17 1BJ, UK
   \and Kavli Institute for Astronomy and Astrophysics, Peking University, Beijing 100871, People's Republic of China}

   \date{Received May 5, 2026}

  \abstract
   {}
   {Heavily reddened quasars (HRQs) are luminous, dust-obscured broad-line quasars thought to trace a short-lived phase of intense black hole growth and feedback. Previous studies have been limited by small sample sizes, restricting robust statistical analyses of their properties. We expand the census of the most luminous HRQs to enable population-level studies, aiming to connect their SEDs to other luminous quasar populations and place them in context within an evolutionary sequence for massive galaxy and black hole formation.} 
   {We assemble multi-wavelength broad-band photometry for the brightest ($K_{\rm AB}<18$ mag) HRQ candidates and select AGN candidates with red near infra-red colours of $(J-K)_{\rm AB}>$1.6. We employ \spx spectrophotometry to confirm the HRQs and determine redshifts. Detailed SED fitting allows us to compare these HRQs to other luminous quasar populations, including a control sample of hyper-luminous, unobscured \textit{unWISE}-Gaia (Quaia) quasars, as well as luminous Hot Dust-Obscured Galaxies (Hot DOGs).}
   {We confirm 76 new HRQs with redshifts $1.5<z_{\rm sys}<3.9$, dust-corrected optical continuum luminosities of log$_{10}({\lambda L_\lambda (3000\mathring{\rm A}) [\rm erg\;s^{-1}]})>47.0$ and line-of-sight extinctions $0.4<\rm E(B-V)<1.6$ (A$_{\rm{V}}\sim1-5$ mag). This more than doubles the number of confirmed HRQs at $z_{\rm sys}>1.5$ including the first 7 HRQs ever identified at $z>3$. A UV excess consistent with scattered quasar emission is detected in 76 per cent of \spx HRQs. We conclusively demonstrate that HRQs are hot-dust poor compared to blue quasars of similar luminosity and redshift. We find that the $6\mu m$ continuum luminosities of HRQs are systematically fainter at fixed 3000\AA\, continuum luminosity relative to blue Quaia quasars, suggesting that HRQs are deficient in both hot and warm dust components. This combination of depleted torus-scale dust reservoirs and higher luminosities compared to Hot DOGs and blue quasars supports a scenario in which HRQs represent a blow-out phase, when strong feedback has begun to clear the central regions of obscuring material.}
   {}
   

   \maketitle
\nolinenumbers

\section{Introduction}

Dust-obscured black hole accretion likely marks an important stage in the co-evolution of galaxies and their central supermassive black holes (SMBHs) \citep{1988ApJ...325...74S,2005ApJ...630..705H,Hopkins_2008,2005Natur.433..604D}. This reddened quasar phase is thought to be a direct result of major-mergers, which peak in frequency during cosmic noon \citep[][]{2009MNRAS.394L..51B,2012ApJ...744...85M} - a period defined by rapid black-hole growth and star formation. In this paradigm, gas inflows fuel powerful initially dust-obscured, active galactic nuclei (AGN) while simultaneously inducing a starburst \citep{Granato_2004,2005Natur.433..604D,HOP_ELV}. The presence of this dust can make radiatively driven outflows more efficient \citep[e.g.][]{2018MNRAS.479.2079C,Ishibashi:17,2022MNRAS.516.4963I}, meaning dusty/obscured quasar populations likely trace periods of extreme quasar-driven feedback.

Evidence of dust-driven feedback in red quasar populations is observable across multiple wavelengths. Extreme outflows in both the broad- and narrow-line regions have been identified in Extremely Red Quasars (ERQs; \citealt{Zakamska:16,2017MNRAS.464.3431H,2019MNRAS.488.4126P,2024MNRAS.527..950G}), Hot Dust-Obscured Galaxies \citep[Hot DOGs;][]{Eisenhardt_2012,2012ApJ...756...96W,2020ApJ...888..110J,2020ApJ...905...16F,2025ApJ...989..230V} and Heavily Reddened Quasars (HRQs; \citealt{2019MNRAS.487.2594T,2024MNRAS.533.2948S}), with kinetic powers comparable to or exceeding those of unobscured systems at similar luminosities. At radio wavelengths, compact emission correlates with dust extinction and, together with steep spectral slopes, favours an origin in wind-driven shocks within a dusty obscuring medium \citep{2023MNRAS.525.5575F,2025MNRAS.537.2003F}. Furthermore, high accretion rates, often reaching or exceeding their Eddington limits \citep[e.g.][]{2024ApJ...971...40L,S26a}, further suggest that red quasars are caught during a brief phase of rapid black hole growth producing strong AGN feedback.

Numerous studies have also been devoted to understanding the spectral energy distribution (SED) properties of red quasar populations. For example, blue photometric colours at rest-UV wavelengths, arising from an excess in the continuum emission relative to the dust-obscured AGN, has been detected in Hot DOGs \citep[][]{2016ApJ...819..111A,2020ApJ...897..112A}) and HRQs \citep[e.g.][]{2018MNRAS.475.3682W,S26a}. While the origin of the UV excess is unclear, the detection of broad emission lines and polarised light in some Hot DOGs \citep[e.g.][]{2016ApJ...819..111A,2020ApJ...897..112A, 2022ApJ...934..101A,2025A&A...702A.124A,2024ApJ...971...40L} suggests that scattered quasar emission may contribute significantly. In comparison, observations of the HRQ ULASJ2315+0143 with VLT X-Shooter reveal both broad emission lines and host galaxy signatures, indicating that the UV-optical continuum may also arise, at least in part, from active star-forming regions within the quasar host galaxy \citep{2024MNRAS.533.2948S}. Furthermore, red quasars exhibit a wide diversity in dust properties. The SEDs of Hot DOGs seem to be dominated by hot and warm dust components \citep[e.g.][]{2012ApJ...756...96W,2015ApJ...805...90T,2023ApJ...958..162L}, whereas the SEDs of HRQs are comparatively hot-dust poor \citep{S26a}. 

Concerted efforts have been made to link the multi-wavelength SED properties of red quasars to their outflow and accretion characteristics in the context of galaxy evolution \citep[e.g.][]{2019MNRAS.487.2594T,2021A&A...649A.102C,2024MNRAS.533.2948S,S26a}. However, the relatively modest sample sizes of the reddest and most heavily obscured quasars limit our ability to robustly establish relationships between the different populations. For example, traditional ultraviolet (UV) and optical selection techniques used to select quasars for wide-field spectroscopic follow up as part of e.g. the Sloan Digital Sky Survey (SDSS; \citealt{York_2000}) and Dark Energy Spectroscopic Instrument (DESI; \citealt{2023AJ....165..124A}) generally do not detect quasars with line-of-sight dust-extinctions $\rm E(B-V)\gtrsim 0.5$ \citep{Richards:03,2020MNRAS.494.4802F}. Hence, assembling large, spectroscopically confirmed samples of the reddest quasars has remained a significant challenge. 

Spectroscopic confirmation of luminous HRQs has previously been conducted through single-object NIR spectroscopic follow-up of candidates selected from wide-field infrared surveys \citep{2012MNRAS.427.2275B,2013MNRAS.429L..55B,2015MNRAS.447.3368B,2019MNRAS.487.2594T}. As the most luminous and massive galaxies are rare with low sky-densities \citep{2001MNRAS.326..255C,Benson_2003,2003ApJ...584..203H,somerville_08,2015MNRAS.446..521S}, luminous HRQ candidates are distributed across the entire sky, making confirmation of statistical samples expensive due to large slew times between targets. Fortunately, in the new era of wide-field/all-sky near-infrared spectroscopic/spectrophotometric surveys - e.g. Euclid \citep{2025A&A...697A...2E,2025A&A...697A...3E,2025A&A...697A...1E} and the Spectro-Photometer for the History of the Universe, Epoch of Reionization, and Ices Explorer \citep[\spx;][]{2014arXiv1412.4872D} - it is now feasible to assemble large, spectroscopically confirmed samples of highly dust-obscured quasars.

In this paper, we exploit the all-sky coverage of \spx to confirm a new sample of hyper-luminous HRQs via spectrophotometry, enabling a comprehensive multi-wavelength SED analysis of the largest HRQ population assembled to date. This study demonstrates the potential of \spx to conduct population studies of obscured quasars and marks the beginning of an ongoing effort to characterise the luminous, obscured quasar population. The structure of the paper is as follows; in Section \ref{sec:HRQs}, we describe the assembly of the multi-wavelength data used to select HRQ candidates. Source confirmation and redshift estimation are presented in Section \ref{sec:HRQs-Confirmation}, followed by a description of the SED modelling in Section \ref{sec:SED_model}. In Section \ref{sec:cQSO}, we introduce the hyper-luminous blue quasar sample used for comparison to the HRQs. We then examine the general properties of the \spx-confirmed HRQ sample in Section \ref{sec:demo}, including comparisons with other quasar populations at similar luminosities and redshifts. We summarise our conclusions in Section \ref{sec:conc}. Here we report vacuum wavelengths throughout and adopt a $\Lambda \textnormal{CDM}$ cosmology with $h_0$ = 0.71, $\Omega_\textnormal{M}$ = 0.27 and $\Omega_\Lambda$ = 0.73 for calculating luminosities.

\section{Heavily Reddened Quasars (HRQs)} \label{sec:HRQs}
\subsection{Infrared Selection} \label{sec:HRQs-Selection}

The selection of the original HRQ population is based on observed optical, near- and mid-infrared photometric colours and is discussed at length in \citet{2012MNRAS.427.2275B}, \citet{2013MNRAS.429L..55B}, \citet{2015MNRAS.447.3368B} and \citet{2019MNRAS.487.2594T} - and summarised in \citet{S26a}. The primary selection of the HRQs is based on their red, observed $(J-K)$ colours. As the objective of this paper is to build a more complete sample of HRQs at the highest quasar luminosities, we make a number of modifications to the original HRQ selection criteria. 

Firstly, we modify the original $K$-band magnitude limit ($K_{\rm AB} < 20.3$) such that only the brightest ($K_{\rm AB} < 18$) sources are selected. Second, given that $>80$ per cent of the original HRQ sample host a statistically significant rest-UV excess with respect to their rest-optical emission (\citealt{S26a} - see also \citealt{2018MNRAS.475.3682W}), we omit the $i$-band magnitude limit ($i_{\rm AB}>20.5$) and $(i-K)_{\rm AB} > 2.5$ colour selection used in \citet{2012MNRAS.427.2275B} to select the original HRQ sample. Finally, given also that the HRQ population is hot-dust poor when compared to luminous blue quasars of similar luminosity and redshift \citep[e.g.][]{2024MNRAS.533.2948S,S26a}, we relax the mid-infrared colour selection to be bluer, and use the \citet{Assef18:WISE} R90 \textit{WISE} AGN catalogue; corresponding to a 90 per cent reliability limit of all Wide-field Infrared Sky Explorer (\textit{WISE}) AGN. The result is a more complete sample of luminous HRQ candidates, whose revised selection criteria are as follows;

\begin{itemize}
    \item $K_{\rm AB} < 18$
    \item $(J-K)_{\rm AB} > 1.6$
    \item $(W1-W2)_{\rm{Vega}}>0.5$
    \item $kclass=-1$ (point-source morphology in the $K/K_S$-band)
\end{itemize}

To gain near-infrared coverage across the northern \emph{and} southern skies, we match the R90 \textit{WISE} AGN catalogue to the United Kingdom Infrared Deep Sky Survey's (UKIDSS) Large Area Survey \citep[UKIDSS-LAS DR10;][]{2007MNRAS.379.1599L}, the United Kingdom Infrared Telescope's Hemisphere Survey \citep[UHS DR2;][]{2018MNRAS.473.5113D} and the Visible and Infrared Survey Telescope for Astronomy's (VISTA) Hemisphere Survey \citep[VHS DR5;][]{2013Msngr.154...35M,2021yCat.2367....0M} using a matching radius of 1$\arcsec$\footnote{Cone searches are conducted with v4.10-5 of the Tool for OPerations on Catalogues And Tables \citep[TOPCAT;][]{2005ASPC..347...29T} software.}. Given that HRQs have compact morphologies in the rest-optical (by selection), we use 2$\arcsec$ diameter aperture magnitudes from UKIDSS-LAS, UHS and VHS for our study, and profile fitting magnitudes from \textit{WISE}. 

Applying the above selection criteria yields 413 HRQ candidates spanning $\simeq20,000\deg^{2}$. Nine sources are already spectroscopically confirmed HRQs from \citet{2012MNRAS.427.2275B, 2015MNRAS.447.3368B} and are therefore removed. An additional source - ULASJ2148-0011 - was previously identified in \citet{2012MNRAS.427.2275B}, but not spectroscopically confirmed via Gemini or VLT-SINFONI, therefore, this source remains in our candidate list. The total number of luminous HRQ candidates is therefore 404. 

\subsection{Multi-wavelength photometric data}

A primary objective of this work is to build the full multi-wavelength spectral energy distributions (SEDs) of the new HRQ sample. Hence, we assemble a catalogue of optical photometry, probing the rest-frame UV SEDs, to complement our near- and mid-infrared data. We match all 404 luminous HRQ candidates to the second data release of the DECam Local Volume Exploration Survey \citep[DELVE-DR2;][]{2022ApJS..261...38D}\footnote{\url{https://delve-survey.github.io}} with a 1$\arcsec$ search radius via the NOIRlab Astro Data Lab Service \citep[][]{10.1117/12.2057445,NIKUTTA2020100411,9347681} - resulting in 332 matches. To extend our coverage at northern latitudes, we also match the sample to the Sloan Digital Sky Survey \citep[SDSS-DR16;][]{2020ApJS..249....3A}. This yields 215 matches, resulting in rest-UV/optical coverage for 389 HRQ candidates via DELVE, SDSS, UKIDSS-LAS, UHS, VHS and \textit{WISE}. Given that many HRQs have extended morphologies in the rest-frame UV \citep[e.g.][]{2018MNRAS.475.3682W,2024MNRAS.533.2948S}, we use MAG$\_$AUTO photometric extractions for both DELVE and SDSS. Since UKIDSS-LAS, UHS and VHS avoid the galactic plane, we do not perform a Milky Way galactic-extinction correction to the optical bands. While this may introduce an additional source of uncertainty to the SED fitting, the error budget will still be dominated by uncertainties in the photometric extraction (particularly in fainter sources) as well as the relative simplicity of our SED model.

\section{Spectrophotometric confirmation with \spx} \label{sec:HRQs-Confirmation}

\subsection{\spx data acquisition}
\spx is a National Aeronautics and Space Administration (NASA) Explorer satellite poised to complete the first all-sky spectral survey at near-infrared wavelengths \citep{2014arXiv1412.4872D,2025arXiv251102985B}. \spx observes in 102 dedicated spectral channels whose wavelength coverage spans $0.75-5.0\mu m$, with estimated $5\sigma$ depths of $\sim19.5-19.9$ AB magnitudes at $0.75<\lambda<3.2\mu m$ by the time of the survey's completion, and a $6\arcsec$ pixel scale \citep{2014arXiv1412.4872D,2025arXiv251102985B}. While \spx has already been utilised for cosmological and interstellar ice studies \citep[e.g.][]{2025arXiv251007684X,2025arXiv251020919B,2025arXiv251207318L}, this work represents one of the first studies in which \spx has been used to identify dust-obscured broad-line quasars, although the suitability of \spx for this application was first highlighted by \citet{2024ApJ...972..171K}. Recently \citet{2026arXiv260310135D} have also used \spx to confirm 87 new luminous quasars at $4.0 <\rm z_{sys} < 5.7$ as well as lower-redshift interlopers.

To acquire \spx spectra for our 404 HRQ candidates, we use the Infrared Science Archive's (IRSA's) Spectrophotometry Tool\footnote{\url{https://irsa.ipac.caltech.edu/applications/spherex/tool-spectrophotometry}}, which implements a multi-stage data processing pipeline to perform photometric extractions, photometric/wavelength calibrations, and background subtractions of the \spx data in real time \citep{2025arXiv251115823A}. In all cases, we used a 15 pixel aperture for background estimation and a point-source morphology for the forced photometric extraction\footnote{The \spx spectra used in this work were downloaded between January and February 2026, however, the \spx data is still accumulating as of the publication date of this manuscript. Therefore, some objects may not have been observed in all spectral channels. This could result in a lower spectral resolution than reported in some objects, and consequently, increased uncertainties in our redshift estimates.}. 

\subsection{Estimating quasar redshifts with \spx} \label{sec:redshifts}

While the \spx Science Team will eventually build high-level \spx data products which include redshift estimates \citep[e.g. level 4 processing;][]{2025arXiv251115823A}, they are not publicly available at present. Hence, we employ a cross-correlation approach similar to \citet{10.1111/j.1365-2966.2010.16648.x} - and subsequently \citet{2020MNRAS.492.4553R} and \citet{2023MNRAS.524.5497S} - to determine the redshifts of the HRQ candidates. Our approach consists of two steps; (i) the definition of a smooth pseudo-continuum from which continuum-subtracted \spx spectra can be defined and (ii) a cross correlation algorithm that uses a resampled quasar template spectrum to estimate the redshift of each source.

\subsubsection{Defining the pseudo-continuum} \label{sec:pseudocontinuum}

\begin{figure}
\centering
 \includegraphics[scale=0.64, clip]{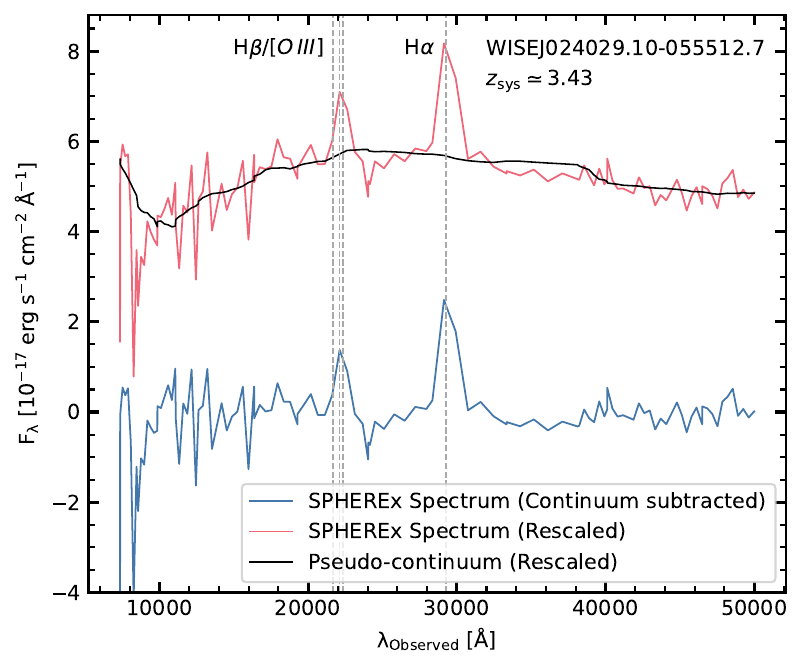} 
  \caption{The continuum-subtracted \spx spectrum of the confirmed HRQ - WISEJ024029.10-055512.7 - is presented in blue. To aid readability the original \spx spectrum (red) and corresponding 29-pixel median-filtered pseudo-continuum (black) have been shifted +4 units in the y-direction. The positions of the broad Balmer emission lines are marked in grey.} 
 \label{fig:SPHEREx_Confirmation}
\end{figure}

Given that the shape of the quasar continuum can vary depending on the SED of the quasar, we define the pseudo-continuum empirically following \citet{10.1111/j.1365-2966.2010.16648.x}. Specifically, the quasar pseudo-continuum is defined by a median-filtered version of the original \spx spectrum, which is reflected at the blue and red ends and stitched together to negate edge effects. We find that the ideal filtering window for the \spx spectra is 29 channels, beyond which there is no significant improvement to the pseudo-continuum. An illustration of how our continuum subtraction recipe operates in practice is presented in Fig. \ref{fig:SPHEREx_Confirmation}.

\subsubsection{Cross-correlation analysis}

Our cross-correlation analysis scheme is similar to that described in section 4.2 of \citet{10.1111/j.1365-2966.2010.16648.x} with two key differences. First, the template used for the cross-correlation scheme is generated by \textsc{qsogen}\footnote{\url{https://github.com/MJTemple/qsogen}}, a Python package that implements an empirically-motivated parametric model to simulate quasar colours, magnitudes and SEDs \citep{2021MNRAS.508..737T}. To produce the template spectrum, we generate a \textsc{qsogen} quasar model (with all free parameters left at their default), which reproduces the spectrum of the average SDSS quasar. The reddening primarily affects the shape of the quasar continuum, which is subtracted before the cross-correlation is conducted. We resample this template spectrum onto the \spx spectral resolution, and subtract the continuum in the same manner as detailed in Section \ref{sec:pseudocontinuum}. The cross-correlation between each template and an individual HRQ spectrum is performed with the HRQ redshift as a free parameter, with bad \spx pixels masked beforehand. Hence, cross-correlation coefficients are calculated across redshift-space, with the redshift solutions varying in 0.01 increments - i.e. $\rm \Delta z_{sys} =0.01$. 

To estimate redshift uncertainties, each \spx spectrum is perturbed by Gaussian noise consistent with the corresponding noise array, and the mean and standard deviation after 100 iterations are used to determine our final results. To ensure that our uncertainty estimates are robust, we performed a cross-correlation analysis on the \spx spectra of nine already confirmed HRQs. We find that the average difference in redshift between the \spx estimates and those from higher-resolution ground-based spectroscopy is $\langle \Delta\rm z_{sys}\rangle \simeq0.011$, broadly consistent with the average Monte-Carlo uncertainties calculated for the remaining sample - e.g. $\langle \Delta\rm z_{sys}\rangle \simeq 0.006$. As the uncertainties associated with the \spx spectral channels are likely underestimated \citep[e.g.][]{2026arXiv260310135D}, the slightly underestimated Monte-Carlo errors are expected. 

The redshift solution that maximises the cross-correlation coefficient for a given \spx spectrum is assigned as the systemic redshift provided that: (i) the cross-correlation coefficient meets the threshold, $\rm cc(z_{sys}) \geq0.5$ and (ii) the redshift solution meets the threshold, $\rm z_{sys} \geq1.5$. The former condition is determined empirically and is more conservative than the $\rm cc(z_{sys}) \geq0.2$ used in \citet{10.1111/j.1365-2966.2010.16648.x} who relied on the SDSS spectroscopic pipeline already minimizing catastrophic redshift misidentifications. The $z>1.5$ cut is imposed because the HRQ selection results in impure quasar samples at low redshift ($\rm z_{sys} \leq0.6$) with many low redshift HRQ candidates being morphologically compact star-forming galaxies \citep[e.g. SDSSJ210050.13-005752.5;][]{2015MNRAS.447.3368B}. Additionally, at $0.6\leq\rm z_{sys} \leq1.5$, contributions from the host galaxy at $\sim1\mu m$ can bias the SED fitting at near-infrared wavelengths (e.g. \citealt{S26a}). Hereafter; when referring to HRQs we consider only the cosmic noon sub-sample from \citet{S26a} and this work. 

\begin{figure} [ht!]
\centering
\begin{tabular}{r}
 \includegraphics[scale=0.545, trim={0cm, 0.82cm, 0cm, 0cm}, clip]{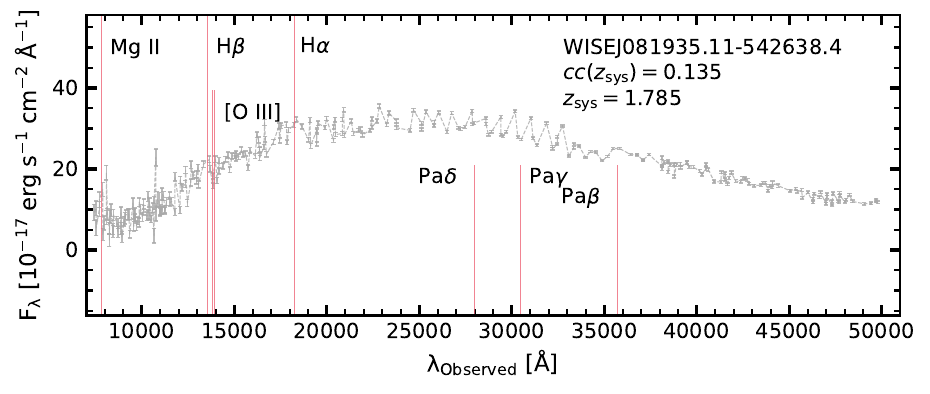} \\
 \includegraphics[scale=0.545, trim={0cm, 0.82cm, 0cm, 0.2cm}, clip]{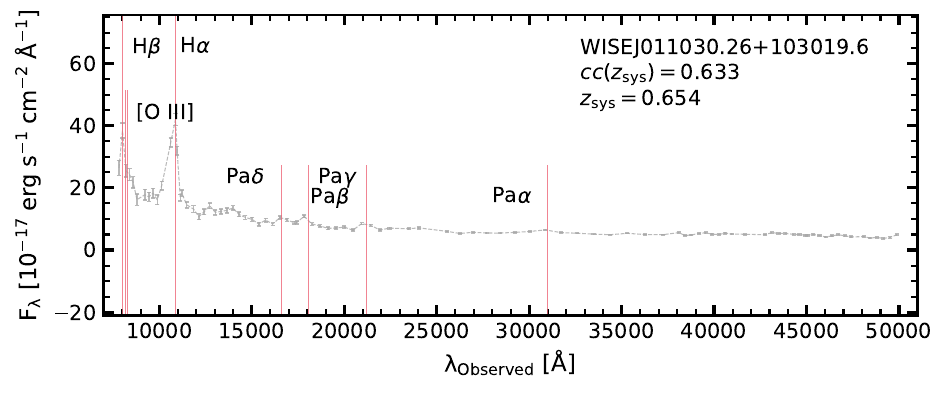} \\
 \includegraphics[scale=0.545, trim={0cm, 0cm, 0cm, 0.2cm}, clip]{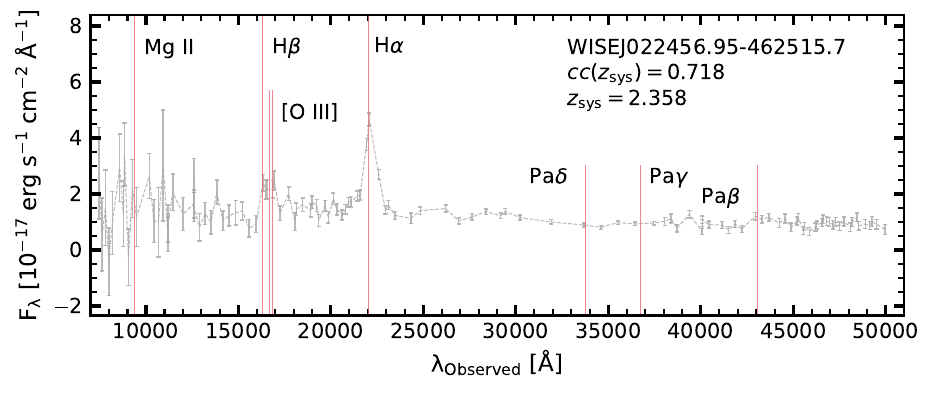} \\
 \end{tabular}
  \caption{We present example continuum-subtracted \spx spectra for sources that; fail to meet the $\rm cc(z_{sys}) \geq0.5$ threshold (top), fails to meet the $\rm z_{sys} \geq1.5$ threshold (middle) and meet both criteria (bottom). The best redshifts and cross-correlation coefficients are shown in the upper right. In the middle/bottom panels the broad H$\alpha$ emission line is clearly visible. In the top panel, the \spx spectrum is featureless, and hence the cross-correlation analysis cannot robustly determine a redshift.}
 \label{fig:SPHEREx_cc}
\end{figure}

There are 305 \spx objects for which our cross-correlation analysis yields coefficients that are too low to meet criterion (i), and a further 20 objects whose optimum redshift is too low to meet criterion (ii). Hence, the result is 76 SPHERE\textsc{x}-confirmed HRQs with redshifts $1.5<\rm z_{sys}<3.9$ including, for the first time, 7 HRQs with $\rm z_{sys}>3.0$. \autoref{fig:SPHEREx_cc} shows typical example spectra for \spx objects that fail either criterion (i) or (ii), in addition to an object that satisfies both. Objects that fail criterion (i) generally have spectra that is either noise-dominated or featureless. The featureless sources may be morphologically compact galaxies or narrow-line AGN, whose nebular emission lines are too narrow and/or weak to be detected at the resolution of \spx - resulting in poor redshift estimates. Objects that fail only criterion (ii) have significant H$\alpha$ emission ($\rm S/N>3$) and have robust redshifts, however, they trace a separate population to the cosmic noon HRQs due to selection effects \citep[][]{S26a}. 

Given the low spectral resolution in the bluer spectral channels of \spx ($R\simeq40$ at $0.75-3.8 \mu m$; \citealt{2014arXiv1412.4872D}), robust H$\alpha$-derived black hole mass measurements are unfeasible at the redshifts probed by this study and are therefore not attempted, however, when fit with a single Gaussian the full-width at half-maximum of the H$\alpha$ line profiles range between $6000-8000\,\rm  km\,s^{-1}$. Our final HRQ sample, with some example \spx spectra, is presented in Appendix \ref{App:SPHEREx_Spectra}.

\section{HRQ spectral energy distribution modelling} \label{sec:SED_model}

In this section, we describe the spectral energy distribution (SED) model used to fit the full suite of multi-wavelength data assembled in Section \ref{sec:HRQs-Selection} for the \spx HRQ sample.

\subsection{The multi-wavelength SED properties of HRQs from broad-band photometry} \label{sec:sed_broadbands}

We model the SEDs with \textsc{qsogen} (see Section \ref{sec:redshifts} and the references within) and adopt a Monte-Carlo approach for the fitting, making use of the \textsc{emcee}\footnote{\url{https://github.com/dfm/emcee}} Python package to explore the likelihood space via the affine-invariant ensemble sampler proposed by \citet{2010CAMCS...5...65G}. Given that many HRQs ($>$80 per cent) host blue photometric colours at rest-UV wavelengths \citep[e.g.][]{2018MNRAS.475.3682W,2024MNRAS.533.2948S,S26a}, we adopt the same two-component model used in \citet{2024MNRAS.533.2948S} and \citet{S26a}. 

The SED model comprises a dust-attenuated quasar component and an additional scattered light component, which we use to reproduce the rest-UV photometry. While this approach is well-motivated \citep[e.g. broad high-ionisation emission lines are detected at rest-UV wavelengths in a VLT-XShooter spectrum of the HRQ - ULASJ2315+0143;][]{2024MNRAS.533.2948S}, the shape of the rest-UV continuum in HRQs remains consistent with both scattered quasar light and on-going star formation in the quasar host \citep[][]{2018MNRAS.475.3682W,2024MNRAS.533.2948S}; therefore, a host galaxy contribution at rest-UV wavelengths cannot be ruled out, indeed, \citet{2024MNRAS.533.2948S} argue that both host galaxy emission and scattered quasar light are likely to contribute to the rest-UV emission of HRQs (see also - Section \ref{sec:uvXS}). Given the degeneracy between the two interpretations of the UV excess, we adopt only the scattered light interpretation for the purposes of fitting the broad-band photometry of HRQs because the model contains fewer free parameters.

At redder wavelengths, the host galaxy can have a measurable contribution to the total SED in luminous blue quasars. For example, in SDSS quasars, the inclusion of the S0 template from the SWIRE library \citep{2007ApJ...663...81P} improved the \textsc{qsogen} SED fit at $\lambda_{\rm rest}\sim 1\mu m$, with an average contribution to the total SED of 5-10 per cent \citep[section 2.5;][]{2021MNRAS.508..737T}. However, the relative contribution of the host galaxy decreases with increasing quasar luminosity \citep[e.g.][]{2006AJ....131.2766R,2006AJ....131...84V}, with SDSS samples following the relation $L_{\rm galaxy} \propto L_{\rm QSO}^{0.684}$ \citep[section 5;][]{2021MNRAS.508..737T}. Given that the average HRQ is $\sim 2.5$ dex brighter than the average SDSS quasar, this corresponds to a relative host galaxy contribution of $<1$ per cent when extrapolated to HRQ luminosities. Hence, in \citet{S26a}, the inclusion of the S0 host-galaxy template only produced favourable SED fits in lower redshift sources whose luminosities log$_{10}\{\lambda$L$_{\lambda} (3000\mathring{A}) [\rm erg\;s^{-1}]\}<46$. Additionally, when fitting a star-forming host galaxy component to the VLT-XShooter spectrum of the HRQ, ULASJ2315+0143, the host galaxy contribution was $<1$ per cent in the H$\alpha$ emission region \citep[][]{2024MNRAS.533.2948S}. As the rest-optical to NIR emission of HRQs has been shown to be completely dominated by the quasar in previous works, we do not account for contributions from the host galaxy when modelling their SEDs.

Therefore, the free parameters used to model our HRQ SEDs are as follows: (i) the dust-corrected 3000\AA\, continuum luminosity of the quasar, log$_{10}\{\lambda$L$_{\lambda} (3000\mathring{A}) [\rm erg\;s^{-1}]\}$; (ii) the line-of-sight dust extinction, $\rm E(B-V)$, (iii) the ratio in luminosity between the hot-dust blackbody and the tail of the quasar UV/optical continuum slope at 2$\mu m$, $ L_{Dust}/L_{Disk}|_{2\mu m}$, which can be used as a proxy for the amplitude of the hot dust emission \citep[e.g. Figure 1;][]{2021MNRAS.501.3061T}; and (iv) the scattering fraction, $F_{\rm UV}$, which represents a fixed fraction of the total unobscured quasar emission. Since the average UV inflection wavelength of the HRQ SED is $\sim2500$\AA\, \citep{S26a}, we only fit the two-component SED model to objects with observed-optical broad-band photometry. Objects with no rest-UV photometric data are fit with the dust-reddened quasar component only. 

In our SED model, the hot-dust blackbody component is described by a single free parameter. Its effective temperature is fixed at $\rm T_{bb} \simeq 1243\,K$, corresponding to the mean blackbody temperature derived from SED fits to blue SDSS quasars and consistent with the sublimation temperature of silicate dust \citep{article,2021MNRAS.501.3061T}. Consequently, the only parameter that must be constrained is the hot-dust ratio, $ L_{Dust}/L_{Disk}|_{2\mu m}$. For quasars at $1.5<z_{\rm sys}<2.7$, the \textit{W2} ($4.6\mu m$) filter provides sufficient sensitivity to constrain this ratio to within $\Delta L_{Dust}/L_{Disk}|_{2\mu m}= 0.25$, assuming the average \textit{W2} uncertainty of 0.028 mag. Similarly, at $3.2<z_{\rm sys}<4.0$, the \textit{W3} ($12\mu m$) filter constrains the hot-dust ratio to the same precision, assuming the average \textit{W3} uncertainty of 0.054 mag. However, for the seven HRQs with $2.7<z_{\rm sys}<3.2$, the hot-dust ratio may be poorly constrained, since the \textit{W2} filter no longer robustly probes the hot-dust emission while the \textit{W3} filter samples rest-frame wavelengths redward of $3\mu m$.

\begin{figure} 
\centering
\begin{tabular}{r}
 \includegraphics[scale=0.53, trim={0.32cm, 0.82cm, 0.0cm, 0.2cm}]{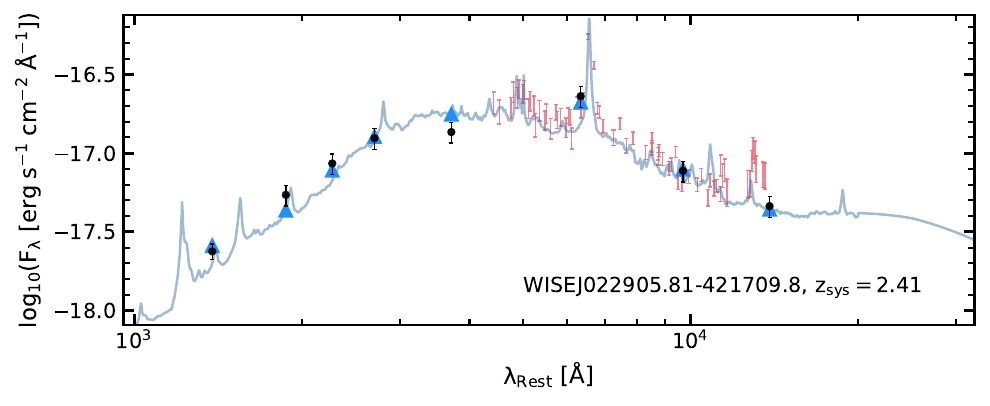} \\
 \includegraphics[scale=0.522, trim={0.32cm, 0.32cm, 0.0cm, 0.2cm}]{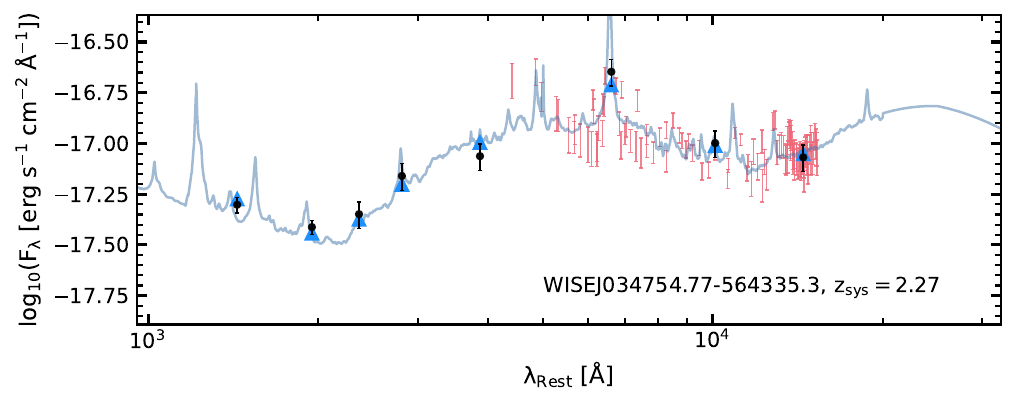} \\
 \end{tabular}
  \caption{The "best-fit" \textsc{qsogen} SEDs for two SPHERE\textsc{x}-confirmed HRQs, one where the inclusion of a scattered component was rejected (top) and one where the scattered component was confirmed (bottom). The broad-band photometric data from DELVE, UKIDSS/VHS and \textit{WISE} are indicated in black with their associated uncertainties. The best-fit SED models are shown as blue lines and triangles while the \spx spectra are overlaid in red.}
 \label{fig:Sed_cases}
\end{figure}

A total of 68 objects have sufficient wavelength coverage to robustly test for a UV excess. We impose the following criteria for the UV excess to be considered statistically significant: (i) the scattered component must contribute at least 50 per cent of the flux to the blue-most photometric band and (ii) the two-component SED model must represent an improvement over the single-component dust-reddened quasar SED at $>99$ per cent confidence, i.e. $\Delta \chi^2> 6.63$ \citep[See Table 1;][]{1976ApJ...210..642A}. An example of an HRQ whose UV excess has been rejected/confirmed is presented in Fig. \ref{fig:Sed_cases}.

\subsection{Leveraging \spx spectra for better constraints in the near-infrared}

The spectral coverage at rest-frame near-infrared wavelengths is limited when relying solely on broad-band \textit{WISE} photometry. The inclusion of \spx spectra enables significantly stronger constraints on the hot-dust properties of HRQs. Given the $6\arcsec$ spatial resolution of \spx \citep[][]{2014arXiv1412.4872D}, we restrict our analysis to sources whose nearest $K$-band neighbour lies at an angular separation greater than $6\arcsec$, thereby minimizing contamination. Hence, we are able to conduct additional SED fitting including \spx spectra for 70/76 HRQs, which is consistent with the $\sim10$ per cent contamination rate from interlopers reported in \citet{2026arXiv260310135D}.

For the fitting, we adopt Gaussian priors informed by the results of Section \ref{sec:sed_broadbands} and fit the dust-reddened quasar + scattered-light SED model to the \spx spectra of each source. Only spectral channels with signal-to-noise ratios $\mathrm{S/N} > 5$ are included in the fitting procedure. Owing to degeneracies among the free parameters \citep[see][for a full discussion]{2024MNRAS.533.2948S}, we perform a final round of SED fitting using the broad-band photometry alone, imposing strict Gaussian priors on the hot-dust amplitude. This final fitting step accounts for any variations in the inferred luminosity, line-of-sight dust extinction, and scattering fraction induced by adjustments to the hot-dust component, particularly for sources where the \spx spectral coverage does not extend sufficiently blueward to constrain the rest-frame UV emission. On average, this approach improves constraints on the hot-dust amplitude by a factor of seven.

A final summary of our SED analysis, in addition to example SED fits, are presented in Appendix \ref{App:Sed}. Table \ref{Tab:SED_Results} will also be made available as online supplementary material. We find that the median reduced chi squared statistic for the HRQ SED fitting is $\overline{\chi^2_\nu}|_{\rm Med} = 1.6$, broadly consistent with the performance of our SED model on the original HRQ sample (e.g. $\overline{\chi^2_\nu}|_{\rm Med} = 1.9$; \citealt{S26a}).

\section{A luminous control sample of blue quasars}  \label{sec:cQSO}

As shown in \citet[Figure 1;][]{S26a}, many objects in the original HRQ sample already exhibit substantially higher rest-frame optical continuum luminosities after correcting for dust, compared to blue SDSS quasars which have been fit using the same SED models (e.g. \citealt{2021MNRAS.501.3061T}, \citealt{2023MNRAS.523..646T} and \citealt{2023MNRAS.524.5497S}. This is because SDSS samples are incomplete at the highest quasar luminosities (log$_{10}\{\lambda$L$_{\lambda} (3000\mathring{A}) [\rm erg\;s^{-1}]\} \gtrsim 46.5$) due to an $i_{\rm AB} > 15$ magnitude limit, which was imposed to prevent saturation and cross-talk in the spectrograph \citep[][]{2006AJ....131.2766R,2010AJ....139.2360S}.  Since the selection presented in Section \ref{sec:HRQs-Selection} features a brighter $K_{\rm AB}$-band magnitude limit than previous work \citep[e.g.][]{2012MNRAS.427.2275B,2013MNRAS.429L..55B,2015MNRAS.447.3368B,2019MNRAS.487.2594T}, it is likely that our study will probe only the brightest HRQs. Consequently, we must construct a new control sample of blue quasars with luminosities comparable to the high-luminosity end of the HRQ distribution.

We begin with the bright ($G_{\rm AB}<20$ mag) Gaia-\textit{unWISE} Quasar Catalog \citep[Quaia;][]{2024ApJ...964...69S} - which contains $\sim750,000$ quasar candidates with Gaia $G$, $BP$, and $RP$ measurements, \textit{unWISE} $W1$ and $W2$ observations, and Gaia-estimated quasar classifier (QSOC) redshifts\footnote{The QSOC module estimates redshifts by using a chi-squared approach to compare BP/RP spectra for each quasar candidate to an SDSS composite spectrum, a more detailed discussion is presented in \citet{2024ApJ...964...69S} and the references within}. To determine a magnitude limit that ensures our blue quasar control sample has luminosities consistent with the HRQ sample, we use \textsc{qsogen} to simulate 10,000 blue quasar SEDs whose free-parameters are defined as follows; log$_{10}\{\lambda$L$_{\lambda} (3000\mathring{A}) [\rm erg\;s^{-1}]\}=46.5-47.5$\footnote{Consistent with the original HRQ sample \citet{S26a}.}, $\rm E(B-V) = -0.1-0.2$\footnote{Consistent with the range in line-of-sight dust-extinction observed amongst blue SDSS quasars \citep[][]{2021MNRAS.501.3061T}.} and $1.5\leq\rm z_{sys} \leq4.0$. Next, we compute the Gaia $G$-band magnitudes for each simulated SED and adopt a $G_{\rm AB}<18$ magnitude limit to ensure sufficiently bright sources, yielding an initial sample of 30,170 objects - 9,112 of which have redshifts $1.5\leq\rm z_{sys} \leq4.0$. 

To enable a full multi-wavelength SED analysis of the cosmic noon Quaia control sample, we cross-match the $G_{\rm AB}<18$ Quaia sample with DELVE-DR2 \citep[][]{2022ApJS..261...38D}, SDSS-DR16 \citep[][]{2020ApJS..249....3A}, UKIDSS-LAS \citep[][]{2007MNRAS.379.1599L}, UHS \citep[][]{2018MNRAS.473.5113D}, and VHS \citep[][]{2013Msngr.154...35M}, following the procedure outlined in Section \ref{sec:HRQs-Selection}. This yields 5,329 sources that have coverage in the optical (either from DELVE or SDSS) and NIR (from either VHS, UHS or UKIDSS-LAS). We then match to R90 \citep[e.g.][]{Assef18:WISE} to gain $W3$ coverage and ensure that the rest-NIR SED is constrained across all redshifts. This final sample contains 4,711 sources, 1,618 of which are in SDSS DR16Q and 3 of which overlap with the \textit{WISE}-SDSS selected hyper-luminous quasar sample \citep[WISSH;][]{2012ApJ...761..184W}. When fitting the broad-band photometry via \textsc{qsogen} we use the following free parameters; log$_{10}\{\lambda$L$_{\lambda} (3000\mathring{A}) [\rm erg\;s^{-1}]\}$, $\rm E(B-V)$ and $L_{Dust}/L_{Disk}|_{2\mu m}$. The median reduced chi squared statistic for the Quaia SED fits is $\overline{\chi^2_\nu}|_{\rm Med} = 1.4$. 

The accuracy of the Gaia-estimated QSOC redshifts increases with flux, i.e. $|\Delta z_{\rm sys}|<0.1$ for 80 per cent of quasars with $G_{\rm AB}<18.5$ \citep[][]{2024ApJ...964...69S}. To better understand the quality of the Quaia QSOC redshifts in our blue control sample, we obtain \spx spectrophotometry for the 500 most luminous Quaia quasars (i.e. the regime in which SDSS was not used to verify the Gaia QSOC redshifts) and perform a cross-correlation analysis, following the same procedure outlined in Section \ref{sec:HRQs-Confirmation}. The average discrepancy between redshift estimates is $\langle\Delta z_{\rm sys}\rangle=0.08$, with $|\Delta z_{\rm sys}|<0.2$, or equivalently $|\Delta z_{\rm sys}/(1+z_{\rm sys})|\lesssim0.1$, for 95 per cent of the sample. Hence, for the purposes of broad-band SED fitting we assume that the Gaia-estimated QSOC redshifts are reliable. The result of SED-fitting to the bright Quaia control sample, are consistently measured quasar luminosities and hot-dust amplitudes for the HRQs and the blue quasars. 

\section{Connecting the different populations of luminous quasars at cosmic noon} \label{sec:demo}

In this section, we place our SED fitting results in the broader context, taking advantage of the substantially expanded sample constructed via our \spx analysis. We compare the HRQ population to the luminous blue Quaia sample (Section \ref{sec:cQSO}) as well as Hot Dust-Obscured Galaxies \citep[e.g. Hot DOGs;][]{Eisenhardt_2012,2015ApJ...804...27A,2020ApJ...897..112A} - another class of very luminous obscured AGN, but with typically higher continuum extinctions than the HRQs. 

\subsection{A unified sample of hyper-luminous HRQs}

\begin{figure}
\centering
\includegraphics[scale=0.73, clip]{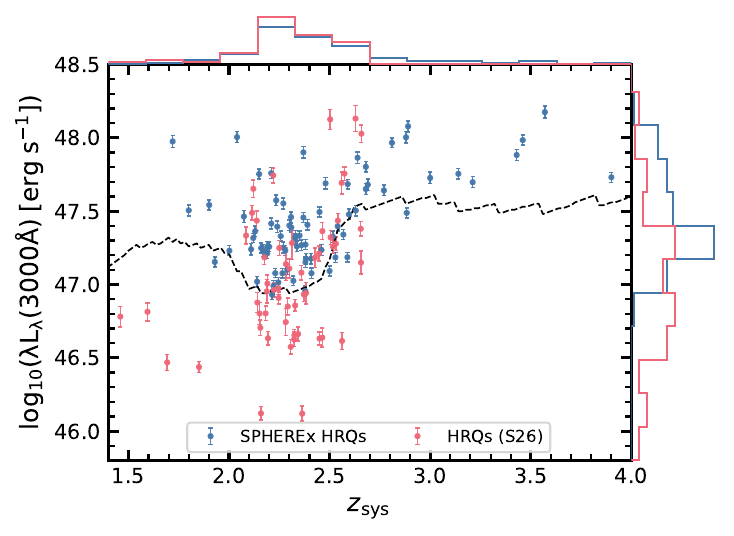} 
 \caption{The dust-corrected 3000\AA\, continuum luminosity of the quasar vs redshift. The marginalised distributions are presented as histograms on their corresponding axis. \spx HRQs are shown in blue and the original $\rm z_{sys}\geq1.5$ HRQ sample (\citealt{S26a}) is shown in red. The \spx HRQs are biased to higher luminosities (due to our brighter $K$-band selection) and extend to higher redshifts. The black dashed line represents the minimum luminosity required for a quasar to meet the $K_{\rm AB} < 18$ magnitude limit at the minimum dust-extinction that satisfies the $(J-K)_{\rm AB} > 1.6$ colour selection.} 
 \label{fig:zsys_vs_L3000}
\end{figure}

\autoref{fig:zsys_vs_L3000} illustrates the redshift and dust-corrected 3000\AA\, quasar continuum luminosity distributions for both the original and \spx HRQ populations. The \spx HRQs are generally consistent with the original HRQ samples, however, they tend towards higher luminosities and extend to higher redshifts. The luminosity bias between the two HRQ samples owes to the brighter $K$-band magnitude selection imposed on the \spx HRQ sample (see Section \ref{sec:HRQs-Selection}). The larger redshifts stem from the broader and more complete wavelength coverage of \spx \citep[e.g. $0.75-5.0\mu m$;][]{2014arXiv1412.4872D,2025arXiv251102985B} compared to VLT-SINFONI \citep[e.g. $1.10-2.45\mu m$;][]{2003SPIE.4841.1548E} and Gemini-GNIRS, which are both ground-based and are therefore affected by atmospheric band gaps \citep[][]{article_gemini}.

In Fig. \ref{fig:zsys_vs_L3000}, we also see that lower 3000\AA\, continuum luminosities are required to satisfy our HRQ selection at $2.0\lesssim\rm z_{sys}\lesssim2.5$. This is due to the presence of the broad H$\alpha$ emission in the $K$-band at these redshifts, which has the effect of boosting the $K$-band flux. This results in a significantly larger number of HRQs with redshifts of $\rm 2.0\lesssim z_{sys}\lesssim2.5$ in both the \spx and original samples \citep[e.g.][]{2012MNRAS.427.2275B,2013MNRAS.429L..55B,2015MNRAS.447.3368B,2019MNRAS.487.2594T}. In addition, the newly confirmed \spx HRQs have line-of-sight dust extinctions, $\rm E(B-V)$, consistent with the original HRQs, with a median $\rm E(B-V)|_{\rm Med} = 0.8$ in both $\rm z_{sys}\geq1.5$ samples (\citealt{S26a}). Taken together, we conclude that the \spx sample is consistent with previous work and therefore comprises true HRQs. Henceforth, we treat the \spx and original HRQ samples as a single unified population comprised of 127 objects at  $\rm z_{sys}\gtrsim1.5$. 

\begin{figure}
\centering
\includegraphics[scale=0.73, clip]{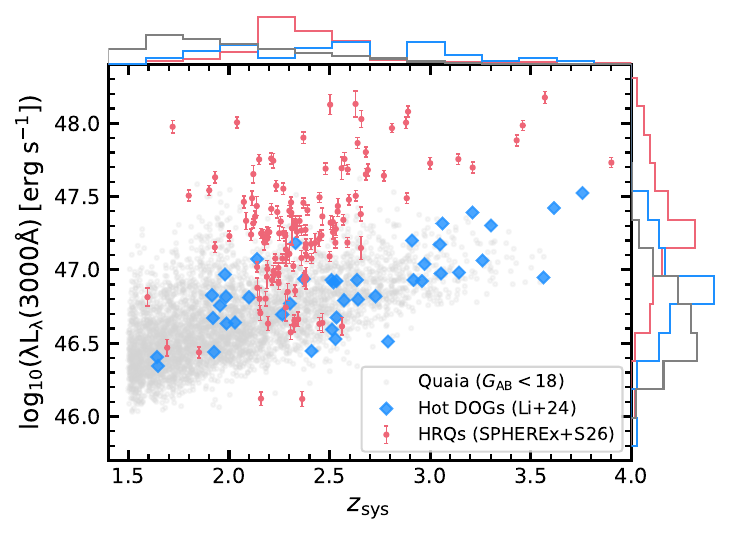} 
 \caption{The dust-corrected 3000\AA\, continuum luminosity vs redshift. The combined HRQ sample from this paper and \citet{S26a} is shown in red and our Quaia $G_{\rm AB}<18$ mag control quasars \citep[e.g.][]{2024ApJ...964...69S} are shown in grey. Blue Hot DOGs \citep[e.g.][]{2024ApJ...971...40L} are represented by blue diamonds. HRQs tend towards higher luminosities than both blue quasars and Blue Hot DOGs.} 
 \label{fig:zsys_vs_L3000_2}
\end{figure}

In Fig. \ref{fig:zsys_vs_L3000_2}, we find that HRQs extend to higher luminosities than the Quaia control sample, although there is significant overlap at more modest HRQ luminosities. While we expect that the \spx HRQs probe the most extreme end of the HRQ luminosity distribution, there is a clear deficit of blue Quaia counterparts at log$_{10}\{\lambda$L$_{\lambda} (3000\mathring{A}) [\rm erg\;s^{-1}]\} > 47.5$. This suggests that the number densities of red quasars may exceed that of their blue counterparts in the highest luminosity regime. This result is consistent with \citet{2015MNRAS.447.3368B}, although a more comprehensive investigation is deferred to future work; where we will push to fainter, more complete and much larger statistical samples with \spx (e.g. Stepney et al., in prep.). In addition to blue quasars, the optical luminosities of our brightest HRQs appear to exceed that of the Hot DOG population\footnote{To calculate the 3000\AA\, continuum luminosities of the Hot DOG sample, we apply the \citet{2019MNRAS.488.5185N} luminosity-dependent corrections to the 5100\AA\, continuum luminosities calculated in \citet{2024ApJ...971...40L}.}. While this result may, at least in part, be as a result of differences between the SED modelling methodologies, it is consistent with reports that HRQs generally have higher accretion rates than their Hot DOG counterparts \citep[e.g.][]{2024ApJ...971...40L,S26a} and may suggest that Hot DOGs represent an earlier stage in the black hole feeding process.

\subsection{Depleted hot-dust reservoirs as evidence for feedback} \label{sec:bbnorm}

A key result from \citet{S26a} is that HRQs represent a phase of black hole accretion characterised by a substantial depletion of the hot-dust reservoir. In this section, we test whether this result persists with our now significantly larger sample of HRQs. In Section \ref{sec:HRQs-Selection}, we re-defined the HRQ selection by relaxing the mid-infrared colour threshold to $(W1-W2)_{\rm{Vega}}>0.5$. In Appendix \ref{app:Hot-dust_Complete}, we confirm that the modification to our selection criteria now enables a complete study of HRQ hot-dust properties. We measure an average hot-dust amplitude of $\langle\rm L_{Dust}/L_{Disk}|_{2\mu m} \rangle_{HRQ} = 1.3\pm0.9$ for all HRQs - significantly lower than the Quaia average of $\langle\rm L_{Dust}/L_{Disk}|_{2\mu m} \rangle_{Quaia} = 4.3\pm1.3$. Furthermore, by using the definition for 'hot-dust poorness' invoked by \citet{2010Natur.464..380J} and \citet{2013ApJ...779..104J} in their respective studies of blue quasars, we find that 21/127 HRQs (17 per cent) are formally "hot-dust poor". This represents a significantly higher fraction than the blue Quaia sample, with just 15/4711 (0.3 per cent) of blue quasars meeting the \citet{2010Natur.464..380J} condition. Hence, with a larger and more complete sample of HRQs, our results are consistent with the conclusions presented in \citet{S26a}.

\begin{figure*} 
\centering
\begin{tabular}{c c}
     \includegraphics[scale=0.66, trim={0.3cm, 0.22cm, 0.0cm, 0.2cm}]{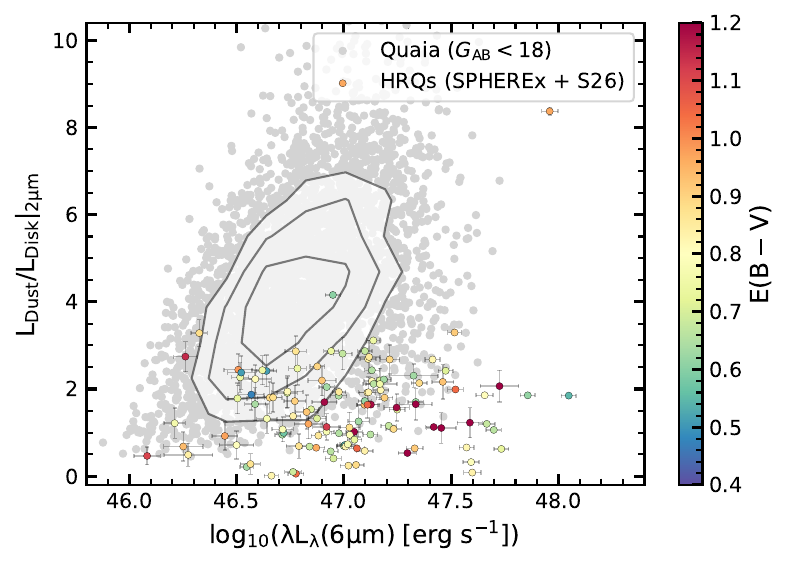} &  \includegraphics[scale=0.64, trim={0cm, 0cm, 0cm, 0cm}, clip]{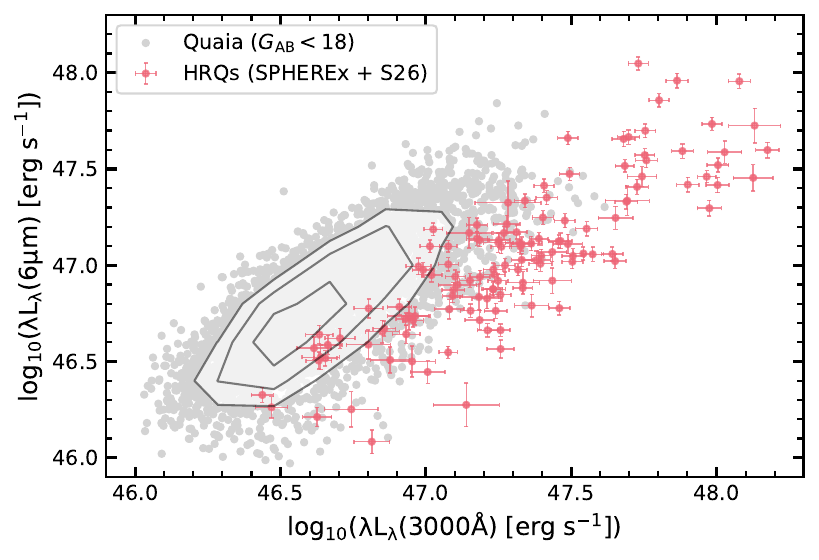} \\
\end{tabular}
 
  \caption{The 2$\mu$m hot-dust emission amplitude as a function of 6$\mu m$ continuum luminosity (left) for luminous Quaia quasars (grey) and HRQs (coloured circles). Density contours encircle 25, 50 and 68 per cent of the Quaia sample respectively. The hot-dust amplitudes of HRQs are systematically lower than Quaia quasars at a given MIR luminosity and show no clear dependence on the line-of-sight dust extinction, $\rm E(B-V)$. We also present the 6$\mu m$ continuum luminosity as a function of the 3000\AA\, continuum luminosity (right) for Quaia quasars (grey) and HRQs (red). HRQs are systematically off-set to lower MIR luminosities at a given accretion disk luminosity when compared to Quaia quasars and do not represent an extension of the Quaia distributions to higher luminosities. Taken together, this shows that HRQs are poor in both 2$\mu m$ and 6$\mu m$ dust relative to Quaia quasars of similar luminosity.}
 \label{fig:l6um_v_hotdust}
\end{figure*}

\autoref{fig:l6um_v_hotdust} (left) shows how the hot-dust amplitudes (when possible to constrain) of both blue Quaia quasars (Section \ref{sec:cQSO}) and HRQs vary as a function of $6\mu m$ continuum luminosity, which probes warm dust components with effective temperatures of $T_{\rm Eff}\simeq450$K. The $6\mu m$ luminosities are calculated by conducting a log-linear interpolation between the \textit{W3} ($12\mu m$) and \textit{W4} ($22\mu m$) photometry of each source. HRQs have systematically lower hot-dust amplitudes compared to their blue Quaia counterparts at any given MIR luminosity. In Fig. \ref{fig:l6um_v_hotdust} (right), we show that HRQs are systematically off-set to lower MIR luminosities than Quaia quasars of similar optical continuum luminosity, and do not simply represent a continuation of the blue Quaia distribution to the extreme luminosity regime. Taken together, these results demonstrate that the weaker hot and warm dust emission observed amongst HRQs relative to blue Quaia quasars is not a luminosity bias \citep[e.g. the 'receding torus'][see also the references within]{2013ApJ...772...26A} but instead suggests that HRQs represent an evolutionary phase in which the torus-scale dust has been depleted.

\citet{S26a} proposed that the depletion of the torus-scale reservoir observed in HRQs is most likely driven by a blow-out phase \citep[e.g.][]{2005ApJ...630..705H,Hopkins_2008}, in which strong radiative feedback ejects or destroys the hottest inner torus component. In this scenario, the remaining dust may be distributed in a much more extended polar rather than equatorial configuration \citep[e.g.][]{Hoenig:19, 2024MNRAS.533.2948S}. This interpretation is also supported by the lack of correlation between the line-of-sight dust extinction, $\rm E(B-V)$, and the hot-dust amplitude in Fig. \ref{fig:l6um_v_hotdust} (left), which suggests that the obscuring medium is not primarily distributed on nuclear scales, but rather, extended to galaxy-wide scales instead. 

\subsection{The origin of the UV excess in luminous, obscured quasars} \label{sec:uvXS}

In this section, we limit our analysis to the 117 HRQs for which it was possible to robustly fit our two-component SED model. In the \spx HRQ sample, we detect a UV excess in 50/66 or equivalently 76 per cent of objects - consistent with the 82 per cent reported in \citet{S26a}. The total number of HRQs with blue rest-UV photometric colours is therefore 92/117, or equivalently 79 per cent - much higher than reported in the Hot DOG population \citep[e.g. 10-25 per cent;][]{2016ApJ...819..111A,2024ApJ...971...40L}. 

\begin{figure} 
\centering
 \includegraphics[scale=0.64, trim={0cm, 0.3cm, 0cm, 0cm}, clip]{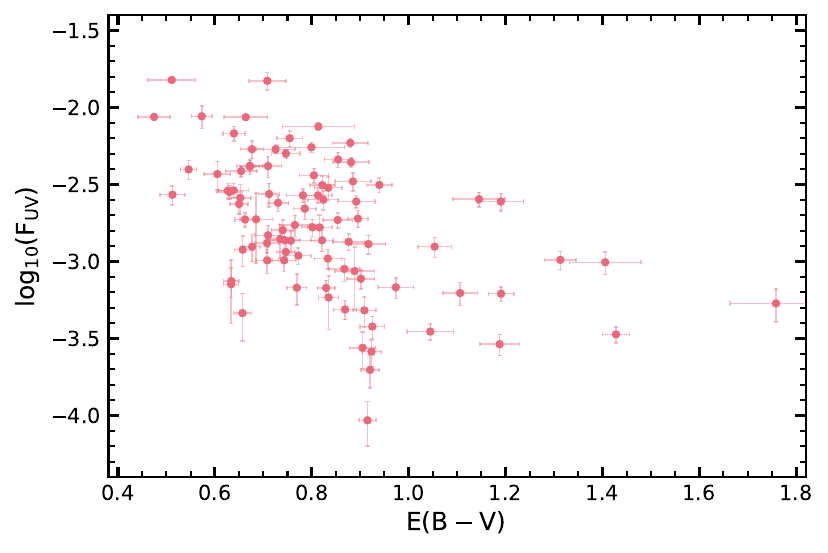} \\
  \caption{Logarithm of the scattering fraction as a function of the line-of-sight dust-extinction. We observe an anti-correlation (e.g. $R_{PCC} = -0.41$ and a $p$-value = $10^{-4}$), which suggests that the UV excess suffers modest dust-extinction in HRQs.}
 \label{fig:ebvFuv}
\end{figure}

The average scattering fraction of HRQs is $\langle F_{\rm{UV}}\rangle_{\rm HRQ} = 0.26$ per cent, an order of magnitude lower than that observed among blue Hot DOGs \citep[e.g. $\langle F_{\rm{UV}}\rangle_{\rm BHDs} = 2.62$ per cent;][]{2024ApJ...971...40L}. In \citet{S26a}, the comparatively modest scattering fractions observed in HRQs were attributed to selection effects driven by the $(i-K)_{\rm AB} > 2.5$ colour criterion employed to select the original HRQ sample \citep[e.g.][]{2012MNRAS.427.2275B,2013MNRAS.429L..55B,2015MNRAS.447.3368B,2019MNRAS.487.2594T}, which excludes objects with blue rest-UV/optical continua. However, as discussed in Section \ref{sec:HRQs-Selection}, the rest-frame optical to near-infrared colour-selection was not applied when defining the \spx sample, and hence the $(i-K)_{\rm AB}$ colour-selection is not responsible for the faint UV excesses observed in the HRQ population. 

One interpretation of the weaker scattering fractions observed in HRQs is that the scattered component in the HRQ model may itself suffer from dust-extinction. For example, the Hot DOG SED model has the flexibility to produce scattered components that suffer moderate dust-extinctions \citep[e.g. $\rm E(B-V)_{Scattered}=0-0.3$;][]{2015ApJ...804...27A,2020ApJ...897..112A,2024ApJ...971...40L} which are generally favoured by the SED fitting\footnote{Although the additional complexity of the Hot DOG SED model could yield improved fits for some HRQs, many objects lack sufficient optical photometric coverage to constrain such a model. For consistency, we therefore adopt the simpler description of the scattered component for all HRQs.}. The anti-correlation observed between the line-of-sight dust extinction and scattering fraction of HRQs (e.g. Fig. \ref{fig:ebvFuv}; $R_{PCC} = -0.41$ and a $p$-value = $10^{-4}$) is consistent with this behaviour, and hence, the HRQ scattering fractions may be systematically under-estimated. However, the lack of HRQs with scattering fractions consistent with the Hot DOG population could also imply that HRQs host intrinsically fainter rest-frame UV excesses, perhaps as a result of differing physical origins for the rest-UV continuum, e.g. star-formation in the host galaxy vs scattered quasar light \citep[][]{2020ApJ...897..112A,2024MNRAS.533.2948S}, differing scattering medium optical depths or different viewing angles. 

\begin{figure} 
\centering
 \includegraphics[scale=0.73, trim={0cm, 0.3cm, 0cm, 0.1cm}, clip]{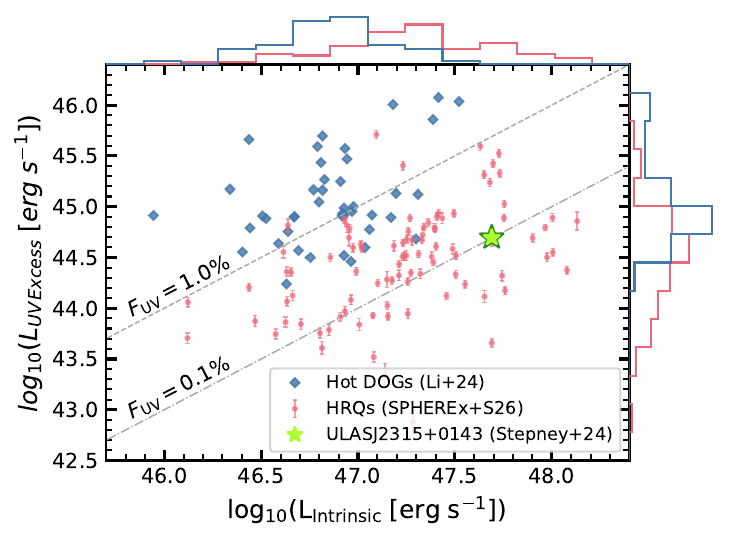} \\
  \caption{The dust-corrected 3000\AA\, continuum luminosity of the UV excess vs the intrinsic quasar component. The combined HRQ sample is shown in red and Hot DOGs \citep[e.g.][]{2024ApJ...971...40L} are represented by blue diamonds. Fixed scattering fractions are illustrated by the grey dashed lines. Hot DOGs tend towards higher luminosity scattered components than HRQs. The green star represents the HRQ ULASJ2315+0143, whose VLT-XShooter spectrum has both scattered light and host galaxy signatures in the rest-UV \citep[][]{2024MNRAS.533.2948S}.}
 \label{fig:l_fuv}
\end{figure}

In Fig. \ref{fig:l_fuv}, we see that the luminosity of the UV excess is lower in HRQs than in Hot DOGs, despite the intrinsic AGN itself having significantly higher continuum luminosities after correcting for dust. A potential explanation for this result is the post blow-out phase depletion of the hot-dust reservoir observed in HRQs, which may reduce the scattering efficiency of HRQs vs Hot DOGs. In Hot DOGs, the scattering medium has high column densities and is primarily composed of hot, graphite-rich dust \citep[][]{2022ApJ...934..101A,2025A&A...702A.124A}. Given also their higher X-ray-derived gas column densities compared to HRQs \citep[e.g. $\rm N_H \sim 10^{23-24} \,cm^{-2}$ vs $\rm N_H \sim 10^{22} \,cm^{-2}$;][]{2015ApJ...804...27A,2020MNRAS.495.2652L}, Hot DOGs likely represent an earlier stage in the evolutionary sequence in which merger-driven black hole growth is in an earlier phase than in HRQs. 

Conversely, in HRQs, enhanced feedback processes powered by super-Eddington accretion may destroy or disperse the hot ($T_{\rm Eff}>1000$K) dust grains which comprise the scattering medium, revealing sight-lines in which star-forming regions in the quasar host can dominate weakly scattered quasar emission. This is particularly true of HRQs with lower luminosity UV excesses with respect to Hot DOGs; such as ULASJ2315+0143 (Fig. \ref{fig:l_fuv}; green star), spectral analysis of which shows evidence for the rest-UV emission to contain both host galaxy and scattered quasar light signatures. The rest-UV continuum luminosity of ULASJ2315+0143 is consistent with a UV star-formation rate of $\rm SFR_{UV}$ $\simeq 100\, M_\odot \,yr^{-1}$ \citep[][]{2024MNRAS.533.2948S}, although ALMA observations suggest that the unobscured SFR of HRQs can exceed $\simeq1000\, M_\odot \,yr^{-1}$ \citep[][]{2021MNRAS.503.5583B}. In this scenario, HRQs represent a later stage in the evolutionary sequence than Hot DOGs, with HRQs exhibiting a lower luminosity UV excess than ULASJ2315+0143 most likely having host-dominated UV continua.

Crucially, should the scattering mechanism in HRQs and Hot DOGs be the same, with HRQs simply representing a phase where the column density of the scattering medium has been reduced, the polarisation angle of the scattered light in these populations would remain consistent. Therefore, future work should employ image polarimetry techniques \citep[e.g.][]{2022ApJ...934..101A,2025A&A...702A.124A} to verify the composition of the scattering medium as well as the polarisation angle of any scattered light in HRQs. Furthermore, spectropolarimetry studies in the most luminous sources will be viable, enabling direct comparisons to other dust obscured quasar populations such as ERQs \citep[e.g.][]{2018MNRAS.479.4936A, 2023MNRAS.525.2716Z}.

\section{Conclusions} \label{sec:conc}

Using \spx spectrophotometry, we have confirmed a new sample of 76 bright ($K_{AB} <18$ mag) heavily reddened quasars (HRQs) with redshifts $1.5\lesssim z_{\rm sys}\lesssim3.9$ and line-of-sight dust-extinctions $0.4<\rm E(B-V)<1.6$ (A$_{\rm{V}}\sim1-5$ mag). To infer the rest-UV to near-infrared spectral energy distribution (SED) properties of the new HRQ sample, we conduct SED fitting on the 76 SPHERE\textsc{x}-confirmed HRQs utilising multi-wavelength surveys spanning the $ugriz-YJHK-W1W2$ filters - which correspond to a rest-frame wavelength coverage of 0.1-3.0$\mu m$ at $z_{\rm sys}\simeq2$. We show that in most cases a two-component SED model featuring a dust-attenuated quasar component and an additional blue scattered light component provides a good fit to the photometric data - consistent with previous work on HRQs (e.g. \citealt{S26a}). Our main findings are as follows;

\begin{itemize}
    \item Using cross-correlation analysis, we show that the spectral resolution of \spx is sufficient to robustly confirm and estimate the redshifts of luminous, dust-obscured, broad-line quasars. We confirm 76 new HRQs with redshifts $1.5\lesssim z_{\rm sys}\lesssim3.9$ including the first 7 HRQs above $z=3$. Given that \spx is space-based and features a broad wavelength coverage \citep[e.g. $\lambda \simeq 0.75-5.0 \mu m$;][]{2014arXiv1412.4872D,2025arXiv251102985B}, a larger dynamic range in redshift can be probed via \spx than with ground-based facilities such as VLT-SINFONI and Gemini-GNIRS with the highest redshift HRQ at $z=3.9$.

    \item We find that having accounted for selection effects, the SED properties of the \spx HRQ sample is consistent with HRQ samples studied previously \citep[e.g.][]{2012MNRAS.427.2275B,2015MNRAS.447.3368B,2019MNRAS.487.2594T,S26a}. Hence, the SPHERE\textsc{x}-confirmed sample represents 76 new bonafide HRQs, increasing the total sample-size to 127 objects spanning redshifts $1.5\lesssim z_{\rm sys}\lesssim3.9$.

    \item Having relaxed the original HRQ mid-infrared colour-selection for the \spx sample to $(W1-W2)_{\rm Vega}>0.5$ mag, we demonstrate that the new sample is complete in terms of their hot-dust properties. We conclusively demonstrate that HRQs are hot-dust poor with respect to blue Gaia-selected quasars of similar luminosity and redshift - $\langle \rm L_{Dust}/L_{Disk}|_{2\mu m} \rangle = 1.3\pm0.9$ versus $4.3\pm1.3$ for HRQs and blue quasars, respectively and show that this result cannot be explained by a luminosity bias \citep[e.g. the 'receding torus'][]{2013ApJ...772...26A} Furthermore, we find that the total fraction of HRQs that meet the \citet{2010Natur.464..380J} condition for 'hot-dust poorness' is 17 per cent, compared to only 0.3 per cent of blue quasars satisfying the same criteria. 

    \item We detect a rest-UV excess in 50/66 \spx HRQs with photometric data covering the rest-UV. The excess emission can be modelled using a scattered AGN light component, however, contributions to the UV continuum from star-formation in the quasar host galaxy cannot be ruled out. The proportion of HRQs with a UV excess in the \spx sample is consistent with previous work, but significantly higher than observed amongst HRQs \citep[e.g.][]{2016ApJ...819..111A,2024ApJ...971...40L}. The average scattering fraction for all 117 HRQs currently known with sufficient photometric coverage in the rest-UV is $\langle F_{\rm{UV}}\rangle_{\rm HRQs} = 0.26$ per cent.  

    \item The scattering fractions in HRQs are low compared to Hot DOGs. This can partially be explained by selection effects. The scattered component used to fit Hot DOG SEDs is free to suffer modest dust extinction, whereas, in HRQs the scattered component is completely unobscured. However, despite having lower intrinsic AGN luminosities, the scattered luminosity of Hot DOGs still exceeds that of an average HRQ. This suggests a heterogeneous origin of the UV excess in HRQs, where a lower column-density of obscuring material very close to the black hole potentially results in unobscured sight-lines to star-forming regions in the quasar host galaxy. Thus the host galaxy may dominate the UV excess in some HRQs relative to the scattered light. 

\end{itemize}

Overall, this work demonstrates that \spx spectra are suitable for rapid and efficient confirmation of large numbers of high-redshift, dust-obscured, broad-line quasars enabling their properties to be put in context with other luminous quasar populations. In future work we will expand the samples even further, exploiting the all-sky capabilities of \spx to understand the number densities and multi-wavelength properties of the ``hidden monsters" that appear to dominate black hole growth at cosmic noon.

\section*{Data Availability}

Table \ref{Tab:SED_Results}, as well as the photometric data used in this study, are only available in electronic form at the CDS via anonymous ftp to cdsarc.u-strasbg.fr (130.79.128.5) or via http://cdsweb.u-strasbg.fr/cgi-bin/qcat?J/A+A/.

\begin{acknowledgements}

We gratefully acknowledge funding from ANID CATA BASAL FB210003 (MS, FEB, RJA). MB acknowledges funding from The Royal Society. FEB acknowledges funding from FONDECYT Regular 1241005. RJA acknowledges funding from FONDECYT Regular 1231718. 

The UHS is a partnership between the UK STFC, The University of Hawaii, The University of Arizona, Lockheed Martin and NASA. This research uses services or data provided by the Astro Data Lab, which is part of the Community Science and Data Center (CSDC) Program of NSF NOIRLab. NOIRLab is operated by the Association of Universities for Research in Astronomy (AURA), Inc. under a cooperative agreement with the U.S. National Science Foundation. This publication also makes use of data products from the Spectro-Photometer for the History of the Universe, Epoch of Reionization and Ices Explorer (SPHERE\textsc{x}), which is a joint project of the Jet Propulsion Laboratory and the California Institute of Technology, and is funded by the National Aeronautics and Space Administration. Finally, the authors would like to give special thanks to Timothy Y. Brooke and the rest of the IPAC Helpdesk team for their support with this project.

\end{acknowledgements}

\bibliographystyle{bibtex/aa} 
\bibliography{bibtex/biblio} 

@article{S26a,
    author = {Stepney, Matthew and Banerji, Manda and Tang, Shenli and Temple, Matthew J and Hewett, Paul C},
    title = {The rest-ultraviolet to infrared spectral energy distributions of heavily reddened quasars are ‘V-shaped’ and hot-dust poor},
    journal = {Monthly Notices of the Royal Astronomical Society},
    volume = {546},
    number = {4},
    pages = {stag191},
    year = {2026},
    month = {01},
    issn = {0035-8711},
    doi = {10.1093/mnras/stag191},
    url = {https://doi.org/10.1093/mnras/stag191},
    eprint = {https://academic.oup.com/mnras/article-pdf/546/4/stag191/66590290/stag191.pdf},
}

@ARTICLE{2013ApJ...772...26A,
       author = {{Assef}, R.~J. and {Stern}, D. and {Kochanek}, C.~S. and {Blain}, A.~W. and {Brodwin}, M. and {Brown}, M.~J.~I. and {Donoso}, E. and {Eisenhardt}, P.~R.~M. and {Jannuzi}, B.~T. and {Jarrett}, T.~H. and {Stanford}, S.~A. and {Tsai}, C.-W. and {Wu}, J. and {Yan}, L.},
        title = "{Mid-infrared Selection of Active Galactic Nuclei with the Wide-field Infrared Survey Explorer. II. Properties of WISE-selected Active Galactic Nuclei in the NDWFS Bo{\"o}tes Field}",
      journal = {\apj},
         year = 2013,
        month = jul,
       volume = {772},
       number = {1},
          eid = {26},
        pages = {26},
          doi = {10.1088/0004-637X/772/1/26},
archivePrefix = {arXiv},
       eprint = {1209.6055},
 primaryClass = {astro-ph.CO},
       adsurl = {https://ui.adsabs.harvard.edu/abs/2013ApJ...772...26A}
}

@ARTICLE{2012ApJ...756...96W,
       author = {{Wu}, Jingwen and {Tsai}, Chao-Wei and {Sayers}, Jack and {Benford}, Dominic and {Bridge}, Carrie and {Blain}, Andrew and {Eisenhardt}, Peter R.~M. and {Stern}, Daniel and {Petty}, Sara and {Assef}, Roberto and {Bussmann}, Shane and {Comerford}, Julia M. and {Cutri}, Roc and {Evans}, II, Neal J. and {Griffith}, Roger and {Jarrett}, Thomas and {Lake}, Sean and {Lonsdale}, Carol and {Rho}, Jeonghee and {Stanford}, S. Adam and {Weiner}, Benjamin and {Wright}, Edward L. and {Yan}, Lin},
        title = "{Submillimeter Follow-up of WISE-selected Hyperluminous Galaxies}",
      journal = {\apj},
         year = 2012,
        month = sep,
       volume = {756},
       number = {1},
          eid = {96},
        pages = {96},
          doi = {10.1088/0004-637X/756/1/96},
archivePrefix = {arXiv},
       eprint = {1208.5518},
 primaryClass = {astro-ph.CO},
       adsurl = {https://ui.adsabs.harvard.edu/abs/2012ApJ...756...96W}
}

@ARTICLE{2020ApJ...888..110J,
       author = {{Jun}, Hyunsung D. and {Assef}, Roberto J. and {Bauer}, Franz E. and {Blain}, Andrew and {D{\'\i}az-Santos}, Tanio and {Eisenhardt}, Peter R.~M. and {Stern}, Daniel and {Tsai}, Chao-Wei and {Wright}, L., Edward and {Wu}, Jingwen},
        title = "{Spectral Classification and Ionized Gas Outflows in z {\ensuremath{\sim}} 2 WISE-selected Hot Dust-obscured Galaxies}",
      journal = {\apj},
         year = 2020,
        month = jan,
       volume = {888},
       number = {2},
          eid = {110},
        pages = {110},
          doi = {10.3847/1538-4357/ab5e7b},
archivePrefix = {arXiv},
       eprint = {1911.09828},
 primaryClass = {astro-ph.GA},
       adsurl = {https://ui.adsabs.harvard.edu/abs/2020ApJ...888..110J}
}

@ARTICLE{2006AJ....131...84V,
       author = {{Vanden Berk}, Daniel E. and {Shen}, Jiajian and {Yip}, Ching-Wa and {Schneider}, Donald P. and {Connolly}, Andrew J. and {Burton}, Ross E. and {Jester}, Sebastian and {Hall}, Patrick B. and {Szalay}, Alex S. and {Brinkmann}, John},
        title = "{Spectral Decomposition of Broad-Line AGNs and Host Galaxies}",
      journal = {\aj},
         year = 2006,
        month = jan,
       volume = {131},
       number = {1},
        pages = {84-99},
          doi = {10.1086/497973},
archivePrefix = {arXiv},
       eprint = {astro-ph/0509332},
 primaryClass = {astro-ph},
       adsurl = {https://ui.adsabs.harvard.edu/abs/2006AJ....131...84V}
}

@ARTICLE{2023ApJ...958..162L,
       author = {{Li}, Guodong and {Tsai}, Chao-Wei and {Stern}, Daniel and {Wu}, Jingwen and {Assef}, Roberto J. and {Blain}, Andrew W. and {D{\'\i}az-Santos}, Tanio and {Eisenhardt}, Peter R.~M. and {Griffith}, Roger L. and {Jarrett}, Thomas H. and {Jun}, Hyunsung D. and {Lake}, Sean E. and {Saade}, M. Lynne},
        title = "{Discovery of a Low-redshift Hot Dust-obscured Galaxy}",
      journal = {\apj},
         year = 2023,
        month = dec,
       volume = {958},
       number = {2},
          eid = {162},
        pages = {162},
          doi = {10.3847/1538-4357/ace25b},
archivePrefix = {arXiv},
       eprint = {2305.13739},
 primaryClass = {astro-ph.GA},
       adsurl = {https://ui.adsabs.harvard.edu/abs/2023ApJ...958..162L}
}

@dataset{2021yCat.2367....0M,
       author = {{McMahon}, R.~G. and {Banerji}, M. and {Gonzalez}, E. and {Koposov}, S.~E. and {Bejar}, V.~J. and {Lodieu}, N. and {Rebolo}, R. and {VHS Collaboration}},
        title = "{VizieR Online Data Catalog: The VISTA Hemisphere Survey (VHS) catalog DR5 (McMahon+, 2020)}",
 howpublished = {VizieR On-line Data Catalog: II/367.  Originally published in: 2013Msngr.154...35M},
         year = 2021,
        month = jan,
          eid = {II/367},
       adsurl = {https://ui.adsabs.harvard.edu/abs/2021yCat.2367....0M}
}

@ARTICLE{2020ApJ...905...16F,
       author = {{Finnerty}, Luke and {Larson}, Kirsten and {Soifer}, B.~T. and {Armus}, Lee and {Matthews}, Keith and {Jun}, Hyunsung D. and {Moon}, Dae-Sik and {Melbourne}, Jason and {Gomez}, Percy and {Tsai}, Chao-Wei and {D{\'\i}az-Santos}, Tanio and {Eisenhardt}, Peter and {Cushing}, Michael},
        title = "{Fast Outflows in Hot Dust-obscured Galaxies Detected with Keck/NIRES}",
      journal = {\apj},
         year = 2020,
        month = dec,
       volume = {905},
       number = {1},
          eid = {16},
        pages = {16},
          doi = {10.3847/1538-4357/abc3bf},
archivePrefix = {arXiv},
       eprint = {2010.10641},
 primaryClass = {astro-ph.GA},
       adsurl = {https://ui.adsabs.harvard.edu/abs/2020ApJ...905...16F}
}

@ARTICLE{2025ApJ...989..230V,
       author = {{Vayner}, Andrey and {D{\'\i}az-Santos}, Tanio and {Eisenhardt}, Peter R.~M. and {Stern}, Daniel and {Armus}, Lee and {Angl{\'e}s-Alc{\'a}zar}, Daniel and {Assef}, Roberto J. and {Fern{\'a}ndez Aranda}, Rom{\'a}n and {Blain}, Andrew W. and {Jun}, Hyunsung D. and {Tsai}, Chao-Wei and {Roy}, Niranjan Chandra and {Brisbin}, Drew and {Ferkinhoff}, Carl D. and {Aravena}, Manuel and {Gonz{\'a}lez-L{\'o}pez}, Jorge and {Li}, Guodong and {Liao}, Mai and {Shobhana}, Devika and {Wu}, Jingwen and {Zewdie}, Dejene},
        title = "{Powerful Nuclear Outflows and Circumgalactic Medium Shocks Driven by the Most Luminous Known Obscured Quasar in the Universe}",
      journal = {\apj},
         year = 2025,
        month = aug,
       volume = {989},
       number = {2},
          eid = {230},
        pages = {230},
          doi = {10.3847/1538-4357/addbdd},
archivePrefix = {arXiv},
       eprint = {2412.02862},
 primaryClass = {astro-ph.GA},
       adsurl = {https://ui.adsabs.harvard.edu/abs/2025ApJ...989..230V}
}

@ARTICLE{2015ApJ...805...90T,
       author = {{Tsai}, Chao-Wei and {Eisenhardt}, Peter R.~M. and {Wu}, Jingwen and {Stern}, Daniel and {Assef}, Roberto J. and {Blain}, Andrew W. and {Bridge}, Carrie R. and {Benford}, Dominic J. and {Cutri}, Roc M. and {Griffith}, Roger L. and {Jarrett}, Thomas H. and {Lonsdale}, Carol J. and {Masci}, Frank J. and {Moustakas}, Leonidas A. and {Petty}, Sara M. and {Sayers}, Jack and {Stanford}, S. Adam and {Wright}, Edward L. and {Yan}, Lin and {Leisawitz}, David T. and {Liu}, Fengchuan and {Mainzer}, Amy K. and {McLean}, Ian S. and {Padgett}, Deborah L. and {Skrutskie}, Michael F. and {Gelino}, Christopher R. and {Beichman}, Charles A. and {Juneau}, St{\'e}phanie},
        title = "{The Most Luminous Galaxies Discovered by WISE}",
      journal = {\apj},
         year = 2015,
        month = jun,
       volume = {805},
       number = {2},
          eid = {90},
        pages = {90},
          doi = {10.1088/0004-637X/805/2/90},
archivePrefix = {arXiv},
       eprint = {1410.1751},
 primaryClass = {astro-ph.GA},
       adsurl = {https://ui.adsabs.harvard.edu/abs/2015ApJ...805...90T}
}

@ARTICLE{2018MNRAS.473.5113D,
       author = {{Dye}, S. and {Lawrence}, A. and {Read}, M.~A. and {Fan}, X. and {Kerr}, T. and {Varricatt}, W. and {Furnell}, K.~E. and {Edge}, A.~C. and {Irwin}, M. and {Hambly}, N. and {Lucas}, P. and {Almaini}, O. and {Chambers}, K. and {Green}, R. and {Hewett}, P. and {Liu}, M.~C. and {McGreer}, I. and {Best}, W. and {Zhang}, Z. and {Sutorius}, E. and {Froebrich}, D. and {Magnier}, E. and {Hasinger}, G. and {Lederer}, S.~M. and {Bold}, M. and {Tedds}, J.~A.},
        title = "{The UKIRT Hemisphere Survey: definition and J-band data release}",
      journal = {\mnras},
         year = 2018,
        month = feb,
       volume = {473},
       number = {4},
        pages = {5113-5125},
          doi = {10.1093/mnras/stx2622},
archivePrefix = {arXiv},
       eprint = {1707.09975},
 primaryClass = {astro-ph.IM},
       adsurl = {https://ui.adsabs.harvard.edu/abs/2018MNRAS.473.5113D}
}

@inproceedings{10.1117/12.2057445,
author = {Michael J. Fitzpatrick and Knut Olsen and Frossie Economou and Elizabeth B. Stobie and T. C. Beers and Mark Dickinson and Patrick Norris and Abi Saha and Robert Seaman and David R. Silva and Robert A. Swaters and Brian Thomas and Francisco Valdes},
title = {{The NOAO Data Laboratory: a conceptual overview}},
volume = {9149},
booktitle = {Observatory Operations: Strategies, Processes, and Systems V},
editor = {Alison B. Peck and Chris R. Benn and Robert L. Seaman},
organization = {International Society for Optics and Photonics},
publisher = {SPIE},
pages = {91491T},
year = {2014},
doi = {10.1117/12.2057445},
URL = {https://doi.org/10.1117/12.2057445}
}

@article{NIKUTTA2020100411,
title = {Data Lab—A community science platform},
journal = {Astronomy and Computing},
volume = {33},
pages = {100411},
year = {2020},
issn = {2213-1337},
doi = {https://doi.org/10.1016/j.ascom.2020.100411},
url = {https://www.sciencedirect.com/science/article/pii/S2213133720300652},
author = {R. Nikutta and M. Fitzpatrick and A. Scott and B.A. Weaver}
}

@ARTICLE{9347681,
  author={Juneau, Stéphanie and Olsen, Knut and Nikutta, Robert and Jacques, Alice and Bailey, Stephen},
  journal={Computing in Science I\& Engineering}, 
  title={Jupyter-Enabled Astrophysical Analysis Using Data-Proximate Computing Platforms}, 
  year={2021},
  volume={23},
  number={2},
  pages={15-25},
  doi={10.1109/MCSE.2021.3057097}}

@ARTICLE{2022ApJS..261...38D,
       author = {{Drlica-Wagner}, A. and {Ferguson}, P.~S. and {Adam{\'o}w}, M. and {Aguena}, M. and {Allam}, S. and {Andrade-Oliveira}, F. and {Bacon}, D. and {Bechtol}, K. and {Bell}, E.~F. and {Bertin}, E. and {Bilaji}, P. and {Bocquet}, S. and {Bom}, C.~R. and {Brooks}, D. and {Burke}, D.~L. and {Carballo-Bello}, J.~A. and {Carlin}, J.~L. and {Carnero Rosell}, A. and {Carrasco Kind}, M. and {Carretero}, J. and {Castander}, F.~J. and {Cerny}, W. and {Chang}, C. and {Choi}, Y. and {Conselice}, C. and {Costanzi}, M. and {Crnojevi{\'c}}, D. and {da Costa}, L.~N. and {de Vicente}, J. and {Desai}, S. and {Esteves}, J. and {Everett}, S. and {Ferrero}, I. and {Fitzpatrick}, M. and {Flaugher}, B. and {Friedel}, D. and {Frieman}, J. and {Garc{\'\i}a-Bellido}, J. and {Gatti}, M. and {Gaztanaga}, E. and {Gerdes}, D.~W. and {Gruen}, D. and {Gruendl}, R.~A. and {Gschwend}, J. and {Hartley}, W.~G. and {Hernandez-Lang}, D. and {Hinton}, S.~R. and {Hollowood}, D.~L. and {Honscheid}, K. and {Hughes}, A.~K. and {Jacques}, A. and {James}, D.~J. and {Johnson}, M.~D. and {Kuehn}, K. and {Kuropatkin}, N. and {Lahav}, O. and {Li}, T.~S. and {Lidman}, C. and {Lin}, H. and {March}, M. and {Marshall}, J.~L. and {Mart{\'\i}nez-Delgado}, D. and {Mart{\'\i}nez-V{\'a}zquez}, C.~E. and {Massana}, P. and {Mau}, S. and {McNanna}, M. and {Melchior}, P. and {Menanteau}, F. and {Miller}, A.~E. and {Miquel}, R. and {Mohr}, J.~J. and {Morgan}, R. and {Mutlu-Pakdil}, B. and {Mu{\~n}oz}, R.~R. and {Neilsen}, E.~H. and {Nidever}, D.~L. and {Nikutta}, R. and {Nilo Castellon}, J.~L. and {No{\"e}l}, N.~E.~D. and {Ogando}, R.~L.~C. and {Olsen}, K.~A.~G. and {Pace}, A.~B. and {Palmese}, A. and {Paz-Chinch{\'o}n}, F. and {Pereira}, M.~E.~S. and {Pieres}, A. and {Plazas Malag{\'o}n}, A.~A. and {Prat}, J. and {Riley}, A.~H. and {Rodriguez-Monroy}, M. and {Romer}, A.~K. and {Roodman}, A. and {Sako}, M. and {Sakowska}, J.~D. and {Sanchez}, E. and {S{\'a}nchez}, F.~J. and {Sand}, D.~J. and {Santana-Silva}, L. and {Santiago}, B. and {Schubnell}, M. and {Serrano}, S. and {Sevilla-Noarbe}, I. and {Simon}, J.~D. and {Smith}, M. and {Soares-Santos}, M. and {Stringfellow}, G.~S. and {Suchyta}, E. and {Suson}, D.~J. and {Tan}, C.~Y. and {Tarle}, G. and {Tavangar}, K. and {Thomas}, D. and {To}, C. and {Tollerud}, E.~J. and {Troxel}, M.~A. and {Tucker}, D.~L. and {Varga}, T.~N. and {Vivas}, A.~K. and {Walker}, A.~R. and {Weller}, J. and {Wilkinson}, R.~D. and {Wu}, J.~F. and {Yanny}, B. and {Zaborowski}, E. and {Zenteno}, A. and {Delve Collaboration} and {Des Collaboration} and {Astro Data Lab}},
        title = "{The DECam Local Volume Exploration Survey Data Release 2}",
      journal = {\apjs},
         year = 2022,
        month = aug,
       volume = {261},
       number = {2},
          eid = {38},
        pages = {38},
          doi = {10.3847/1538-4365/ac78eb},
archivePrefix = {arXiv},
       eprint = {2203.16565},
 primaryClass = {astro-ph.IM},
       adsurl = {https://ui.adsabs.harvard.edu/abs/2022ApJS..261...38D}
}

@ARTICLE{2006AJ....131.2766R,
       author = {{Richards}, Gordon T. and {Strauss}, Michael A. and {Fan}, Xiaohui and {Hall}, Patrick B. and {Jester}, Sebastian and {Schneider}, Donald P. and {Vanden Berk}, Daniel E. and {Stoughton}, Chris and {Anderson}, Scott F. and {Brunner}, Robert J. and {Gray}, Jim and {Gunn}, James E. and {Ivezi{\'c}}, {\v{Z}}eljko and {Kirkland}, Margaret K. and {Knapp}, G.~R. and {Loveday}, Jon and {Meiksin}, Avery and {Pope}, Adrian and {Szalay}, Alexander S. and {Thakar}, Anirudda R. and {Yanny}, Brian and {York}, Donald G. and {Barentine}, J.~C. and {Brewington}, Howard J. and {Brinkmann}, J. and {Fukugita}, Masataka and {Harvanek}, Michael and {Kent}, Stephen M. and {Kleinman}, S.~J. and {Krzesi{\'n}ski}, Jurek and {Long}, Daniel C. and {Lupton}, Robert H. and {Nash}, Thomas and {Neilsen}, Jr., Eric H. and {Nitta}, Atsuko and {Schlegel}, David J. and {Snedden}, Stephanie A.},
        title = "{The Sloan Digital Sky Survey Quasar Survey: Quasar Luminosity Function from Data Release 3}",
      journal = {\aj},
         year = 2006,
        month = jun,
       volume = {131},
       number = {6},
        pages = {2766-2787},
          doi = {10.1086/503559},
archivePrefix = {arXiv},
       eprint = {astro-ph/0601434},
 primaryClass = {astro-ph},
       adsurl = {https://ui.adsabs.harvard.edu/abs/2006AJ....131.2766R}
}

@ARTICLE{2010AJ....139.2360S,
       author = {{Schneider}, Donald P. and {Richards}, Gordon T. and {Hall}, Patrick B. and {Strauss}, Michael A. and {Anderson}, Scott F. and {Boroson}, Todd A. and {Ross}, Nicholas P. and {Shen}, Yue and {Brandt}, W.~N. and {Fan}, Xiaohui and {Inada}, Naohisa and {Jester}, Sebastian and {Knapp}, G.~R. and {Krawczyk}, Coleman M. and {Thakar}, Anirudda R. and {Vanden Berk}, Daniel E. and {Voges}, Wolfgang and {Yanny}, Brian and {York}, Donald G. and {Bahcall}, Neta A. and {Bizyaev}, Dmitry and {Blanton}, Michael R. and {Brewington}, Howard and {Brinkmann}, J. and {Eisenstein}, Daniel and {Frieman}, Joshua A. and {Fukugita}, Masataka and {Gray}, Jim and {Gunn}, James E. and {Hibon}, Pascale and {Ivezi{\'c}}, {\v{Z}}eljko and {Kent}, Stephen M. and {Kron}, Richard G. and {Lee}, Myung Gyoon and {Lupton}, Robert H. and {Malanushenko}, Elena and {Malanushenko}, Viktor and {Oravetz}, Dan and {Pan}, K. and {Pier}, Jeffrey R. and {Price}, III, Ted N. and {Saxe}, David H. and {Schlegel}, David J. and {Simmons}, Audry and {Snedden}, Stephanie A. and {SubbaRao}, Mark U. and {Szalay}, Alexander S. and {Weinberg}, David H.},
        title = "{The Sloan Digital Sky Survey Quasar Catalog. V. Seventh Data Release}",
      journal = {\aj},
         year = 2010,
        month = jun,
       volume = {139},
       number = {6},
          eid = {2360},
        pages = {2360},
          doi = {10.1088/0004-6256/139/6/2360},
archivePrefix = {arXiv},
       eprint = {1004.1167},
 primaryClass = {astro-ph.CO},
       adsurl = {https://ui.adsabs.harvard.edu/abs/2010AJ....139.2360S}
}

@ARTICLE{2025arXiv251115823A,
       author = {{Akeson}, Rachel and {Dubois-Felsmann}, Gregory P. and {Crill}, Brendan P. and {Faisst}, Andreas L. and {Fatahi}, Tamim and {Fazar}, Candice M. and {Goldina}, Tatiana and {Masters}, Daniel C. and {Nelson}, Christina and {Paladini}, Roberta and {Teplitz}, Harry I. and {Torrini}, Gabriela and {Velicheti}, Phani and {Ashby}, Matthew L.~N. and {Avner}, Dan and {Bach}, Yoonsoo P. and {Bock}, James J. and {Bruton}, Sean and {Bryan}, Sean A. and {Chang}, Tzu-Ching and {Chen}, Shuang-Shuang and {Cukierman}, Ari J. and {Dore}, O. and {Dowell}, C. Darren and {Everett}, Spencer and {Feder}, Richard M. and {Huai}, Zhaoyu and {Hui}, Howard and {Jeong}, Woong-Seob and {Jo}, Young-Soo and {Korngut}, Phil M. and {Kwon}, Yuna G. and {Lee}, Bomee and {Melnick}, Gary J. and {Murgia}, Giulia and {Nguyen}, Chi H. and {Pourrahmani}, Milad and {Rustamkulov}, Zafar and {Tolls}, Volker and {Wang}, Pao-Yu and {Yang}, Yujin and {Zemcov}, Michael},
        title = "{The SPHEREx Image and Spectrophotometry Processing Pipeline}",
      journal = {arXiv e-prints},
         year = 2025,
        month = nov,
          eid = {arXiv:2511.15823 },
        pages = {arXiv:2511.15823 (submitted to the ApJ Supplement Series)},
          doi = {10.48550/arXiv.2511.15823},
archivePrefix = {arXiv},
       eprint = {2511.15823},
 primaryClass = {astro-ph.IM},
       adsurl = {https://ui.adsabs.harvard.edu/abs/2025arXiv251115823A}
}

@ARTICLE{2025A&A...702A.124A,
       author = {{Assef}, R.~J. and {Stalevski}, M. and {Armus}, L. and {Bauer}, F.~E. and {Blain}, A. and {Brightman}, M. and {D{\'\i}az-Santos}, T. and {Eisenhardt}, P.~R.~M. and {Fern{\'a}ndez-Aranda}, R. and {Jun}, H.~D. and {Liao}, M. and {Li}, G. and {Martin}, L.~R. and {Shablovinskaia}, E. and {Shobhana}, D. and {Stern}, D. and {Tsai}, C.-W. and {Vayner}, A. and {Walton}, D.~J. and {Wu}, J. and {Zewdie}, D.},
        title = "{A massive gas outflow outside the line of sight: Imaging polarimetry of the blue excess Hot Dust-Obscured Galaxy W0204─0506}",
      journal = {\aap},
         year = 2025,
        month = oct,
       volume = {702},
          eid = {A124},
        pages = {A124},
          doi = {10.1051/0004-6361/202555245},
archivePrefix = {arXiv},
       eprint = {2504.15913},
 primaryClass = {astro-ph.GA},
       adsurl = {https://ui.adsabs.harvard.edu/abs/2025A&A...702A.124A}
}

@ARTICLE{2024MNRAS.533.2948S,
       author = {{Stepney}, Matthew and {Banerji}, Manda and {Tang}, Shenli and {Hewett}, Paul C. and {Temple}, Matthew J. and {Wethers}, Clare F. and {Puglisi}, Annagrazia and {Molyneux}, Stephen J.},
        title = "{A big red dot: scattered light, host galaxy signatures, and multiphase gas flows in a luminous, heavily reddened quasar at cosmic noon}",
      journal = {\mnras},
         year = 2024,
        month = sep,
       volume = {533},
       number = {3},
        pages = {2948-2965},
          doi = {10.1093/mnras/stae1970},
archivePrefix = {arXiv},
       eprint = {2408.10403},
 primaryClass = {astro-ph.GA},
       adsurl = {https://ui.adsabs.harvard.edu/abs/2024MNRAS.533.2948S}
}

@ARTICLE{2022MNRAS.516.4963I,
       author = {{Ishibashi}, W. and {Fabian}, A.~C.},
        title = "{What powers galactic outflows: nuclear starbursts or AGN?}",
      journal = {\mnras},
         year = 2022,
        month = nov,
       volume = {516},
       number = {4},
        pages = {4963-4970},
          doi = {10.1093/mnras/stac2614},
archivePrefix = {arXiv},
       eprint = {2209.10580},
 primaryClass = {astro-ph.GA},
       adsurl = {https://ui.adsabs.harvard.edu/abs/2022MNRAS.516.4963I}
}

@ARTICLE{2021MNRAS.508..737T,
       author = {{Temple}, Matthew J. and {Hewett}, Paul C. and {Banerji}, Manda},
        title = "{Modelling type 1 quasar colours in the era of Rubin and Euclid}",
      journal = {\mnras},
         year = 2021,
        month = nov,
       volume = {508},
       number = {1},
        pages = {737-754},
          doi = {10.1093/mnras/stab2586},
archivePrefix = {arXiv},
       eprint = {2109.04472},
 primaryClass = {astro-ph.GA},
       adsurl = {https://ui.adsabs.harvard.edu/abs/2021MNRAS.508..737T}
}

@article{emcee,
   author = {{Foreman-Mackey}, D. and {Hogg}, D.~W. and {Lang}, D. and {Goodman}, J.},
    title = {emcee: The MCMC Hammer},
  journal = {PASP},
     year = 2013,
   volume = 125,
    pages = {306-312},
   eprint = {1202.3665},
      doi = {10.1086/670067}
}

@ARTICLE{2020ApJ...897..112A,
       author = {{Assef}, R.~J. and {Brightman}, M. and {Walton}, D.~J. and {Stern}, D. and {Bauer}, F.~E. and {Blain}, A.~W. and {D{\'\i}az-Santos}, T. and {Eisenhardt}, P.~R.~M. and {Hickox}, R.~C. and {Jun}, H.~D. and {Psychogyios}, A. and {Tsai}, C. -W. and {Wu}, J.~W.},
        title = "{Hot Dust-obscured Galaxies with Excess Blue Light}",
      journal = {\apj},
         year = 2020,
        month = jul,
       volume = {897},
       number = {2},
          eid = {112},
        pages = {112},
          doi = {10.3847/1538-4357/ab9814},
archivePrefix = {arXiv},
       eprint = {1905.04320},
 primaryClass = {astro-ph.GA},
       adsurl = {https://ui.adsabs.harvard.edu/abs/2020ApJ...897..112A}
}

@ARTICLE{2023MNRAS.524.5497S,
       author = {{Stepney}, Matthew and {Banerji}, Manda and {Hewett}, Paul C. and {Temple}, Matthew J. and {Rankine}, Amy L. and {Matthews}, James H. and {Richards}, Gordon T.},
        title = "{No redshift evolution in the rest-frame ultraviolet emission line properties of quasars from z = 1.5 to z = 4.0}",
      journal = {\mnras},
         year = 2023,
        month = oct,
       volume = {524},
       number = {4},
        pages = {5497-5513},
          doi = {10.1093/mnras/stad2060},
archivePrefix = {arXiv},
       eprint = {2307.02962},
 primaryClass = {astro-ph.GA},
       adsurl = {https://ui.adsabs.harvard.edu/abs/2023MNRAS.524.5497S}
}

@ARTICLE{2019MNRAS.487.2594T,
       author = {{Temple}, Matthew J. and {Banerji}, Manda and {Hewett}, Paul C. and {Coatman}, Liam and {Maddox}, Natasha and {Peroux}, Celine},
        title = "{[O III] Emission line properties in a new sample of heavily reddened quasars at z > 2}",
      journal = {\mnras},
         year = 2019,
        month = aug,
       volume = {487},
       number = {2},
        pages = {2594-2613},
          doi = {10.1093/mnras/stz1420},
archivePrefix = {arXiv},
       eprint = {1905.08198},
 primaryClass = {astro-ph.GA},
       adsurl = {https://ui.adsabs.harvard.edu/abs/2019MNRAS.487.2594T}
}

@ARTICLE{2018MNRAS.475.3682W,
       author = {{Wethers}, C.~F. and {Banerji}, M. and {Hewett}, P.~C. and {Lemon}, C.~A. and {McMahon}, R.~G. and {Reed}, S.~L. and {Shen}, Y. and {Abdalla}, F.~B. and {Benoit-L{\'e}vy}, A. and {Brooks}, D. and {Buckley-Geer}, E. and {Capozzi}, D. and {Carnero Rosell}, A. and {CarrascoKind}, M. and {Carretero}, J. and {Cunha}, C.~E. and {D'Andrea}, C.~B. and {da Costa}, L.~N. and {DePoy}, D.~L. and {Desai}, S. and {Doel}, P. and {Flaugher}, B. and {Fosalba}, P. and {Frieman}, J. and {Garc{\'\i}a-Bellido}, J. and {Gerdes}, D.~W. and {Gruen}, D. and {Gruendl}, R.~A. and {Gschwend}, J. and {Gutierrez}, G. and {Honscheid}, K. and {James}, D.~J. and {Jeltema}, T. and {Kuehn}, K. and {Kuhlmann}, S. and {Kuropatkin}, N. and {Lima}, M. and {Maia}, M.~A.~G. and {Marshall}, J.~L. and {Martini}, P. and {Menanteau}, F. and {Miquel}, R. and {Nichol}, R.~C. and {Nord}, B. and {Plazas}, A.~A. and {Romer}, A.~K. and {Sanchez}, E. and {Scarpine}, V. and {Schindler}, R. and {Schubnell}, M. and {Sevilla-Noarbe}, I. and {Smith}, M. and {Smith}, R.~C. and {Soares-Santos}, M. and {Sobreira}, F. and {Suchyta}, E. and {Tarle}, G. and {Walker}, A.~R.},
        title = "{UV-luminous, star-forming hosts of z {\ensuremath{\sim}} 2 reddened quasars in the Dark Energy Survey}",
      journal = {\mnras},
         year = 2018,
        month = apr,
       volume = {475},
       number = {3},
        pages = {3682-3699},
          doi = {10.1093/mnras/stx3332},
archivePrefix = {arXiv},
       eprint = {1801.02635},
 primaryClass = {astro-ph.GA},
       adsurl = {https://ui.adsabs.harvard.edu/abs/2018MNRAS.475.3682W}
}

@ARTICLE{2021MNRAS.503.5583B,
       author = {{Banerji}, Manda and {Jones}, Gareth C. and {Carniani}, Stefano and {DeGraf}, Colin and {Wagg}, Jeff},
        title = "{Resolving discs and mergers in z {\ensuremath{\sim}} 2 heavily reddened quasars and their companion galaxies with ALMA}",
      journal = {\mnras},
         year = 2021,
        month = jun,
       volume = {503},
       number = {4},
        pages = {5583-5599},
          doi = {10.1093/mnras/stab852},
archivePrefix = {arXiv},
       eprint = {2103.10174},
 primaryClass = {astro-ph.GA},
       adsurl = {https://ui.adsabs.harvard.edu/abs/2021MNRAS.503.5583B}
}

@ARTICLE{2024MNRAS.527..950G,
       author = {{Gillette}, Jarred and {Hamann}, Fred and {Lau}, Marie Wingyee and {Perrotta}, Serena},
        title = "{Accurate systemic redshifts and outflow speeds for extremely red quasars (ERQs)}",
      journal = {\mnras},
         year = 2024,
        month = jan,
       volume = {527},
       number = {1},
        pages = {950-958},
          doi = {10.1093/mnras/stad2890},
archivePrefix = {arXiv},
       eprint = {2305.11223},
 primaryClass = {astro-ph.GA},
       adsurl = {https://ui.adsabs.harvard.edu/abs/2024MNRAS.527..950G}
}

@ARTICLE{2020MNRAS.495.2652L,
       author = {{Lansbury}, G.~B. and {Banerji}, M. and {Fabian}, A.~C. and {Temple}, M.~J.},
        title = "{X-ray observations of luminous dusty quasars at z \ensuremath{>} 2}",
      journal = {\mnras},
         year = 2020,
        month = jan,
       volume = {495},
       number = {3},
        pages = {2652-2663},
          doi = {10.1093/mnras/staa1220},
archivePrefix = {arXiv},
       eprint = {1910.00602},
 primaryClass = {astro-ph.HE},
       adsurl = {https://ui.adsabs.harvard.edu/abs/2020MNRAS.495.2652L}
}

@ARTICLE{2019MNRAS.488.5185N,
       author = {{Netzer}, Hagai},
        title = "{Bolometric correction factors for active galactic nuclei}",
      journal = {\mnras},
         year = 2019,
        month = oct,
       volume = {488},
       number = {4},
        pages = {5185-5191},
          doi = {10.1093/mnras/stz2016},
archivePrefix = {arXiv},
       eprint = {1907.09534},
 primaryClass = {astro-ph.GA},
       adsurl = {https://ui.adsabs.harvard.edu/abs/2019MNRAS.488.5185N}
}

@ARTICLE{2019MNRAS.488.4126P,
       author = {{Perrotta}, S. and {Hamann}, F. and {Zakamska}, N.~L. and {Alexandroff}, R.~M. and {Rupke}, D. and {Wylezalek}, D.},
        title = "{ERQs are the BOSS of quasar samples: the highest velocity [O III] quasar outflows}",
      journal = {\mnras},
         year = 2019,
        month = sep,
       volume = {488},
       number = {3},
        pages = {4126-4148},
          doi = {10.1093/mnras/stz1993},
archivePrefix = {arXiv},
       eprint = {1906.00980},
 primaryClass = {astro-ph.GA},
       adsurl = {https://ui.adsabs.harvard.edu/abs/2019MNRAS.488.4126P}
}

@ARTICLE{2023AJ....165..124A,
       author = {{Alexander}, David M. and {Davis}, Tamara M. and {Chaussidon}, E. and {Fawcett}, V.~A. and {X. Gonzalez-Morales}, Alma and {Lan}, Ting-Wen and {Y{\`e}che}, Christophe and {Ahlen}, S. and {Aguilar}, J.~N. and {Armengaud}, E. and {Bailey}, S. and {Brooks}, D. and {Cai}, Z. and {Canning}, R. and {Carr}, A. and {Chabanier}, S. and {Cousinou}, Marie-Claude and {Dawson}, K. and {de la Macorra}, A. and {Dey}, A. and {Dey}, Biprateep and {Dhungana}, G. and {Edge}, A.~C. and {Eftekharzadeh}, S. and {Fanning}, K. and {Farr}, James and {Font-Ribera}, A. and {Garcia-Bellido}, J. and {Garrison}, Lehman and {Gazta{\~n}aga}, E. and {A Gontcho}, Satya Gontcho and {Gordon}, C. and {Medellin Gonzalez}, Stefany Guadalupe and {Guy}, J. and {Herrera-Alcantar}, Hiram K. and {Jiang}, L. and {Juneau}, S. and {Kara{\c{c}}ayl{\i}}, N.~G. and {Kehoe}, R. and {Kisner}, T. and {Kov{\'a}cs}, A. and {Landriau}, M. and {Levi}, Michael E. and {Magneville}, C. and {Martini}, P. and {Meisner}, Aaron M. and {Mezcua}, M. and {Miquel}, R. and {Camacho}, P. Montero and {Moustakas}, J. and {Mu{\~n}oz-Guti{\'e}rrez}, Andrea and {Myers}, Adam D. and {Nadathur}, S. and {Napolitano}, L. and {Nie}, J.~D. and {Palanque-Delabrouille}, N. and {Pan}, Z. and {Percival}, W.~J. and {P{\'e}rez-R{\`a}fols}, I. and {Poppett}, C. and {Prada}, F. and {Ram{\'\i}rez-P{\'e}rez}, C{\'e}sar and {Ravoux}, C. and {Rosario}, D.~J. and {Schubnell}, M. and {Tarl{\'e}}, Gregory and {Walther}, M. and {Weiner}, B. and {Youles}, S. and {Zhou}, Zhimin and {Zou}, H. and {Zou}, Siwei},
        title = "{The DESI Survey Validation: Results from Visual Inspection of the Quasar Survey Spectra}",
      journal = {\aj},
         year = 2023,
        month = mar,
       volume = {165},
       number = {3},
          eid = {124},
        pages = {124},
          doi = {10.3847/1538-3881/acacfc},
archivePrefix = {arXiv},
       eprint = {2208.08517},
 primaryClass = {astro-ph.GA},
       adsurl = {https://ui.adsabs.harvard.edu/abs/2023AJ....165..124A}
}

@ARTICLE{2023MNRAS.525.5575F,
       author = {{Fawcett}, V.~A. and {Alexander}, D.~M. and {Brodzeller}, A. and {Edge}, A.~C. and {Rosario}, D.~J. and {Myers}, A.~D. and {Aguilar}, J. and {Ahlen}, S. and {Alfarsy}, R. and {Brooks}, D. and {Canning}, R. and {Circosta}, C. and {Dawson}, K. and {de la Macorra}, A. and {Doel}, P. and {Fanning}, K. and {Font-Ribera}, A. and {Forero-Romero}, J.~E. and {Gontcho A Gontcho}, S. and {Guy}, J. and {Harrison}, C.~M. and {Honscheid}, K. and {Juneau}, S. and {Kehoe}, R. and {Kisner}, T. and {Kremin}, A. and {Landriau}, M. and {Manera}, M. and {Meisner}, A.~M. and {Miquel}, R. and {Moustakas}, J. and {Nie}, J. and {Percival}, W.~J. and {Poppett}, C. and {Pucha}, R. and {Rossi}, G. and {Schlegel}, D. and {Siudek}, M. and {Tarl{\'e}}, G. and {Weaver}, B.~A. and {Zhou}, Z. and {Zou}, H.},
        title = "{A striking relationship between dust extinction and radio detection in DESI QSOs: evidence for a dusty blow-out phase in red QSOs}",
      journal = {\mnras},
         year = 2023,
        month = nov,
       volume = {525},
       number = {4},
        pages = {5575-5596},
          doi = {10.1093/mnras/stad2603},
archivePrefix = {arXiv},
       eprint = {2308.14790},
 primaryClass = {astro-ph.GA},
       adsurl = {https://ui.adsabs.harvard.edu/abs/2023MNRAS.525.5575F}
}

@ARTICLE{2013MNRAS.429L..55B,
       author = {{Banerji}, M. and {McMahon}, R.~G. and {Hewett}, P.~C. and {Gonzalez-Solares}, E. and {Koposov}, S.~E.},
        title = "{Hyperluminous reddened broad-line quasars at z \raisebox{-0.5ex}\textasciitilde 2 from the VISTA  hemisphere survey and WISE all-sky survey.}",
      journal = {\mnras},
         year = 2013,
        month = feb,
       volume = {429},
        pages = {L55-L59},
          doi = {10.1093/mnrasl/sls023},
archivePrefix = {arXiv},
       eprint = {1210.6668},
 primaryClass = {astro-ph.CO},
       adsurl = {https://ui.adsabs.harvard.edu/abs/2013MNRAS.429L..55B}
}

@ARTICLE{2020ApJS..249....3A,
       author = {{Ahumada}, Romina and {Allende Prieto}, Carlos and {Almeida}, Andr{\'e}s and {Anders}, Friedrich and {Anderson}, Scott F. and {Andrews}, Brett H. and {Anguiano}, Borja and {Arcodia}, Riccardo and {Armengaud}, Eric and {Aubert}, Marie and {Avila}, Santiago and {Avila-Reese}, Vladimir and {Badenes}, Carles and {Balland}, Christophe and {Barger}, Kat and {Barrera-Ballesteros}, Jorge K. and {Basu}, Sarbani and {Bautista}, Julian and {Beaton}, Rachael L. and {Beers}, Timothy C. and {Benavides}, B. Izamar T. and {Bender}, Chad F. and {Bernardi}, Mariangela and {Bershady}, Matthew and {Beutler}, Florian and {Bidin}, Christian Moni and {Bird}, Jonathan and {Bizyaev}, Dmitry and {Blanc}, Guillermo A. and {Blanton}, Michael R. and {Boquien}, M{\'e}d{\'e}ric and {Borissova}, Jura and {Bovy}, Jo and {Brandt}, W.~N. and {Brinkmann}, Jonathan and {Brownstein}, Joel R. and {Bundy}, Kevin and {Bureau}, Martin and {Burgasser}, Adam and {Burtin}, Etienne and {Cano-D{\'\i}az}, Mariana and {Capasso}, Raffaella and {Cappellari}, Michele and {Carrera}, Ricardo and {Chabanier}, Sol{\`e}ne and {Chaplin}, William and {Chapman}, Michael and {Cherinka}, Brian and {Chiappini}, Cristina and {Doohyun Choi}, Peter and {Chojnowski}, S. Drew and {Chung}, Haeun and {Clerc}, Nicolas and {Coffey}, Damien and {Comerford}, Julia M. and {Comparat}, Johan and {da Costa}, Luiz and {Cousinou}, Marie-Claude and {Covey}, Kevin and {Crane}, Jeffrey D. and {Cunha}, Katia and {Ilha}, Gabriele da Silva and {Dai}, Yu Sophia and {Damsted}, Sanna B. and {Darling}, Jeremy and {Davidson}, James W., Jr. and {Davies}, Roger and {Dawson}, Kyle and {De}, Nikhil and {de la Macorra}, Axel and {De Lee}, Nathan and {Queiroz}, Anna B{\'a}rbara de Andrade and {Deconto Machado}, Alice and {de la Torre}, Sylvain and {Dell'Agli}, Flavia and {du Mas des Bourboux}, H{\'e}lion and {Diamond-Stanic}, Aleksandar M. and {Dillon}, Sean and {Donor}, John and {Drory}, Niv and {Duckworth}, Chris and {Dwelly}, Tom and {Ebelke}, Garrett and {Eftekharzadeh}, Sarah and {Davis Eigenbrot}, Arthur and {Elsworth}, Yvonne P. and {Eracleous}, Mike and {Erfanianfar}, Ghazaleh and {Escoffier}, Stephanie and {Fan}, Xiaohui and {Farr}, Emily and {Fern{\'a}ndez-Trincado}, Jos{\'e} G. and {Feuillet}, Diane and {Finoguenov}, Alexis and {Fofie}, Patricia and {Fraser-McKelvie}, Amelia and {Frinchaboy}, Peter M. and {Fromenteau}, Sebastien and {Fu}, Hai and {Galbany}, Llu{\'\i}s and {Garcia}, Rafael A. and {Garc{\'\i}a-Hern{\'a}ndez}, D.~A. and {Garma Oehmichen}, Luis Alberto and {Ge}, Junqiang and {Geimba Maia}, Marcio Antonio and {Geisler}, Doug and {Gelfand}, Joseph and {Goddy}, Julian and {Gonzalez-Perez}, Violeta and {Grabowski}, Kathleen and {Green}, Paul and {Grier}, Catherine J. and {Guo}, Hong and {Guy}, Julien and {Harding}, Paul and {Hasselquist}, Sten and {Hawken}, Adam James and {Hayes}, Christian R. and {Hearty}, Fred and {Hekker}, S. and {Hogg}, David W. and {Holtzman}, Jon A. and {Horta}, Danny and {Hou}, Jiamin and {Hsieh}, Bau-Ching and {Huber}, Daniel and {Hunt}, Jason A.~S. and {Ider Chitham}, J. and {Imig}, Julie and {Jaber}, Mariana and {Jimenez Angel}, Camilo Eduardo and {Johnson}, Jennifer A. and {Jones}, Amy M. and {J{\"o}nsson}, Henrik and {Jullo}, Eric and {Kim}, Yerim and {Kinemuchi}, Karen and {Kirkpatrick}, Charles C., IV and {Kite}, George W. and {Klaene}, Mark and {Kneib}, Jean-Paul and {Kollmeier}, Juna A. and {Kong}, Hui and {Kounkel}, Marina and {Krishnarao}, Dhanesh and {Lacerna}, Ivan and {Lan}, Ting-Wen and {Lane}, Richard R. and {Law}, David R. and {Le Goff}, Jean-Marc and {Leung}, Henry W. and {Lewis}, Hannah and {Li}, Cheng and {Lian}, Jianhui and {Lin}, Lihwai and {Long}, Dan and {Longa-Pe{\~n}a}, Pen{\'e}lope and {Lundgren}, Britt and {Lyke}, Brad W. and {Mackereth}, J. Ted and {MacLeod}, Chelsea L. and {Majewski}, Steven R. and {Manchado}, Arturo and {Maraston}, Claudia and {Martini}, Paul and {Masseron}, Thomas and {Masters}, Karen L. and {Mathur}, Savita and {McDermid}, Richard M. and {Merloni}, Andrea and {Merrifield}, Michael and {M{\'e}sz{\'a}ros}, Szabolcs and {Miglio}, Andrea and {Minniti}, Dante and {Minsley}, Rebecca and {Miyaji}, Takamitsu and {Mohammad}, Faizan Gohar and {Mosser}, Benoit and {Mueller}, Eva-Maria and {Muna}, Demitri and {Mu{\~n}oz-Guti{\'e}rrez}, Andrea and {Myers}, Adam D. and {Nadathur}, Seshadri and {Nair}, Preethi and {Nandra}, Kirpal and {Correa do Nascimento}, Janaina and {Nevin}, Rebecca Jean and {Newman}, Jeffrey A. and {Nidever}, David L. and {Nitschelm}, Christian and {Noterdaeme}, Pasquier and {O'Connell}, Julia E. and {Olmstead}, Matthew D. and {Oravetz}, Daniel and {Oravetz}, Audrey and {Osorio}, Yeisson and {Pace}, Zachary J. and {Padilla}, Nelson and {Palanque-Delabrouille}, Nathalie and {Palicio}, Pedro A. and {Pan}, Hsi-An and {Pan}, Kaike and {Parker}, James and {Paviot}, Romain and {Peirani}, Sebastien and {Ram{\'r}ez}, Karla Pe{\~n}a and {Penny}, Samantha and {Percival}, Will J. and {Perez-Fournon}, Ismael and {P{\'e}rez-R{\`a}fols}, Ignasi and {Petitjean}, Patrick and {Pieri}, Matthew M. and {Pinsonneault}, Marc and {Poovelil}, Vijith Jacob and {Povick}, Joshua Tyler and {Prakash}, Abhishek and {Price-Whelan}, Adrian M. and {Raddick}, M. Jordan and {Raichoor}, Anand and {Ray}, Amy and {Rembold}, Sandro Barboza and {Rezaie}, Mehdi and {Riffel}, Rogemar A. and {Riffel}, Rog{\'e}rio and {Rix}, Hans-Walter and {Robin}, Annie C. and {Roman-Lopes}, A. and {Rom{\'a}n-Z{\'u}{\~n}iga}, Carlos and {Rose}, Benjamin and {Ross}, Ashley J. and {Rossi}, Graziano and {Rowlands}, Kate and {Rubin}, Kate H.~R. and {Salvato}, Mara and {S{\'a}nchez}, Ariel G. and {S{\'a}nchez-Menguiano}, Laura and {S{\'a}nchez-Gallego}, Jos{\'e} R. and {Sayres}, Conor and {Schaefer}, Adam and {Schiavon}, Ricardo P. and {Schimoia}, Jaderson S. and {Schlafly}, Edward and {Schlegel}, David and {Schneider}, Donald P. and {Schultheis}, Mathias and {Schwope}, Axel and {Seo}, Hee-Jong and {Serenelli}, Aldo and {Shafieloo}, Arman and {Shamsi}, Shoaib Jamal and {Shao}, Zhengyi and {Shen}, Shiyin and {Shetrone}, Matthew and {Shirley}, Raphael and {Silva Aguirre}, V{\'\i}ctor and {Simon}, Joshua D. and {Skrutskie}, M.~F. and {Slosar}, An{\v{z}}e and {Smethurst}, Rebecca and {Sobeck}, Jennifer and {Sodi}, Bernardo Cervantes and {Souto}, Diogo and {Stark}, David V. and {Stassun}, Keivan G. and {Steinmetz}, Matthias and {Stello}, Dennis and {Stermer}, Julianna and {Storchi-Bergmann}, Thaisa and {Streblyanska}, Alina and {Stringfellow}, Guy S. and {Stutz}, Amelia and {Su{\'a}rez}, Genaro and {Sun}, Jing and {Taghizadeh-Popp}, Manuchehr and {Talbot}, Michael S. and {Tayar}, Jamie and {Thakar}, Aniruddha R. and {Theriault}, Riley and {Thomas}, Daniel and {Thomas}, Zak C. and {Tinker}, Jeremy and {Tojeiro}, Rita and {Toledo}, Hector Hernandez and {Tremonti}, Christy A. and {Troup}, Nicholas W. and {Tuttle}, Sarah and {Unda-Sanzana}, Eduardo and {Valentini}, Marica and {Vargas-Gonz{\'a}lez}, Jaime and {Vargas-Maga{\~n}a}, Mariana and {V{\'a}zquez-Mata}, Jose Antonio and {Vivek}, M. and {Wake}, David and {Wang}, Yuting and {Weaver}, Benjamin Alan and {Weijmans}, Anne-Marie and {Wild}, Vivienne and {Wilson}, John C. and {Wilson}, Robert F. and {Wolthuis}, Nathan and {Wood-Vasey}, W.~M. and {Yan}, Renbin and {Yang}, Meng and {Y{\`e}che}, Christophe and {Zamora}, Olga and {Zarrouk}, Pauline and {Zasowski}, Gail and {Zhang}, Kai and {Zhao}, Cheng and {Zhao}, Gongbo and {Zheng}, Zheng and {Zheng}, Zheng and {Zhu}, Guangtun and {Zou}, Hu},
        title = "{The 16th Data Release of the Sloan Digital Sky Surveys: First Release from the APOGEE-2 Southern Survey and Full Release of eBOSS Spectra}",
      journal = {\apjs},
         year = 2020,
        month = jul,
       volume = {249},
       number = {1},
          eid = {3},
        pages = {3},
          doi = {10.3847/1538-4365/ab929e},
archivePrefix = {arXiv},
       eprint = {1912.02905},
 primaryClass = {astro-ph.GA},
       adsurl = {https://ui.adsabs.harvard.edu/abs/2020ApJS..249....3A}
}

@ARTICLE{1988ApJ...325...74S,
       author = {{Sanders}, D.~B. and {Soifer}, B.~T. and {Elias}, J.~H. and {Madore}, B.~F. and {Matthews}, K. and {Neugebauer}, G. and {Scoville}, N.~Z.},
        title = "{Ultraluminous Infrared Galaxies and the Origin of Quasars}",
      journal = {\apj},
         year = 1988,
        month = feb,
       volume = {325},
        pages = {74},
          doi = {10.1086/165983},
       adsurl = {https://ui.adsabs.harvard.edu/abs/1988ApJ...325...74S}
}

@article{Hopkins_2008,
doi = {10.1086/524362},
url = {https://dx.doi.org/10.1086/524362},
year = {2008},
month = {apr},
publisher = {},
volume = {175},
number = {2},
pages = {356},
author = {Philip F. Hopkins and Lars Hernquist and Thomas J. Cox and Dušan Kereš},
title = {A Cosmological Framework for the Co-Evolution of Quasars, Supermassive Black Holes, and Elliptical Galaxies. I. Galaxy Mergers and Quasar Activity},
journal = {\apjs},
}

@article{somerville_08,
    author = {Somerville, Rachel S. and Hopkins, Philip F. and Cox, Thomas J. and Robertson, Brant E. and Hernquist, Lars},
    title = "{A semi-analytic model for the co-evolution of galaxies, black holes and active galactic nuclei}",
    journal = {\mnras},
    volume = {391},
    number = {2},
    pages = {481-506},
    year = {2008},
    month = {11},
    issn = {0035-8711},
    doi = {10.1111/j.1365-2966.2008.13805.x},
    url = {https://doi.org/10.1111/j.1365-2966.2008.13805.x},
    eprint = {https://academic.oup.com/mnras/article-pdf/391/2/481/5764146/mnras0391-0481.pdf},
}

@ARTICLE{2017MNRAS.464.3431H,
       author = {{Hamann}, Fred and {Zakamska}, Nadia L. and {Ross}, Nicholas and {Paris}, Isabelle and {Alexandroff}, Rachael M. and {Villforth}, Carolin and {Richards}, Gordon T. and {Herbst}, Hanna and {Brandt}, W. Niel and {Cook}, Ben and {Denney}, Kelly D. and {Greene}, Jenny E. and {Schneider}, Donald P. and {Strauss}, Michael A.},
        title = "{Extremely red quasars in BOSS}",
      journal = {\mnras},
         year = 2017,
        month = jan,
       volume = {464},
       number = {3},
        pages = {3431-3463},
          doi = {10.1093/mnras/stw2387},
archivePrefix = {arXiv},
       eprint = {1609.07241},
 primaryClass = {astro-ph.GA},
       adsurl = {https://ui.adsabs.harvard.edu/abs/2017MNRAS.464.3431H}
}

@article{Eisenhardt_2012,
doi = {10.1088/0004-637X/755/2/173},
url = {https://dx.doi.org/10.1088/0004-637X/755/2/173},
year = {2012},
month = {aug},
publisher = {The American Astronomical Society},
volume = {755},
number = {2},
pages = {173},
author = {Peter R. M. Eisenhardt and Jingwen Wu and Chao-Wei Tsai and Roberto Assef and Dominic Benford and Andrew Blain and Carrie Bridge and J. J. Condon and Michael C. Cushing and Roc Cutri and Neal J. Evans and Chris Gelino and Roger L. Griffith and Carl J. Grillmair and Tom Jarrett and Carol J. Lonsdale and Frank J. Masci and Brian S. Mason and Sara Petty and Jack Sayers and S. A. Stanford and Daniel Stern and Edward L. Wright and Lin Yan},
title = {THE FIRST HYPER-LUMINOUS INFRARED GALAXY DISCOVERED BY WISE},
journal = {\apj},

}

@ARTICLE{2015ApJ...804...27A,
       author = {{Assef}, R.~J. and {Eisenhardt}, P.~R.~M. and {Stern}, D. and {Tsai}, C. -W. and {Wu}, J. and {Wylezalek}, D. and {Blain}, A.~W. and {Bridge}, C.~R. and {Donoso}, E. and {Gonzales}, A. and {Griffith}, R.~L. and {Jarrett}, T.~H.},
        title = "{Half of the Most Luminous Quasars May Be Obscured: Investigating the Nature of WISE-Selected Hot Dust-Obscured Galaxies}",
      journal = {\apj},
         year = 2015,
        month = may,
       volume = {804},
       number = {1},
          eid = {27},
        pages = {27},
          doi = {10.1088/0004-637X/804/1/27},
archivePrefix = {arXiv},
       eprint = {1408.1092},
 primaryClass = {astro-ph.GA},
       adsurl = {https://ui.adsabs.harvard.edu/abs/2015ApJ...804...27A}
}

@ARTICLE{2021A&A...649A.102C,
       author = {{Calistro Rivera}, G. and {Alexander}, D.~M. and {Rosario}, D.~J. and {Harrison}, C.~M. and {Stalevski}, M. and {Rakshit}, S. and {Fawcett}, V.~A. and {Morabito}, L.~K. and {Klindt}, L. and {Best}, P.~N. and {Bonato}, M. and {Bowler}, R.~A.~A. and {Costa}, T. and {Kondapally}, R.},
        title = "{The multiwavelength properties of red QSOs: Evidence for dusty winds as the origin of QSO reddening}",
      journal = {\aap},
         year = 2021,
        month = may,
       volume = {649},
          eid = {A102},
        pages = {A102},
          doi = {10.1051/0004-6361/202040214},
archivePrefix = {arXiv},
       eprint = {2103.02610},
 primaryClass = {astro-ph.GA},
       adsurl = {https://ui.adsabs.harvard.edu/abs/2021A&A...649A.102C}
}

@ARTICLE{2022ApJ...934..101A,
       author = {{Assef}, R.~J. and {Bauer}, F.~E. and {Blain}, A.~W. and {Brightman}, M. and {D{\'\i}az-Santos}, T. and {Eisenhardt}, P.~R.~M. and {Jun}, H.~D. and {Stern}, D. and {Tsai}, C. -W. and {Walton}, D.~J. and {Wu}, J.~W.},
        title = "{Imaging Polarization of the Blue-excess Hot Dust-obscured Galaxy WISE J011601.41-050504.0}",
      journal = {\apj},
         year = 2022,
        month = aug,
       volume = {934},
       number = {2},
          eid = {101},
        pages = {101},
          doi = {10.3847/1538-4357/ac77fc},
archivePrefix = {arXiv},
       eprint = {2206.04093},
 primaryClass = {astro-ph.GA},
       adsurl = {https://ui.adsabs.harvard.edu/abs/2022ApJ...934..101A}
}

@ARTICLE{2005Natur.433..604D,
       author = {{Di Matteo}, Tiziana and {Springel}, Volker and {Hernquist}, Lars},
        title = "{Energy input from quasars regulates the growth and activity of black holes and their host galaxies}",
      journal = {\nat},
         year = 2005,
        month = feb,
       volume = {433},
       number = {7026},
        pages = {604-607},
          doi = {10.1038/nature03335},
archivePrefix = {arXiv},
       eprint = {astro-ph/0502199},
 primaryClass = {astro-ph},
       adsurl = {https://ui.adsabs.harvard.edu/abs/2005Natur.433..604D}
}

@ARTICLE{2005ApJ...630..705H,
       author = {{Hopkins}, Philip F. and {Hernquist}, Lars and {Cox}, Thomas J. and {Di Matteo}, Tiziana and {Martini}, Paul and {Robertson}, Brant and {Springel}, Volker},
        title = "{Black Holes in Galaxy Mergers: Evolution of Quasars}",
      journal = {\apj},
         year = 2005,
        month = sep,
       volume = {630},
       number = {2},
        pages = {705-715},
          doi = {10.1086/432438},
archivePrefix = {arXiv},
       eprint = {astro-ph/0504190},
 primaryClass = {astro-ph},
       adsurl = {https://ui.adsabs.harvard.edu/abs/2005ApJ...630..705H}
}

@ARTICLE{2020MNRAS.494.4802F,
       author = {{Fawcett}, V.~A. and {Alexander}, D.~M. and {Rosario}, D.~J. and {Klindt}, L. and {Fotopoulou}, S. and {Lusso}, E. and {Morabito}, L.~K. and {Calistro Rivera}, G.},
        title = "{Fundamental differences in the radio properties of red and blue quasars: enhanced compact AGN emission in red quasars}",
      journal = {\mnras},
         year = 2020,
        month = jun,
       volume = {494},
       number = {4},
        pages = {4802-4818},
          doi = {10.1093/mnras/staa954},
archivePrefix = {arXiv},
       eprint = {2004.01197},
 primaryClass = {astro-ph.GA},
       adsurl = {https://ui.adsabs.harvard.edu/abs/2020MNRAS.494.4802F}
}

@ARTICLE{2012ApJ...761..184W,
       author = {{Weedman}, Daniel and {Sargsyan}, Lusine and {Lebouteiller}, Vianney and {Houck}, James and {Barry}, Donald},
        title = "{Infrared Classification and Luminosities for Dusty Active Galactic Nuclei and the Most Luminous Quasars}",
      journal = {\apj},
         year = 2012,
        month = dec,
       volume = {761},
       number = {2},
          eid = {184},
        pages = {184},
          doi = {10.1088/0004-637X/761/2/184},
archivePrefix = {arXiv},
       eprint = {1211.0683},
 primaryClass = {astro-ph.CO},
       adsurl = {https://ui.adsabs.harvard.edu/abs/2012ApJ...761..184W}
}

@ARTICLE{2016ApJ...819..111A,
       author = {{Assef}, R.~J. and {Walton}, D.~J. and {Brightman}, M. and {Stern}, D. and {Alexander}, D. and {Bauer}, F. and {Blain}, A.~W. and {Diaz-Santos}, T. and {Eisenhardt}, P.~R.~M. and {Finkelstein}, S.~L. and {Hickox}, R.~C. and {Tsai}, C. -W. and {Wu}, J.~W.},
        title = "{Hot Dust Obscured Galaxies with Excess Blue Light: Dual AGN or Single AGN Under Extreme Conditions?}",
      journal = {\apj},
         year = 2016,
        month = mar,
       volume = {819},
       number = {2},
          eid = {111},
        pages = {111},
          doi = {10.3847/0004-637X/819/2/111},
archivePrefix = {arXiv},
       eprint = {1511.05155},
 primaryClass = {astro-ph.GA},
       adsurl = {https://ui.adsabs.harvard.edu/abs/2016ApJ...819..111A}
}

@ARTICLE{2018MNRAS.479.4936A,
       author = {{Alexandroff}, Rachael M. and {Zakamska}, Nadia L. and {Barth}, Aaron J. and {Hamann}, Fred and {Strauss}, Michael A. and {Krolik}, Julian and {Greene}, Jenny E. and {P{\^a}ris}, Isabelle and {Ross}, Nicholas P.},
        title = "{Spectropolarimetry of high-redshift obscured and red quasars}",
      journal = {\mnras},
         year = 2018,
        month = oct,
       volume = {479},
       number = {4},
        pages = {4936-4957},
          doi = {10.1093/mnras/sty1685},
archivePrefix = {arXiv},
       eprint = {1806.10138},
 primaryClass = {astro-ph.GA},
       adsurl = {https://ui.adsabs.harvard.edu/abs/2018MNRAS.479.4936A}
}

@ARTICLE{2021MNRAS.501.3061T,
       author = {{Temple}, Matthew J. and {Banerji}, Manda and {Hewett}, Paul C. and {Rankine}, Amy L. and {Richards}, Gordon T.},
        title = "{Exploring the link between C IV outflow kinematics and sublimation-temperature dust in quasars}",
      journal = {\mnras},
         year = 2021,
        month = feb,
       volume = {501},
       number = {2},
        pages = {3061-3073},
          doi = {10.1093/mnras/staa3842},
archivePrefix = {arXiv},
       eprint = {2012.01425},
 primaryClass = {astro-ph.GA},
       adsurl = {https://ui.adsabs.harvard.edu/abs/2021MNRAS.501.3061T}
}

@ARTICLE{Richards:03,
       author = {{Richards}, Gordon T. and {Hall}, Patrick B. and {Vanden Berk}, Daniel E. and {Strauss}, Michael A. and {Schneider}, Donald P. and {Weinstein}, Michael A. and {Reichard}, Timothy A. and {York}, Donald G. and {Knapp}, G.~R. and {Fan}, Xiaohui and {Ivezi{\'c}}, {\v{Z}}eljko and {Brinkmann}, J. and {Budav{\'a}ri}, Tam{\'a}s and {Csabai}, Istv{\'a}n and {Nichol}, R.~C.},
        title = "{Red and Reddened Quasars in the Sloan Digital Sky Survey}",
      journal = {\aj},
         year = 2003,
        month = sep,
       volume = {126},
       number = {3},
        pages = {1131-1147},
          doi = {10.1086/377014},
archivePrefix = {arXiv},
       eprint = {astro-ph/0305305},
 primaryClass = {astro-ph},
       adsurl = {https://ui.adsabs.harvard.edu/abs/2003AJ....126.1131R}
}

@ARTICLE{2001MNRAS.326..255C,
       author = {{Cole}, Shaun and {Norberg}, Peder and {Baugh}, Carlton M. and {Frenk}, Carlos S. and {Bland-Hawthorn}, Joss and {Bridges}, Terry and {Cannon}, Russell and {Colless}, Matthew and {Collins}, Chris and {Couch}, Warrick and {Cross}, Nicholas and {Dalton}, Gavin and {De Propris}, Roberto and {Driver}, Simon P. and {Efstathiou}, George and {Ellis}, Richard S. and {Glazebrook}, Karl and {Jackson}, Carole and {Lahav}, Ofer and {Lewis}, Ian and {Lumsden}, Stuart and {Maddox}, Steve and {Madgwick}, Darren and {Peacock}, John A. and {Peterson}, Bruce A. and {Sutherland}, Will and {Taylor}, Keith},
        title = "{The 2dF galaxy redshift survey: near-infrared galaxy luminosity functions}",
      journal = {\mnras},
         year = 2001,
        month = sep,
       volume = {326},
       number = {1},
        pages = {255-273},
          doi = {10.1046/j.1365-8711.2001.04591.x},
archivePrefix = {arXiv},
       eprint = {astro-ph/0012429},
 primaryClass = {astro-ph},
       adsurl = {https://ui.adsabs.harvard.edu/abs/2001MNRAS.326..255C}
}

@ARTICLE{2003ApJ...584..203H,
       author = {{Huang}, J. -S. and {Glazebrook}, K. and {Cowie}, L.~L. and {Tinney}, C.},
        title = "{The Hawaii+Anglo-Australian Observatory K-Band Galaxy Redshift Survey. I. The Local K-Band Luminosity Function}",
      journal = {\apj},
         year = 2003,
        month = feb,
       volume = {584},
       number = {1},
        pages = {203-209},
          doi = {10.1086/345619},
archivePrefix = {arXiv},
       eprint = {astro-ph/0209440},
 primaryClass = {astro-ph},
       adsurl = {https://ui.adsabs.harvard.edu/abs/2003ApJ...584..203H}
}

@article{Benson_2003,
doi = {10.1086/379160},
url = {https://dx.doi.org/10.1086/379160},
year = {2003},
month = {dec},
publisher = {},
volume = {599},
number = {1},
pages = {38},
author = {A. J. Benson and R. G. Bower and C. S. Frenk and C. G. Lacey and C. M. Baugh and S. Cole},
title = {What Shapes the Luminosity Function of Galaxies?},
journal = {\apj},
}

@article{Granato_2004,
doi = {10.1086/379875},
url = {https://dx.doi.org/10.1086/379875},
year = {2004},
month = {jan},
publisher = {},
volume = {600},
number = {2},
pages = {580},
author = {Gian Luigi Granato and Gianfranco De Zotti and Laura Silva and Alessandro Bressan and Luigi Danese},
title = {A Physical Model for the Coevolution of QSOs and Their Spheroidal Hosts},
journal = {\apj},
}

@article{HOP_ELV,
    author = {Hopkins, Philip F. and Elvis, Martin},
    title = "{Quasar feedback: more bang for your buck}",
    journal = {\mnras},
    volume = {401},
    number = {1},
    pages = {7-14},
    year = {2009},
    month = {12},
    issn = {0035-8711},
    doi = {10.1111/j.1365-2966.2009.15643.x},
    url = {https://doi.org/10.1111/j.1365-2966.2009.15643.x},
    eprint = {https://academic.oup.com/mnras/article-pdf/401/1/7/18573428/mnras0401-0007.pdf},
}

@ARTICLE{2018MNRAS.479.2079C,
       author = {{Costa}, Tiago and {Rosdahl}, Joakim and {Sijacki}, Debora and {Haehnelt}, Martin G.},
        title = "{Quenching star formation with quasar outflows launched by trapped IR radiation}",
      journal = {\mnras},
         year = 2018,
        month = sep,
       volume = {479},
       number = {2},
        pages = {2079-2111},
          doi = {10.1093/mnras/sty1514},
archivePrefix = {arXiv},
       eprint = {1709.08638},
 primaryClass = {astro-ph.GA},
       adsurl = {https://ui.adsabs.harvard.edu/abs/2018MNRAS.479.2079C}
}

@ARTICLE{2015MNRAS.446..521S,
       author = {{Schaye}, Joop and {Crain}, Robert A. and {Bower}, Richard G. and {Furlong}, Michelle and {Schaller}, Matthieu and {Theuns}, Tom and {Dalla Vecchia}, Claudio and {Frenk}, Carlos S. and {McCarthy}, I.~G. and {Helly}, John C. and {Jenkins}, Adrian and {Rosas-Guevara}, Y.~M. and {White}, Simon D.~M. and {Baes}, Maarten and {Booth}, C.~M. and {Camps}, Peter and {Navarro}, Julio F. and {Qu}, Yan and {Rahmati}, Alireza and {Sawala}, Till and {Thomas}, Peter A. and {Trayford}, James},
        title = "{The EAGLE project: simulating the evolution and assembly of galaxies and their environments}",
      journal = {\mnras},
         year = 2015,
        month = jan,
       volume = {446},
       number = {1},
        pages = {521-554},
          doi = {10.1093/mnras/stu2058},
archivePrefix = {arXiv},
       eprint = {1407.7040},
 primaryClass = {astro-ph.GA},
       adsurl = {https://ui.adsabs.harvard.edu/abs/2015MNRAS.446..521S}
}

@ARTICLE{2009MNRAS.394L..51B,
       author = {{Bluck}, Asa F.~L. and {Conselice}, Christopher J. and {Bouwens}, Rychard J. and {Daddi}, Emanuele and {Dickinson}, Mark and {Papovich}, Casey and {Yan}, Haojing},
        title = "{A surprisingly high pair fraction for extremely massive galaxies at z \raisebox{-0.5ex}\textasciitilde 3 in the GOODS NICMOS survey}",
      journal = {\mnras},
         year = 2009,
        month = mar,
       volume = {394},
       number = {1},
        pages = {L51-L55},
          doi = {10.1111/j.1745-3933.2008.00608.x},
archivePrefix = {arXiv},
       eprint = {0812.0926},
 primaryClass = {astro-ph},
       adsurl = {https://ui.adsabs.harvard.edu/abs/2009MNRAS.394L..51B}
}

@ARTICLE{2012ApJ...744...85M,
       author = {{Man}, Allison W.~S. and {Toft}, Sune and {Zirm}, Andrew W. and {Wuyts}, Stijn and {van der Wel}, Arjen},
        title = "{The Pair Fraction of Massive Galaxies at 0 \ensuremath{\leq} z \ensuremath{\leq} 3}",
      journal = {\apj},
         year = 2012,
        month = jan,
       volume = {744},
       number = {2},
          eid = {85},
        pages = {85},
          doi = {10.1088/0004-637X/744/2/85},
archivePrefix = {arXiv},
       eprint = {1109.2895},
 primaryClass = {astro-ph.CO},
       adsurl = {https://ui.adsabs.harvard.edu/abs/2012ApJ...744...85M}
}

@ARTICLE{Ishibashi:17,
       author = {{Ishibashi}, W. and {Banerji}, M. and {Fabian}, A.~C.},
        title = "{AGN radiative feedback in dusty quasar populations}",
      journal = {\mnras},
         year = 2017,
        month = aug,
       volume = {469},
       number = {2},
        pages = {1496-1501},
          doi = {10.1093/mnras/stx921},
archivePrefix = {arXiv},
       eprint = {1704.03712},
 primaryClass = {astro-ph.GA},
       adsurl = {https://ui.adsabs.harvard.edu/abs/2017MNRAS.469.1496I}
}

@article{York_2000,
doi = {10.1086/301513},
url = {https://dx.doi.org/10.1086/301513},
year = {2000},
month = {sep},
publisher = {},
volume = {120},
number = {3},
pages = {1579},
author = {York, Donald G. and Adelman, J. and Anderson, Jr., John E. and Anderson, Scott F. and Annis, James and Bahcall, Neta A. and Bakken, J. A. and Barkhouser, Robert and Bastian, Steven and Berman, Eileen and Boroski, William N. and Bracker, Steve and Briegel, Charlie and Briggs, John W. and Brinkmann, J. and Brunner, Robert and Burles, Scott and Carey, Larry and Carr, Michael A. and Castander, Francisco J. and Chen, Bing and Colestock, Patrick L. and Connolly, A. J. and Crocker, J. H. and Csabai, István and Czarapata, Paul C. and Davis, John Eric and Doi, Mamoru and Dombeck, Tom and Eisenstein, Daniel and Ellman, Nancy and Elms, Brian R. and Evans, Michael L. and Fan, Xiaohui and Federwitz, Glenn R. and Fiscelli, Larry and Friedman, Scott and Frieman, Joshua A. and Fukugita, Masataka and Gillespie, Bruce and Gunn, James E. and Gurbani, Vijay K. and de Haas, Ernst and Haldeman, Merle and Harris, Frederick H. and Hayes, J. and Heckman, Timothy M. and Hennessy, G. S. and Hindsley, Robert B. and Holm, Scott and Holmgren, Donald J. and Huang, Chi-hao and Hull, Charles and Husby, Don and Ichikawa, Shin-Ichi and Ichikawa, Takashi and Ivezić, Željko and Kent, Stephen and Kim, Rita S. J. and Kinney, E. and Klaene, Mark and Kleinman, A. N. and Kleinman, S. and Knapp, G. R. and Korienek, John and Kron, Richard G. and Kunszt, Peter Z. and Lamb, D. Q. and Lee, B. and Leger, R. French and Limmongkol, Siriluk and Lindenmeyer, Carl and Long, Daniel C. and Loomis, Craig and Loveday, Jon and Lucinio, Rich and Lupton, Robert H. and MacKinnon, Bryan and Mannery, Edward J. and Mantsch, P. M. and Margon, Bruce and McGehee, Peregrine and McKay, Timothy A. and Meiksin, Avery and Merelli, Aronne and Monet, David G. and Munn, Jeffrey A. and Narayanan, Vijay K. and Nash, Thomas and Neilsen, Eric and Neswold, Rich and Newberg, Heidi Jo and Nichol, R. C. and Nicinski, Tom and Nonino, Mario and Okada, Norio and Okamura, Sadanori and Ostriker, Jeremiah P. and Owen, Russell and Pauls, A. George and Peoples, John and Peterson, R. L. and Petravick, Donald and Pier, Jeffrey R. and Pope, Adrian and Pordes, Ruth and Prosapio, Angela and Rechenmacher, Ron and Quinn, Thomas R. and Richards, Gordon T. and Richmond, Michael W. and Rivetta, Claudio H. and Rockosi, Constance M. and Ruthmansdorfer, Kurt and Sandford, Dale and Schlegel, David J. and Schneider, Donald P. and Sekiguchi, Maki and Sergey, Gary and Shimasaku, Kazuhiro and Siegmund, Walter A. and Smee, Stephen and Smith, J. Allyn and Snedden, S. and Stone, R. and Stoughton, Chris and Strauss, Michael A. and Stubbs, Christopher and SubbaRao, Mark and Szalay, Alexander S. and Szapudi, Istvan and Szokoly, Gyula P. and Thakar, Anirudda R. and Tremonti, Christy and Tucker, Douglas L. and Uomoto, Alan and Vanden Berk, Dan and Vogeley, Michael S. and Waddell, Patrick and Wang, Shu-i and Watanabe, Masaru and Weinberg, David H. and Yanny, Brian and Yasuda, Naoki},
title = {The Sloan Digital Sky Survey: Technical Summary},
journal = {The Astronomical Journal}
}

@ARTICLE{2023MNRAS.525.2716Z,
       author = {{Zakamska}, Nadia L. and {Alexandroff}, Rachael M.},
        title = "{Torus skin outflow in a near-Eddington quasar revealed by spectropolarimetry}",
      journal = {\mnras},
         year = 2023,
        month = oct,
       volume = {525},
       number = {2},
        pages = {2716-2730},
          doi = {10.1093/mnras/stad2474},
archivePrefix = {arXiv},
       eprint = {2306.06303},
 primaryClass = {astro-ph.GA},
       adsurl = {https://ui.adsabs.harvard.edu/abs/2023MNRAS.525.2716Z}
}

@ARTICLE{2020MNRAS.492.4553R,
       author = {{Rankine}, Amy L. and {Hewett}, Paul C. and {Banerji}, Manda and {Richards}, Gordon T.},
        title = "{BAL and non-BAL quasars: continuum, emission, and absorption properties establish a common parent sample}",
      journal = {\mnras},
         year = 2020,
        month = mar,
       volume = {492},
       number = {3},
        pages = {4553-4575},
          doi = {10.1093/mnras/staa130},
archivePrefix = {arXiv},
       eprint = {1912.08700},
 primaryClass = {astro-ph.GA},
       adsurl = {https://ui.adsabs.harvard.edu/abs/2020MNRAS.492.4553R}
}

@article{10.1111/j.1365-2966.2010.16648.x,
    author = {Hewett, Paul C. and Wild, Vivienne},
    title = "{Improved redshifts for SDSS quasar spectra}",
    journal = {\mnras},
    volume = {405},
    number = {4},
    pages = {2302-2316},
    year = {2010},
    month = {07},
    issn = {0035-8711},
    doi = {10.1111/j.1365-2966.2010.16648.x},
    url = {https://doi.org/10.1111/j.1365-2966.2010.16648.x},
    eprint = {https://academic.oup.com/mnras/article-pdf/405/4/2302/18745236/mnras0405-2302.pdf},
}

@ARTICLE{2010CAMCS...5...65G,
       author = {{Goodman}, Jonathan and {Weare}, Jonathan},
        title = "{Ensemble samplers with affine invariance}",
      journal = {Communications in Applied Mathematics and Computational Science},
         year = 2010,
        month = jan,
       volume = {5},
       number = {1},
        pages = {65-80},
          doi = {10.2140/camcos.2010.5.65},
       adsurl = {https://ui.adsabs.harvard.edu/abs/2010CAMCS...5...65G}
}

@article{article,
author = {Rosa, Gisella and Venemans, Bram and Decarli, Roberto and Gennaro, Mario and Simcoe, Robert and Dietrich, Matthias and Peterson, Bradley and Walter, Fabian and Frank, Stephan and McMahon, Richard and Hewett, Paul and Mortlock, Daniel and Simpson, Chris and Warren, S.},
year = {2013},
month = {11},
pages = {},
title = {Black hole mass estimates and emission line properties of a sample of redshift Z \ensuremath{>} 6:5 quasars},
volume = {790},
journal = {\apj},
doi = {10.1088/0004-637X/790/2/145}
}

@ARTICLE{2012MNRAS.427.2275B,
       author = {{Banerji}, Manda and {McMahon}, Richard G. and {Hewett}, Paul C. and {Alaghband-Zadeh}, Susannah and {Gonzalez-Solares}, Eduardo and {Venemans}, Bram P. and {Hawthorn}, Melanie J.},
        title = "{Heavily reddened quasars at z \ensuremath{\sim} 2 in the UKIDSS Large Area Survey: a transitional phase in AGN evolution}",
      journal = {\mnras},
         year = 2012,
        month = dec,
       volume = {427},
       number = {3},
        pages = {2275-2291},
          doi = {10.1111/j.1365-2966.2012.22099.x},
archivePrefix = {arXiv},
       eprint = {1203.5530},
 primaryClass = {astro-ph.CO},
       adsurl = {https://ui.adsabs.harvard.edu/abs/2012MNRAS.427.2275B}
}

@ARTICLE{2015MNRAS.447.3368B,
       author = {{Banerji}, Manda and {Alaghband-Zadeh}, S. and {Hewett}, Paul C. and {McMahon}, Richard G.},
        title = "{Heavily reddened type 1 quasars at z \ensuremath{>} 2 - I. Evidence for significant obscured black hole growth at the highest quasar luminosities}",
      journal = {\mnras},
         year = 2015,
        month = mar,
       volume = {447},
       number = {4},
        pages = {3368-3389},
          doi = {10.1093/mnras/stu2649},
archivePrefix = {arXiv},
       eprint = {1501.00815},
 primaryClass = {astro-ph.GA},
       adsurl = {https://ui.adsabs.harvard.edu/abs/2015MNRAS.447.3368B}
}

@ARTICLE{2024ApJ...964...69S,
       author = {{Storey-Fisher}, Kate and {Hogg}, David W. and {Rix}, Hans-Walter and {Eilers}, Anna-Christina and {Fabbian}, Giulio and {Blanton}, Michael R. and {Alonso}, David},
        title = "{Quaia, the Gaia-unWISE Quasar Catalog: An All-sky Spectroscopic Quasar Sample}",
      journal = {\apj},
         year = 2024,
        month = mar,
       volume = {964},
       number = {1},
          eid = {69},
        pages = {69},
          doi = {10.3847/1538-4357/ad1328},
archivePrefix = {arXiv},
       eprint = {2306.17749},
 primaryClass = {astro-ph.GA},
       adsurl = {https://ui.adsabs.harvard.edu/abs/2024ApJ...964...69S}
}

@ARTICLE{2023MNRAS.523..646T,
       author = {{Temple}, Matthew J. and {Matthews}, James H. and {Hewett}, Paul C. and {Rankine}, Amy L. and {Richards}, Gordon T. and {Banerji}, Manda and {Ferland}, Gary J. and {Knigge}, Christian and {Stepney}, Matthew},
        title = "{Testing AGN outflow and accretion models with C IV and He II emission line demographics in z {\ensuremath{\approx}} 2 quasars}",
      journal = {\mnras},
         year = 2023,
        month = jul,
       volume = {523},
       number = {1},
        pages = {646-666},
          doi = {10.1093/mnras/stad1448},
archivePrefix = {arXiv},
       eprint = {2301.02675},
 primaryClass = {astro-ph.GA},
       adsurl = {https://ui.adsabs.harvard.edu/abs/2023MNRAS.523..646T}
}

@ARTICLE{1976ApJ...210..642A,
       author = {{Avni}, Y.},
        title = "{Energy spectra of X-ray clusters of galaxies.}",
      journal = {\apj},
         year = 1976,
        month = dec,
       volume = {210},
        pages = {642-646},
          doi = {10.1086/154870},
       adsurl = {https://ui.adsabs.harvard.edu/abs/1976ApJ...210..642A}
}

@ARTICLE{2025MNRAS.537.2003F,
       author = {{Fawcett}, V.~A. and {Harrison}, C.~M. and {Alexander}, D.~M. and {Morabito}, L.~K. and {Kharb}, P. and {Rosario}, D.~J. and {Baghel}, Janhavi and {Ghosh}, Salmoli and {Silpa}, Sasikumar and {Petley}, J. and {Sargent}, C. and {Calistro Rivera}, G.},
        title = "{Connection between steep radio spectral slopes and dust extinction in QSOs: evidence for outflow-driven shocks in dusty QSOs}",
      journal = {\mnras},
         year = 2025,
        month = feb,
       volume = {537},
       number = {2},
        pages = {2003-2023},
          doi = {10.1093/mnras/staf105},
archivePrefix = {arXiv},
       eprint = {2501.10501},
 primaryClass = {astro-ph.GA},
       adsurl = {https://ui.adsabs.harvard.edu/abs/2025MNRAS.537.2003F}
}

@ARTICLE{2007ApJ...663...81P,
       author = {{Polletta}, M. and {Tajer}, M. and {Maraschi}, L. and {Trinchieri}, G. and {Lonsdale}, C.~J. and {Chiappetti}, L. and {Andreon}, S. and {Pierre}, M. and {Le F{\`e}vre}, O. and {Zamorani}, G. and {Maccagni}, D. and {Garcet}, O. and {Surdej}, J. and {Franceschini}, A. and {Alloin}, D. and {Shupe}, D.~L. and {Surace}, J.~A. and {Fang}, F. and {Rowan-Robinson}, M. and {Smith}, H.~E. and {Tresse}, L.},
        title = "{Spectral Energy Distributions of Hard X-Ray Selected Active Galactic Nuclei in the XMM-Newton Medium Deep Survey}",
      journal = {\apj},
         year = 2007,
        month = jul,
       volume = {663},
       number = {1},
        pages = {81-102},
          doi = {10.1086/518113},
archivePrefix = {arXiv},
       eprint = {astro-ph/0703255},
 primaryClass = {astro-ph},
       adsurl = {https://ui.adsabs.harvard.edu/abs/2007ApJ...663...81P}
}

@ARTICLE{2013ApJ...779..104J,
       author = {{Jun}, Hyunsung David and {Im}, Myungshin},
        title = "{Physical Properties of Luminous Dust-poor Quasars}",
      journal = {\apj},
         year = 2013,
        month = dec,
       volume = {779},
       number = {2},
          eid = {104},
        pages = {104},
          doi = {10.1088/0004-637X/779/2/104},
archivePrefix = {arXiv},
       eprint = {1310.3034},
 primaryClass = {astro-ph.CO},
       adsurl = {https://ui.adsabs.harvard.edu/abs/2013ApJ...779..104J}
}

@ARTICLE{2007MNRAS.379.1599L,
       author = {{Lawrence}, A. and {Warren}, S.~J. and {Almaini}, O. and {Edge}, A.~C. and {Hambly}, N.~C. and {Jameson}, R.~F. and {Lucas}, P. and {Casali}, M. and {Adamson}, A. and {Dye}, S. and {Emerson}, J.~P. and {Foucaud}, S. and {Hewett}, P. and {Hirst}, P. and {Hodgkin}, S.~T. and {Irwin}, M.~J. and {Lodieu}, N. and {McMahon}, R.~G. and {Simpson}, C. and {Smail}, I. and {Mortlock}, D. and {Folger}, M.},
        title = "{The UKIRT Infrared Deep Sky Survey (UKIDSS)}",
      journal = {\mnras},
         year = 2007,
        month = aug,
       volume = {379},
       number = {4},
        pages = {1599-1617},
          doi = {10.1111/j.1365-2966.2007.12040.x},
archivePrefix = {arXiv},
       eprint = {astro-ph/0604426},
 primaryClass = {astro-ph},
       adsurl = {https://ui.adsabs.harvard.edu/abs/2007MNRAS.379.1599L}
}

@ARTICLE{2013Msngr.154...35M,
       author = {{McMahon}, R.~G. and {Banerji}, M. and {Gonzalez}, E. and {Koposov}, S.~E. and {Bejar}, V.~J. and {Lodieu}, N. and {Rebolo}, R. and {VHS Collaboration}},
        title = "{First Scientific Results from the VISTA Hemisphere Survey (VHS)}",
      journal = {The Messenger},
         year = 2013,
        month = dec,
       volume = {154},
        pages = {35-37},
       adsurl = {https://ui.adsabs.harvard.edu/abs/2013Msngr.154...35M}
}

@ARTICLE{2010Natur.464..380J,
       author = {{Jiang}, Linhua and {Fan}, Xiaohui and {Brandt}, W.~N. and {Carilli}, Chris L. and {Egami}, Eiichi and {Hines}, Dean C. and {Kurk}, Jaron D. and {Richards}, Gordon T. and {Shen}, Yue and {Strauss}, Michael A. and {Vestergaard}, Marianne and {Walter}, Fabian},
        title = "{Dust-free quasars in the early Universe}",
      journal = {\nat},
         year = 2010,
        month = mar,
       volume = {464},
       number = {7287},
        pages = {380-383},
          doi = {10.1038/nature08877},
archivePrefix = {arXiv},
       eprint = {1003.3432},
 primaryClass = {astro-ph.CO},
       adsurl = {https://ui.adsabs.harvard.edu/abs/2010Natur.464..380J}
}

@ARTICLE{2025A&A...697A...1E,
       author = {{Euclid Collaboration} and {Mellier}, Y. and {Abdurro'uf} and {Acevedo Barroso}, J.~A. and {Ach{\'u}carro}, A. and {Adamek}, J. and {Adam}, R. and {Addison}, G.~E. and {Aghanim}, N. and {Aguena}, M. and {Ajani}, V. and {Akrami}, Y. and {Al-Bahlawan}, A. and {Alavi}, A. and {Albuquerque}, I.~S. and {Alestas}, G. and {Alguero}, G. and {Allaoui}, A. and {Allen}, S.~W. and {Allevato}, V. and {Alonso-Tetilla}, A.~V. and {Altieri}, B. and {Alvarez-Candal}, A. and {Alvi}, S. and {Amara}, A. and {Amendola}, L. and {Amiaux}, J. and {Andika}, I.~T. and {Andreon}, S. and {Andrews}, A. and {Angora}, G. and {Angulo}, R.~E. and {Annibali}, F. and {Anselmi}, A. and {Anselmi}, S. and {Arcari}, S. and {Archidiacono}, M. and {Aric{\`o}}, G. and {Arnaud}, M. and {Arnouts}, S. and {Asgari}, M. and {Asorey}, J. and {Atayde}, L. and {Atek}, H. and {Atrio-Barandela}, F. and {Aubert}, M. and {Aubourg}, E. and {Auphan}, T. and {Auricchio}, N. and {Aussel}, B. and {Aussel}, H. and {Avelino}, P.~P. and {Avgoustidis}, A. and {Avila}, S. and {Awan}, S. and {Azzollini}, R. and {Baccigalupi}, C. and {Bachelet}, E. and {Bacon}, D. and {Baes}, M. and {Bagley}, M.~B. and {Bahr-Kalus}, B. and {Balaguera-Antolinez}, A. and {Balbinot}, E. and {Balcells}, M. and {Baldi}, M. and {Baldry}, I. and {Balestra}, A. and {Ballardini}, M. and {Ballester}, O. and {Balogh}, M. and {Ba{\~n}ados}, E. and {Barbier}, R. and {Bardelli}, S. and {Baron}, M. and {Barreiro}, T. and {Barrena}, R. and {Barriere}, J. -C. and {Barros}, B.~J. and {Barthelemy}, A. and {Bartolo}, N. and {Basset}, A. and {Battaglia}, P. and {Battisti}, A.~J. and {Baugh}, C.~M. and {Baumont}, L. and {Bazzanini}, L. and {Beaulieu}, J. -P. and {Beckmann}, V. and {Belikov}, A.~N. and {Bel}, J. and {Bellagamba}, F. and {Bella}, M. and {Bellini}, E. and {Benabed}, K. and {Bender}, R. and {Benevento}, G. and {Bennett}, C.~L. and {Benson}, K. and {Bergamini}, P. and {Bermejo-Climent}, J.~R. and {Bernardeau}, F. and {Bertacca}, D. and {Berthe}, M. and {Berthier}, J. and {Bethermin}, M. and {Beutler}, F. and {Bevillon}, C. and {Bhargava}, S. and {Bhatawdekar}, R. and {Bianchi}, D. and {Bisigello}, L. and {Biviano}, A. and {Blake}, R.~P. and {Blanchard}, A. and {Blazek}, J. and {Blot}, L. and {Bosco}, A. and {Bodendorf}, C. and {Boenke}, T. and {B{\"o}hringer}, H. and {Boldrini}, P. and {Bolzonella}, M. and {Bonchi}, A. and {Bonici}, M. and {Bonino}, D. and {Bonino}, L. and {Bonvin}, C. and {Bon}, W. and {Booth}, J.~T. and {Borgani}, S. and {Borlaff}, A.~S. and {Borsato}, E. and {Bose}, B. and {Botticella}, M.~T. and {Boucaud}, A. and {Bouche}, F. and {Boucher}, J.~S. and {Boutigny}, D. and {Bouvard}, T. and {Bouwens}, R. and {Bouy}, H. and {Bowler}, R.~A.~A. and {Bozza}, V. and {Bozzo}, E. and {Branchini}, E. and {Brando}, G. and {Brau-Nogue}, S. and {Brekke}, P. and {Bremer}, M.~N. and {Brescia}, M. and {Breton}, M. -A. and {Brinchmann}, J. and {Brinckmann}, T. and {Brockley-Blatt}, C. and {Brodwin}, M. and {Brouard}, L. and {Brown}, M.~L. and {Bruton}, S. and {Bucko}, J. and {Buddelmeijer}, H. and {Buenadicha}, G. and {Buitrago}, F. and {Burger}, P. and {Burigana}, C. and {Busillo}, V. and {Busonero}, D. and {Cabanac}, R. and {Cabayol-Garcia}, L. and {Cagliari}, M.~S. and {Caillat}, A. and {Caillat}, L. and {Calabrese}, M. and {Calabro}, A. and {Calderone}, G. and {Calura}, F. and {Camacho Quevedo}, B. and {Camera}, S. and {Campos}, L. and {Ca{\~n}as-Herrera}, G. and {Candini}, G.~P. and {Cantiello}, M. and {Capobianco}, V. and {Cappellaro}, E. and {Cappelluti}, N. and {Cappi}, A. and {Caputi}, K.~I. and {Cara}, C. and {Carbone}, C. and {Cardone}, V.~F. and {Carella}, E. and {Carlberg}, R.~G. and {Carle}, M. and {Carminati}, L. and {Caro}, F. and {Carrasco}, J.~M. and {Carretero}, J. and {Carrilho}, P. and {Carron Duque}, J. and {Carry}, B.},
        title = "{Euclid: I. Overview of the Euclid mission}",
      journal = {\aap},
         year = 2025,
        month = may,
       volume = {697},
          eid = {A1},
        pages = {A1},
          doi = {10.1051/0004-6361/202450810},
archivePrefix = {arXiv},
       eprint = {2405.13491},
 primaryClass = {astro-ph.CO},
       adsurl = {https://ui.adsabs.harvard.edu/abs/2025A&A...697A...1E}
}

@ARTICLE{2025A&A...697A...2E,
       author = {{Euclid Collaboration} and {Cropper}, M.~S. and {Al-Bahlawan}, A. and {Amiaux}, J. and {Awan}, S. and {Azzollini}, R. and {Benson}, K. and {Berthe}, M. and {Boucher}, J. and {Bozzo}, E. and {Brockley-Blatt}, C. and {Candini}, G.~P. and {Cara}, C. and {Chaudery}, R.~A. and {Cole}, R.~E. and {Danto}, P. and {Denniston}, J. and {Di Giorgio}, A.~M. and {Dryer}, B. and {Dubois}, J. -P. and {Endicott}, J. and {Farina}, M. and {Galli}, E. and {Genolet}, L. and {Gow}, J.~P.~D. and {Guttridge}, P. and {Hailey}, M. and {Hall}, D. and {Harper}, C. and {Hoekstra}, H. and {Holland}, A.~D. and {Horeau}, B. and {Hu}, D. and {James}, R.~E. and {Khalil}, A. and {King}, R. and {Kitching}, T. and {Kohley}, R. and {Larcheveque}, C. and {Lawrenson}, A. and {Liebing}, P. and {Liu}, S.~J. and {Martignac}, J. and {Massey}, R. and {McCracken}, H.~J. and {Miller}, L. and {Murray}, N. and {Nakajima}, R. and {Niemi}, S. -M. and {Nightingale}, J.~W. and {Paltani}, S. and {Pendem}, A. and {Philippon}, A. and {Plana}, C. and {Pool}, P. and {Pottinger}, S. and {Racca}, G.~D. and {Rhodes}, J. and {Rousseau}, A. and {Ruane}, K. and {Salatti}, M. and {Salvignol}, J. -C. and {Sciortino}, A. and {Short}, A. and {Skottfelt}, J. and {Smit}, S.~J.~A. and {Swindells}, I. and {Szafraniec}, M. and {Thomas}, P.~D. and {Thomas}, W. and {Tommasi}, E. and {Tosti}, S. and {Visticot}, F. and {Walton}, D.~M. and {Willis}, G. and {Winter}, B. and {Aghanim}, N. and {Altieri}, B. and {Amara}, A. and {Andreon}, S. and {Auricchio}, N. and {Aussel}, H. and {Baccigalupi}, C. and {Baldi}, M. and {Balestra}, A. and {Bardelli}, S. and {Basset}, A. and {Bender}, R. and {Bernardeau}, F. and {Bodendorf}, C. and {Boenke}, T. and {Bonino}, D. and {Branchini}, E. and {Brescia}, M. and {Brinchmann}, J. and {Camera}, S. and {Capobianco}, V. and {Carbone}, C. and {Cardone}, V.~F. and {Carretero}, J. and {Casas}, R. and {Casas}, S. and {Castander}, F.~J. and {Castellano}, M. and {Castignani}, G. and {Cavuoti}, S. and {Cimatti}, A. and {Colodro-Conde}, C. and {Congedo}, G. and {Conselice}, C.~J. and {Conversi}, L. and {Copin}, Y. and {Courbin}, F. and {Courtois}, H.~M. and {Crocce}, M. and {Cuby}, J. -G. and {Cuillandre}, J. -C. and {Da Silva}, A. and {Degaudenzi}, H. and {De Lucia}, G. and {Dinis}, J. and {Dolding}, C. and {Douspis}, M. and {Duncan}, C.~A.~J. and {Dupac}, X. and {Dusini}, S. and {Ealet}, A. and {Fabricius}, M. and {Farrens}, S. and {Ferriol}, S. and {Fosalba}, P. and {Fotopoulou}, S. and {Frailis}, M. and {Franceschi}, E. and {Franzetti}, P. and {Frugier}, P. -A. and {Fumana}, M. and {Galeotta}, S. and {Garilli}, B. and {George}, K. and {Gillard}, W. and {Gillis}, B. and {Giocoli}, C. and {G{\'o}mez-Alvarez}, P. and {Granett}, B.~R. and {Grazian}, A. and {Grupp}, F. and {Guzzo}, L. and {Haugan}, S.~V.~H. and {Herent}, O. and {Hoar}, J. and {Holliman}, M.~S. and {Holmes}, W. and {Hook}, I. and {Hormuth}, F. and {Hornstrup}, A. and {Hudelot}, P. and {Ili{\'c}}, S. and {Jahnke}, K. and {Jhabvala}, M. and {Joachimi}, B. and {Keih{\"a}nen}, E. and {Kermiche}, S. and {Kiessling}, A. and {Kilbinger}, M. and {Kubik}, B. and {Kuijken}, K. and {K{\"u}mmel}, M. and {Kunz}, M. and {Kurki-Suonio}, H. and {Lahav}, O. and {Laureijs}, R. and {Ligori}, S. and {Lilje}, P.~B. and {Lindholm}, V. and {Lloro}, I. and {Lorenzo Alvarez}, J. and {Mainetti}, G. and {Maino}, D. and {Maiorano}, E. and {Mansutti}, O. and {Marcin}, S. and {Marggraf}, O. and {Markovic}, K. and {Martinelli}, M. and {Martinet}, N. and {Marulli}, F. and {Masters}, D.~C. and {Maurogordato}, S. and {Medinaceli}, E. and {Mei}, S. and {Melchior}, M. and {Mellier}, Y. and {Meneghetti}, M. and {Merlin}, E. and {Meylan}, G. and {Mohr}, J.~J. and {Moresco}, M. and {Moscardini}, L. and {Neissner}, C.},
        title = "{Euclid: II. The VIS instrument}",
      journal = {\aap},
         year = 2025,
        month = may,
       volume = {697},
          eid = {A2},
        pages = {A2},
          doi = {10.1051/0004-6361/202450996},
archivePrefix = {arXiv},
       eprint = {2405.13492},
 primaryClass = {astro-ph.IM},
       adsurl = {https://ui.adsabs.harvard.edu/abs/2025A&A...697A...2E}
}

@ARTICLE{2025A&A...697A...3E,
       author = {{Euclid Collaboration} and {Jahnke}, K. and {Gillard}, W. and {Schirmer}, M. and {Ealet}, A. and {Maciaszek}, T. and {Prieto}, E. and {Barbier}, R. and {Bonoli}, C. and {Corcione}, L. and {Dusini}, S. and {Grupp}, F. and {Hormuth}, F. and {Ligori}, S. and {Martin}, L. and {Morgante}, G. and {Padilla}, C. and {Toledo-Moreo}, R. and {Trifoglio}, M. and {Valenziano}, L. and {Bender}, R. and {Castander}, F.~J. and {Garilli}, B. and {Lilje}, P.~B. and {Rix}, H. -W. and {Andersen}, M.~I. and {Auricchio}, N. and {Balestra}, A. and {Barriere}, J. -C. and {Battaglia}, P. and {Berthe}, M. and {Bodendorf}, C. and {Boenke}, T. and {Bon}, W. and {Bonnefoi}, A. and {Caillat}, A. and {Capobianco}, V. and {Carle}, M. and {Casas}, R. and {Cho}, H. and {Costille}, A. and {Ducret}, F. and {Ferriol}, S. and {Franceschi}, E. and {Gimenez}, J. -L. and {Holmes}, W. and {Hornstrup}, A. and {Jhabvala}, M. and {Kohley}, R. and {Kubik}, B. and {Laureijs}, R. and {Le Mignant}, D. and {Lloro}, I. and {Medinaceli}, E. and {Mellier}, Y. and {Polenta}, G. and {Racca}, G.~D. and {Renzi}, A. and {Salvignol}, J. -C. and {Secroun}, A. and {Seidel}, G. and {Seiffert}, M. and {Sirignano}, C. and {Sirri}, G. and {Strada}, P. and {Smadja}, G. and {Stanco}, L. and {Wachter}, S. and {Anselmi}, S. and {Borsato}, E. and {Caillat}, L. and {Cogato}, F. and {Colodro-Conde}, C. and {Crouzet}, P. -E. and {Conforti}, V. and {D'Alessandro}, M. and {Copin}, Y. and {Cuillandre}, J. -C. and {Davies}, J.~E. and {Davini}, S. and {Derosa}, A. and {Diaz}, J.~J. and {Di Domizio}, S. and {Di Ferdinando}, D. and {Farinelli}, R. and {Ferrari}, A.~G. and {Fornari}, F. and {Gabarra}, L. and {Garcia}, R. and {Gutierrez}, C.~M. and {Giacomini}, F. and {Lagier}, P. and {Gianotti}, F. and {Krause}, O. and {Madrid}, F. and {Laudisio}, F. and {Macias-Perez}, J. and {Naletto}, G. and {Niclas}, M. and {Marpaud}, J. and {Mauri}, N. and {da Silva}, R. and {Passalacqua}, F. and {Paterson}, K. and {Patrizii}, L. and {Risso}, I. and {Solheim}, B.~G.~B. and {Scodeggio}, M. and {Stassi}, P. and {Steinwagner}, J. and {Tenti}, M. and {Testera}, G. and {Travaglini}, R. and {Tosi}, S. and {Troja}, A. and {Tubio}, O. and {Valieri}, C. and {Vescovi}, C. and {Ventura}, S. and {Aghanim}, N. and {Altieri}, B. and {Amara}, A. and {Amiaux}, J. and {Andreon}, S. and {Appleton}, P.~N. and {Aussel}, H. and {Baccigalupi}, C. and {Baldi}, M. and {Bardelli}, S. and {Basset}, A. and {Bonchi}, A. and {Bonino}, D. and {Branchini}, E. and {Brescia}, M. and {Brinchmann}, J. and {Camera}, S. and {Carbone}, C. and {Cardone}, V.~F. and {Carretero}, J. and {Casas}, S. and {Castellano}, M. and {Castignani}, G. and {Cavuoti}, S. and {Chabaud}, P. -Y. and {Cimatti}, A. and {Congedo}, G. and {Conselice}, C.~J. and {Conversi}, L. and {Courbin}, F. and {Courtois}, H.~M. and {Crocce}, M. and {Cropper}, M. and {Cuby}, J. -G. and {Da Silva}, A. and {Degaudenzi}, H. and {De Lucia}, G. and {Di Giorgio}, A.~M. and {Dinis}, J. and {Douspis}, M. and {Dubath}, F. and {Duncan}, C.~A.~J. and {Dupac}, X. and {Fabricius}, M. and {Farina}, M. and {Farrens}, S. and {Faustini}, F. and {Fosalba}, P. and {Fotopoulou}, S. and {Fourmanoit}, N. and {Frailis}, M. and {Franzetti}, P. and {Galeotta}, S. and {George}, K. and {Gillis}, B. and {Giocoli}, C. and {G{\'o}mez-Alvarez}, P. and {Granett}, B.~R. and {Grazian}, A. and {Guzzo}, L. and {Hailey}, M. and {Haugan}, S.~V.~H. and {Hoar}, J. and {Hoekstra}, H. and {Hook}, I. and {Hudelot}, P. and {Ili{\'c}}, S. and {Joachimi}, B. and {Keih{\"a}nen}, E. and {Kermiche}, S. and {Kiessling}, A. and {Kilbinger}, M. and {Kitching}, T. and {K{\"u}mmel}, M. and {Kunz}, M. and {Kurki-Suonio}, H. and {Lahav}, O. and {Liebing}, P. and {Lindholm}, V. and {Lorenzo Alvarez}, J. and {Mainetti}, G.},
        title = "{Euclid: III. The NISP Instrument}",
      journal = {\aap},
         year = 2025,
        month = may,
       volume = {697},
          eid = {A3},
        pages = {A3},
          doi = {10.1051/0004-6361/202450786},
archivePrefix = {arXiv},
       eprint = {2405.13493},
 primaryClass = {astro-ph.IM},
       adsurl = {https://ui.adsabs.harvard.edu/abs/2025A&A...697A...3E}
}

@ARTICLE{Zakamska:16,
       author = {{Zakamska}, Nadia L. and {Hamann}, Fred and {P{\^a}ris}, Isabelle and {Brandt}, W.~N. and {Greene}, Jenny E. and {Strauss}, Michael A. and {Villforth}, Carolin and {Wylezalek}, Dominika and {Alexandroff}, Rachael M. and {Ross}, Nicholas P.},
        title = "{Discovery of extreme [O III] {\ensuremath{\lambda}}5007 {\r{A}} outflows in high-redshift red quasars}",
      journal = {\mnras},
         year = 2016,
        month = jul,
       volume = {459},
       number = {3},
        pages = {3144-3160},
          doi = {10.1093/mnras/stw718},
archivePrefix = {arXiv},
       eprint = {1512.02642},
 primaryClass = {astro-ph.GA},
       adsurl = {https://ui.adsabs.harvard.edu/abs/2016MNRAS.459.3144Z}
}

@ARTICLE{2026arXiv260310135D,
       author = {{Davies}, Frederick B. and {Bosman}, Sarah E.~I. and {Ganguly}, Arpita and {Ba{\~n}ados}, Eduardo and {Belladitta}, Silvia and {Stern}, Daniel and {Acevedo Barroso}, Javier A. and {Yang}, Daming and {Hennawi}, Joseph F. and {Wang}, Feige and {Yang}, Jinyi and {Fan}, Xiaohui},
        title = "{Three Hundred Quasars from the Couch: A first look at high-redshift quasar discovery with SPHEREx}",
      journal = {arXiv e-prints},
         year = 2026,
        month = mar,
          eid = {arXiv:2603.10135},
        pages = {arXiv:2603.10135, (submitted to A\&A)},
archivePrefix = {arXiv},
       eprint = {2603.10135},
 primaryClass = {astro-ph.GA},
       adsurl = {https://ui.adsabs.harvard.edu/abs/2026arXiv260310135D}
}

@ARTICLE{2024ApJ...971...40L,
       author = {{Li}, Guodong and {Assef}, Roberto J. and {Tsai}, Chao-Wei and {Wu}, Jingwen and {Eisenhardt}, Peter R.~M. and {Stern}, Daniel and {D{\'\i}az-Santos}, Tanio and {Blain}, Andrew W. and {Jun}, Hyunsung D. and {Fern{\'a}ndez Aranda}, Rom{\'a}n and {Zewdie}, Dejene},
        title = "{Black Hole Mass and Eddington Ratio Distribution of Hot Dust-obscured Galaxies}",
      journal = {\apj},
         year = 2024,
        month = aug,
       volume = {971},
       number = {1},
          eid = {40},
        pages = {40},
          doi = {10.3847/1538-4357/ad5317},
archivePrefix = {arXiv},
       eprint = {2405.20479},
 primaryClass = {astro-ph.GA},
       adsurl = {https://ui.adsabs.harvard.edu/abs/2024ApJ...971...40L}
}

@article{article_gemini,
author = {Elias, Jonathan and Joyce, Richard and Liang, Ming and Muller, Gary and Hileman, Edward and George, James},
year = {2006},
month = {06},
pages = {},
title = {Design of the Gemini near-infrared spectrograph - art. no. 62694C},
volume = {6269},
journal = {Proc. SPIE},
doi = {10.1117/12.671817}
}

@INPROCEEDINGS{2003SPIE.4841.1548E,
       author = {{Eisenhauer}, Frank and {Abuter}, Roberto and {Bickert}, Klaus and {Biancat-Marchet}, Fabio and {Bonnet}, Henri and {Brynnel}, Joar and {Conzelmann}, Ralf D. and {Delabre}, Bernard and {Donaldson}, Robert and {Farinato}, Jacopo and {Fedrigo}, Enrico and {Genzel}, Reinhard and {Hubin}, Norbert N. and {Iserlohe}, Christof and {Kasper}, Markus E. and {Kissler-Patig}, Markus and {Monnet}, Guy J. and {Roehrle}, Claudia and {Schreiber}, Juergen and {Stroebele}, Stefan and {Tecza}, Matthias and {Thatte}, Niranjan A. and {Weisz}, Harald},
        title = "{SINFONI - Integral field spectroscopy at 50 milli-arcsecond resolution with the ESO VLT}",
    booktitle = {Instrument Design and Performance for Optical/Infrared Ground-based Telescopes},
         year = 2003,
       editor = {{Iye}, Masanori and {Moorwood}, Alan F.~M.},
       series = {Society of Photo-Optical Instrumentation Engineers (SPIE) Conference Series},
       volume = {4841},
        month = mar,
        pages = {1548-1561},
          doi = {10.1117/12.459468},
archivePrefix = {arXiv},
       eprint = {astro-ph/0306191},
 primaryClass = {astro-ph},
       adsurl = {https://ui.adsabs.harvard.edu/abs/2003SPIE.4841.1548E}
}

@INPROCEEDINGS{2005ASPC..347...29T,
       author = {{Taylor}, M.~B.},
        title = "{TOPCAT \& STIL: Starlink Table/VOTable Processing Software}",
    booktitle = {Astronomical Data Analysis Software and Systems XIV},
         year = 2005,
       editor = {{Shopbell}, P. and {Britton}, M. and {Ebert}, R.},
       series = {Astronomical Society of the Pacific Conference Series},
       volume = {347},
        month = dec,
        pages = {29},
       adsurl = {https://ui.adsabs.harvard.edu/abs/2005ASPC..347...29T}
}

@ARTICLE{2025arXiv251020919B,
       author = {{Boersma}, C. and {Maragkoudakis}, A. and {Allamandola}, L.~J. and {Bregman}, J.~D. and {Temi}, P. and {Esposito}, V.~J. and {Fortenberry}, R.~C.},
        title = "{SPHEREx: Aromatics, Aliphatics, and PAH Size across the Iris Nebula}",
      journal = {\apj},
         year = 2026,
        month = feb,
       volume = {997},
       number = {2},
          eid = {239},
        pages = {239},
          doi = {10.3847/1538-4357/ae2a33},
archivePrefix = {arXiv},
       eprint = {2510.20919},
 primaryClass = {astro-ph.GA},
       adsurl = {https://ui.adsabs.harvard.edu/abs/2026ApJ...997..239B}
}

@ARTICLE{2025arXiv251007684X,
       author = {{Xia}, Bin and {Ramachandra}, Nesar and {Wells}, Azton I. and {Habib}, Salman and {Wise}, John},
        title = "{Multi-modal Foundation Model for Cosmological Simulation Data}",
      journal = {arXiv e-prints},
         year = 2025,
        month = oct,
          eid = {arXiv:2510.07684},
        pages = {arXiv:2510.07684},
          doi = {10.48550/arXiv.2510.07684},
archivePrefix = {arXiv},
       eprint = {2510.07684},
 primaryClass = {astro-ph.GA},
       adsurl = {https://ui.adsabs.harvard.edu/abs/2025arXiv251007684X}
}

\begin{appendix}
\onecolumn
\nolinenumbers

\section{Confirming heavily reddened quasar candidates with \spx spectrophotometry} \label{App:SPHEREx_Spectra}

\begin{figure*} [ht!]
\centering
\begin{tabular}{rr}
\includegraphics[scale=0.560, trim={0.3cm, 0.82cm, 0.1cm, 0.2cm}, clip]{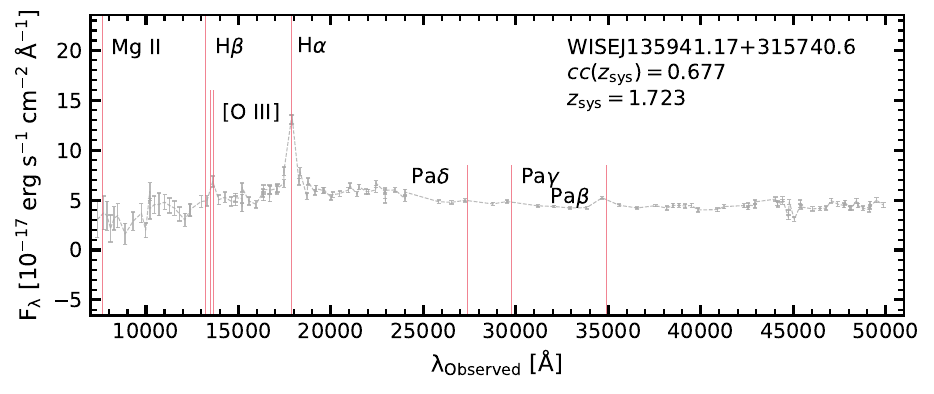} &
\includegraphics[scale=0.560, trim={0.3cm, 0.82cm, 0.1cm, 0.2cm}, clip]{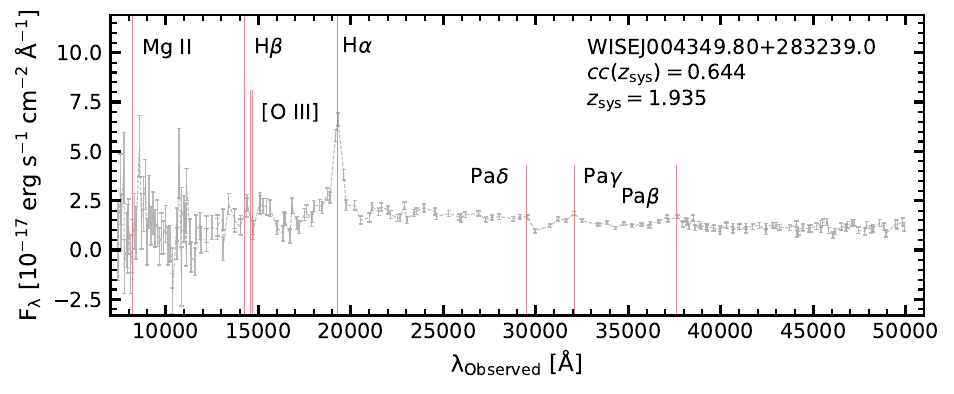} \\
\includegraphics[scale=0.560, trim={0.3cm, 0.82cm, 0.1cm, 0.2cm}, clip]{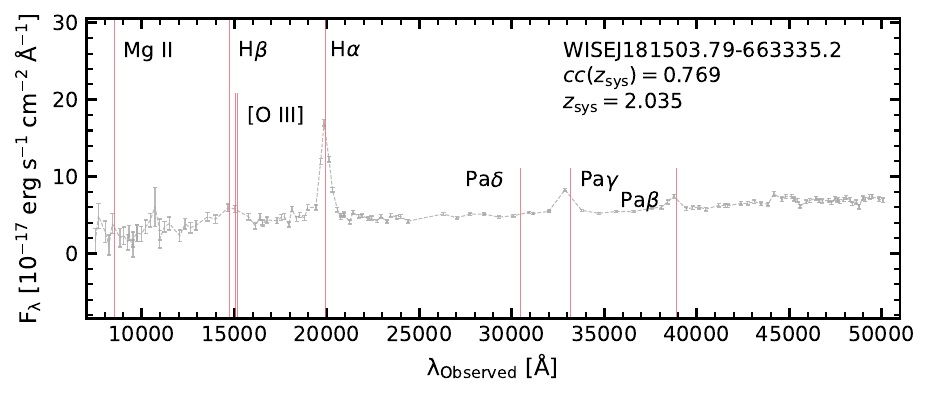} &
 \includegraphics[scale=0.560, trim={0.3cm, 0.82cm, 0.1cm, 0.2cm}, clip]{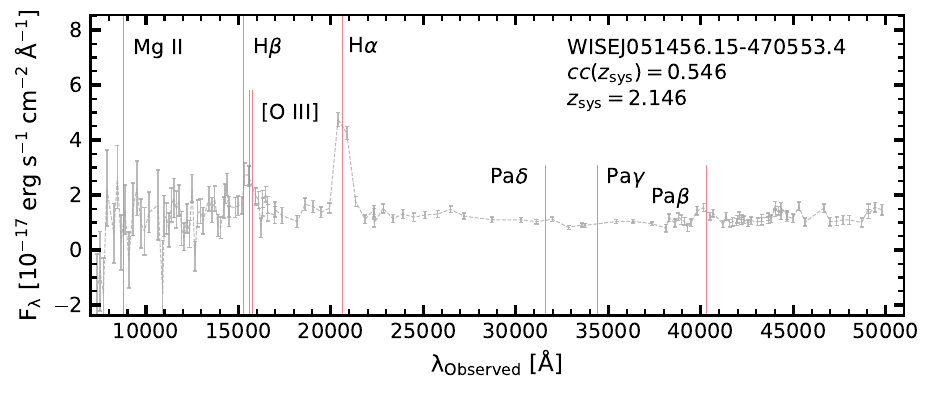} \\
\includegraphics[scale=0.560, trim={0.3cm, 0.82cm, 0.1cm, 0.2cm}, clip]{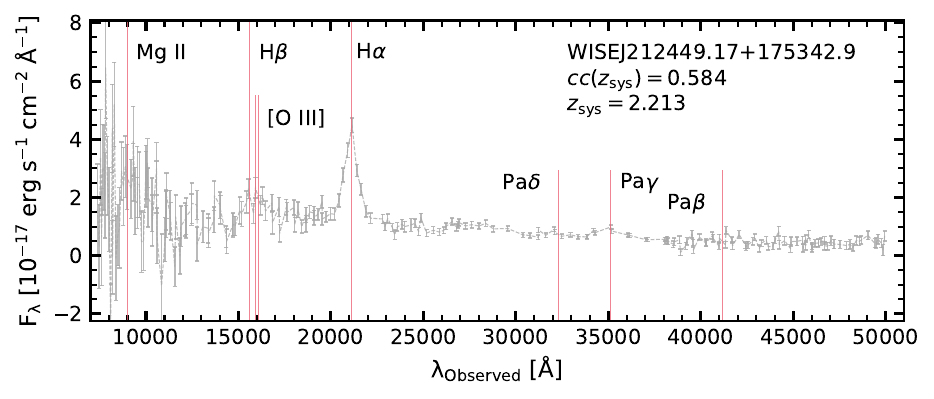} &
 \includegraphics[scale=0.560, trim={0.3cm, 0.82cm, 0.1cm, 0.2cm}, clip]{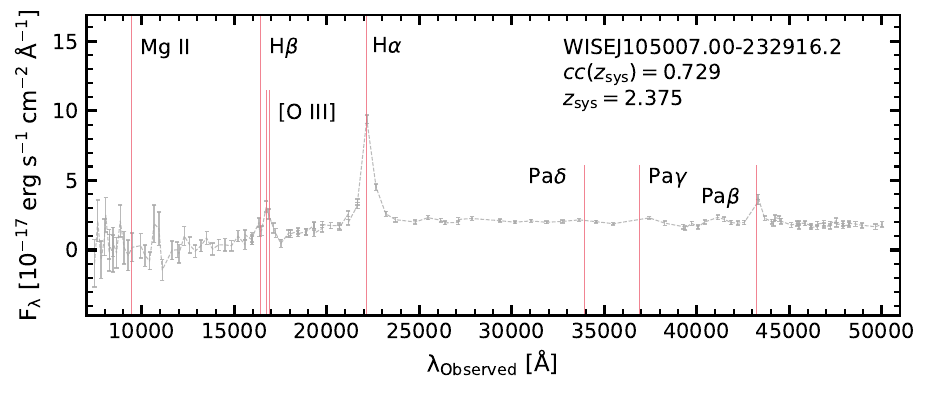} \\
\includegraphics[scale=0.560, trim={0.3cm, 0.82cm, 0.1cm, 0.2cm}, clip]{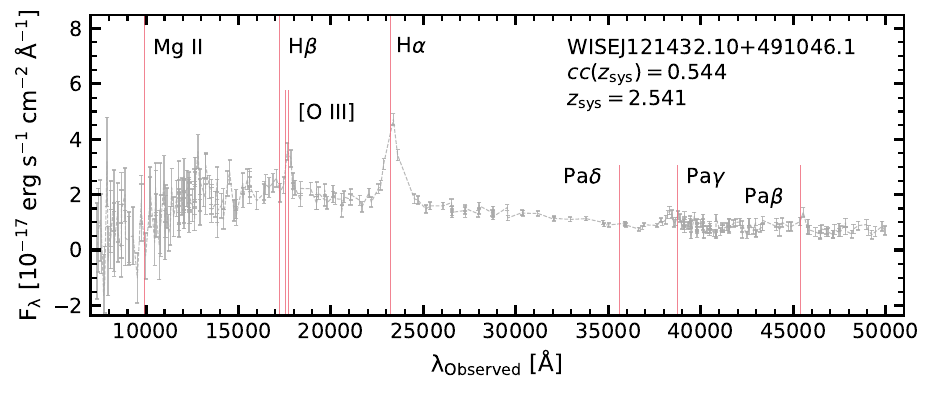} &
 \includegraphics[scale=0.560, trim={0.3cm, 0.82cm, 0.1cm, 0.2cm}, clip]{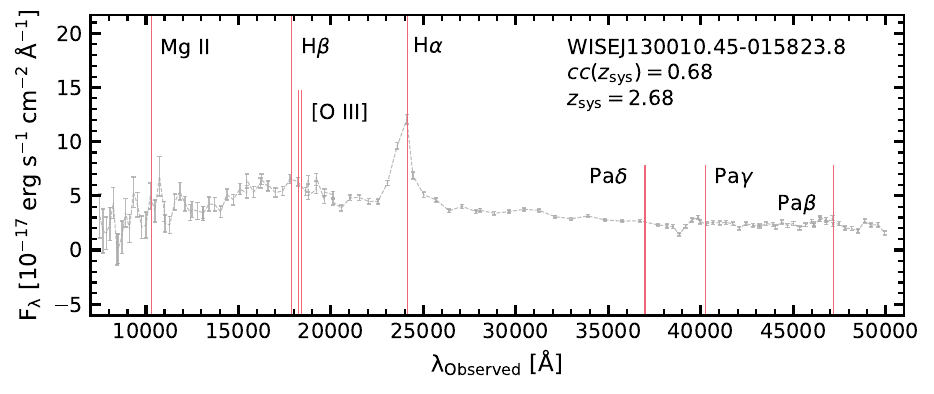} \\
  \includegraphics[scale=0.560, trim={0.3cm, 0.82cm, 0.1cm, 0.2cm}, clip]{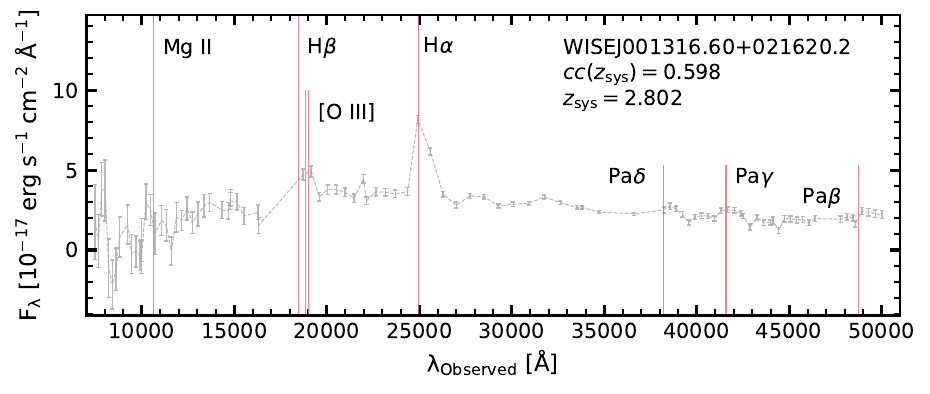} &
 \includegraphics[scale=0.560, trim={0.3cm, 0.82cm, 0.1cm, 0.2cm}, clip]{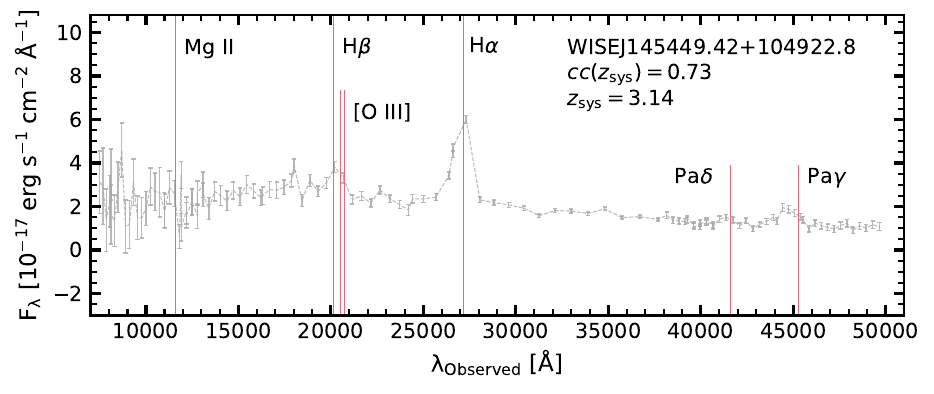} \\
\includegraphics[scale=0.560, trim={0.3cm, 0.22cm, 0.1cm, 0.2cm}, clip]{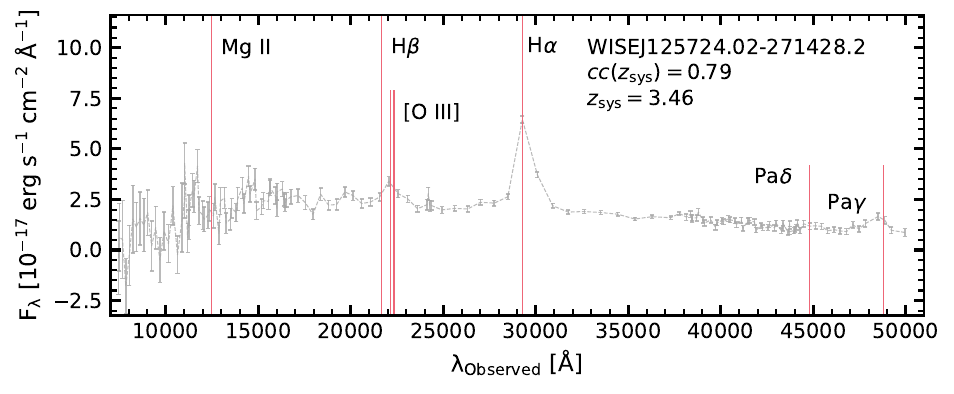} &
 \includegraphics[scale=0.560, trim={0.3cm, 0.22cm, 0.1cm, 0.2cm}, clip]{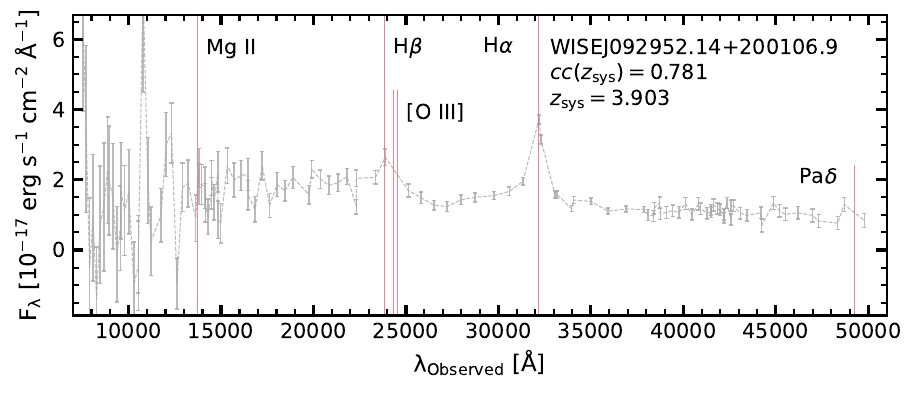} \\

 \end{tabular}
  \caption{We present the \spx spectra for 12 confirmed HRQs which span the full sample range in redshift space. Various emission lines are marked with vertical red lines. The cross-correlation coefficient corresponding to the optimum redshift solution is also presented in each panel.}
 \label{fig:ccl}
\end{figure*}

In Fig. \ref{fig:ccl}, we present 12 example \spx spectra spanning a range in redshift, illustrating the effectiveness of our cross-correlation analysis across the full redshift parameter space. In Table \ref{Tab:z_Results}, we present the redshift results for all 76 HRQs.

\begin{table*} [ht!]
    \centering
    \caption{The co-ordinates, $K$-band magnitudes and redshift solutions for our 76 \spx HRQs.}   \label{Tab:z_Results} 
    \begin{tabular}{l r r c c}
      \hline
      \multicolumn{1}{l} {WISE designation} & \multicolumn{1}{r} {RA [degrees]} & \multicolumn{1}{r}{Dec. [degrees]} & \multicolumn{1}{c}{$K_{\rm Vega}$ [mag]} & \multicolumn{1}{c} {$z_{\rm sys}$} \\
      \hline
        WISEJ0009+3608$^a$ & 2.29665    & 36.14956  & 15.85$\pm$0.03 & 2.639$\pm$0.003   \\
        WISEJ0013+0216     & 3.31920    & 2.27230   & 15.27$\pm$0.01 & 2.809$\pm$0.003   \\
        WISEJ0020+0814     & 5.23227    & 8.24782   & 15.51$\pm$0.02 & 2.770$\pm$0.004   \\
        WISEJ0032+1543     & 8.22131    & 15.71792  & 15.13$\pm$0.01 & 2.368$\pm$0.003   \\
        WISEJ0034-0123     & 8.52866    & -1.39736  & 15.99$\pm$0.02 & 1.801$\pm$0.002   \\
        WISEJ0043+2832     & 10.95753   & 28.54418  & 15.81$\pm$0.03 & 1.931$\pm$0.004   \\
        WISEJ0135+3147     & 23.81188   & 31.79294  & 15.71$\pm$0.03 & 2.380$\pm$0.001   \\
        WISEJ0146+2121     & 26.55067   & 21.36302  & 16.08$\pm$0.04 & 2.88$\pm$0.02     \\
        WISEJ0146-3946     & 26.67365   & -39.77846 & 15.74$\pm$0.02 & 2.210$\pm$0.003   \\
        WISEJ0152+2608     & 28.14810   & 26.13851  & 15.80$\pm$0.03 & 2.266$\pm$0.009   \\
        WISEJ0202+1342     & 30.66227   & 13.71474  & 15.10$\pm$0.01 & 2.891$\pm$0.003   \\
        WISEJ0224-4625     & 36.23730   & -46.42105 & 15.76$\pm$0.02 & 2.363$\pm$0.007   \\
        WISEJ0229-4225     & 37.27422   & -42.28606 & 15.61$\pm$0.02 & 2.413$\pm$0.008   \\
        WISEJ0236+2637     & 39.01852   & 26.62830  & 15.90$\pm$0.03 & 2.500$\pm$0.005   \\
        WISEJ0240-0555     & 40.12126   & -5.92019  & 15.97$\pm$0.02 & 3.43$\pm$0.04     \\
        WISEJ0311-6309     & 47.83019   & -63.15375 & 15.22$\pm$0.02 & 2.479$\pm$0.003   \\
        WISEJ0321+4017     & 50.25450   & 40.28816  & 15.63$\pm$0.03 & 2.26$\pm$0.01     \\
        WISEJ0326+1547     & 51.57182   & 15.79397  & 15.79$\pm$0.03 & 2.308$\pm$0.007   \\
        WISEJ0338-1048     & 54.70936   & -10.80746 & 15.36$\pm$0.02 & 2.165$\pm$0.006   \\
        WISEJ0347-5643     & 56.97823   & -56.72647 & 15.64$\pm$0.02 & 2.270$\pm$0.003   \\
        WISEJ0418-0822     & 64.67986   & -8.36814  & 15.74$\pm$0.02 & 2.299$\pm$0.005   \\
        WISEJ0439-3909     & 69.76437   & -39.15342 & 15.48$\pm$0.02 & 2.240$\pm$0.001   \\
        WISEJ0442+1040     & 70.60530   & 10.66678  & 15.40$\pm$0.02 & 2.383$\pm$0.005   \\
        WISEJ0514-4705     & 78.73398   & -47.09819 & 15.68$\pm$0.02 & 2.145$\pm$0.002   \\
        WISEJ0612+5629     & 93.24742   & 56.48883  & 15.48$\pm$0.02 & 2.590$\pm$0.001   \\
        WISEJ0746+4722     & 115.54396  & 47.36816  & 16.06$\pm$0.04 & 2.34$\pm$0.01     \\
        WISEJ0800+0517     & 120.16923  & 5.28665   & 15.99$\pm$0.04 & 2.351$\pm$0.006   \\
        WISEJ0819+1901     & 124.82929  & 19.01788  & 15.97$\pm$0.04 & 2.00$\pm$0.02     \\
        WISEJ0819+3537     & 124.96216  & 35.62288  & 15.48$\pm$0.02 & 1.902$\pm$0.004   \\
        WISEJ0836+4230     & 129.22868  & 42.51441  & 15.70$\pm$0.03 & 2.460$\pm$0.003   \\
        WISEJ0929+2001     & 142.46728  & 20.0186   & 16.01$\pm$0.04 & 3.901$\pm$0.002   \\
        WISEJ1050-2329     & 162.52919  & -23.48785 & 15.06$\pm$0.01 & 2.374$\pm$0.005   \\
        WISEJ1056+2535     & 164.01755  & 25.59324  & 15.95$\pm$0.03 & 2.320$\pm$0.001   \\
        WISEJ1140+0634     & 175.06440  & 6.57494   & 15.80$\pm$0.02 & 2.164$\pm$0.006   \\
        WISEJ1214+4910     & 183.63378  & 49.17948  & 15.67$\pm$0.03 & 2.540$\pm$0.001   \\
        WISEJ1252+0715     & 193.05381  & 7.25129   & 14.40$\pm$0.01 & 2.150$\pm$0.001   \\
        WISEJ1257-2714     & 194.35010  & -27.24119 & 15.63$\pm$0.03 & 3.460$\pm$0.002   \\
        WISEJ1300-0158     & 195.04357  & -1.97328  & 14.72$\pm$0.01 & 2.68$\pm$0.01     \\
        WISEJ1306-2210     & 196.69636  & -22.16696 & 15.77$\pm$0.03 & 2.330$\pm$0.006   \\
        WISEJ1323-1750     & 200.79116  & -17.84267 & 15.62$\pm$0.03 & 2.375$\pm$0.009   \\
        WISEJ1325+4921     & 201.27454  & -49.35776 & 15.82$\pm$0.03 & 2.875$\pm$0.005   \\
        WISEJ1336+0228     & 204.12161  & 2.47703   & 15.41$\pm$0.02 & 2.282$\pm$0.005   \\
        WISEJ1341+1005     & 205.47987  & 10.09553  & 15.77$\pm$0.02 & 2.39$\pm$0.01     \\
        WISEJ1351+4026     & 207.84937  & 40.44721  & 16.08$\pm$0.03 & 2.59$\pm$0.01     \\
        WISEJ1359+3157     & 209.92155  & 31.9613   & 14.91$\pm$0.01 & 1.720$\pm$0.002   \\
        WISEJ1424+0315     & 216.02611  & 3.25249   & 15.79$\pm$0.02 & 2.683$\pm$0.004   \\
        WISEJ1453+1740     & 223.32730  & 17.66678  & 15.39$\pm$0.02 & 2.309$\pm$0.002   \\
        WISEJ1454+1049     & 223.70595  & 10.82300  & 15.69$\pm$0.02 & 3.14$\pm$0.02     \\
        WISEJ1455+0324     & 223.87093  & 3.40982   & 15.75$\pm$0.02 & 2.627$\pm$0.005   \\
        WISEJ1511+0031     & 227.95365  & 0.51968   & 15.59$\pm$0.02 & 2.690$\pm$0.001   \\
        WISEJ1558+2057     & 239.56130  & 20.96097  & 15.89$\pm$0.02 & 3.000$\pm$0.001   \\
        WISEJ1648+3447     & 252.13162  & 34.79002  & 16.00$\pm$0.03 & 2.379$\pm$0.004   \\
        WISEJ1805-6813     & 271.39146  & -68.22513 & 15.53$\pm$0.03 & 3.570$\pm$0.001   \\
        WISEJ1806+1909     & 271.50784  & 19.15600  & 15.80$\pm$0.03 & 3.21$\pm$0.01     \\
        WISEJ1815-6633     & 273.76583  & -66.55979 & 14.58$\pm$0.01 & 2.040$\pm$0.001   \\
        WISEJ1815+1612     & 273.91352  & 16.20658  & 15.87$\pm$0.03 & 2.596$\pm$0.005   \\
        WISEJ1843+4345     & 280.79209  & 43.76268  & 15.15$\pm$0.02 & 2.235$\pm$0.009   \\
     \hline
\end{tabular}
\label{Tab:cc_Results}
\end{table*}

\begin{table*}
    \centering
    \ContinuedFloat
    \caption{\emph{Continued}}    
    \begin{tabular}{l r r c c}
      \hline
      \multicolumn{1}{l} {WISE designation} & \multicolumn{1}{r} {RA [degrees]} & \multicolumn{1}{r}{Dec. [degrees]} & \multicolumn{1}{c}{$K_{\rm Vega}$ [mag]} & \multicolumn{1}{c} {$z_{\rm sys}$} \\
      \hline
        
        WISEJ1939+4712     & 294.84363  & 47.20423  & 15.68$\pm$0.03 & 2.529$\pm$0.004   \\
        WISEJ1949+0436     & 297.38766  & 4.60399   & 16.06$\pm$0.04 & 2.073$\pm$0.005   \\
        WISEJ1952-5629 & 298.20802  & -56.48433 & 15.32$\pm$0.02 & 2.110$\pm$0.007   \\
        WISEJ2018+5938 & 304.57186  & 59.64462  & 15.84$\pm$0.04 & 2.291$\pm$0.003   \\
        WISEJ2020+5931 & 305.12776  & 59.52840  & 15.60$\pm$0.03 & 2.408$\pm$0.004   \\
        WISEJ2021-7808 & 305.28283  & -78.14237 & 15.37$\pm$0.03 & 2.274$\pm$0.009   \\
        WISEJ2059+1225 & 314.84451  & 12.42702  & 15.61$\pm$0.02 & 2.568$\pm$0.004   \\
        WISEJ2124+1753 & 321.20488  & 17.89526  & 15.91$\pm$0.03 & 2.214$\pm$0.009   \\
        WISEJ2137-2819 & 324.37107  & -28.31847 & 15.57$\pm$0.02 & 2.131$\pm$0.003   \\
        WISEJ2152+2605 & 328.07529  & 26.09234  & 15.83$\pm$0.03 & 2.245$\pm$0.005   \\
        WISEJ2154+0640 & 328.56611  & 6.67735   & 16.00$\pm$0.03 & 1.929$\pm$0.005   \\ 
        WISEJ2213-0534 & 333.35749  & -5.57885  & 15.90$\pm$0.03 & 2.332$\pm$0.008   \\
        WISEJ2213+3437 & 333.44869  & 34.61899  & 15.91$\pm$0.03 & 2.516$\pm$0.008   \\
        WISEJ2222-1746 & 335.73677  & -17.78058 & 15.67$\pm$0.03 & 2.205$\pm$0.005   \\
        WISEJ2316+0938 & 349.19449  & 9.63618   & 15.33$\pm$0.02 & 2.19$\pm$0.01     \\
        WISEJ2320+2919 & 350.08620  & 29.32515  & 16.01$\pm$0.03 & 2.22$\pm$0.01     \\
        WISEJ2342+4149 & 355.60148  & 41.82580  & 15.64$\pm$0.02 & 2.121$\pm$0.003   \\
        WISEJ2349+0636 & 357.47072  & 6.60591   & 15.70$\pm$0.02 & 2.235$\pm$0.009   \\
        WISEJ2359+0640 & 359.82387  & 6.68005   & 15.17$\pm$0.01 & 2.45$\pm$0.01     \\
     \hline        
\end{tabular}
\tablefoot{The J2000 coordinates of each HRQ have been truncated to aid readership. \\ \tablefoottext{a}{This source was also identified in \citet{2026arXiv260310135D}}}
\end{table*}

\section{The SED properties of 76 SPHERE\textsc{x}-confirmed heavily reddened quasars} \label{App:Sed}

\begin{figure*} [ht!]
\centering
\begin{tabular}{rr}
 \includegraphics[scale=0.528, trim={0.3cm, 0.72cm, 0.1cm, 0.0cm}, clip]{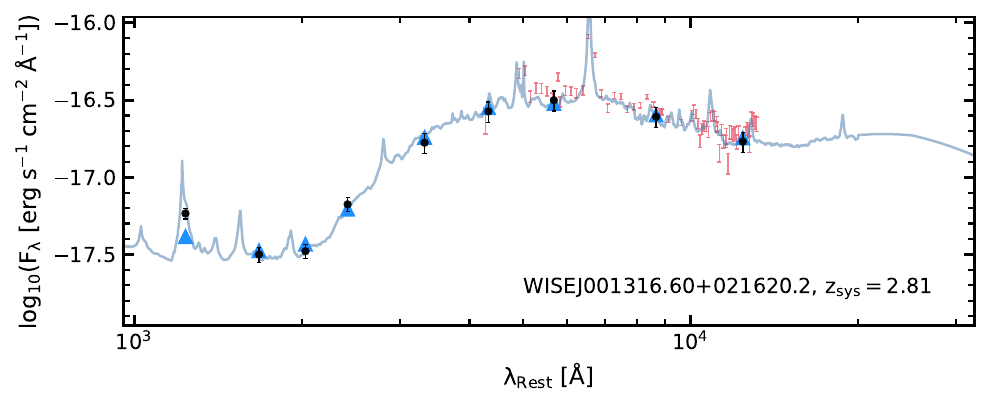} &
 \includegraphics[scale=0.528, trim={0.3cm, 0.72cm, 0.1cm, 0.0cm}, clip]{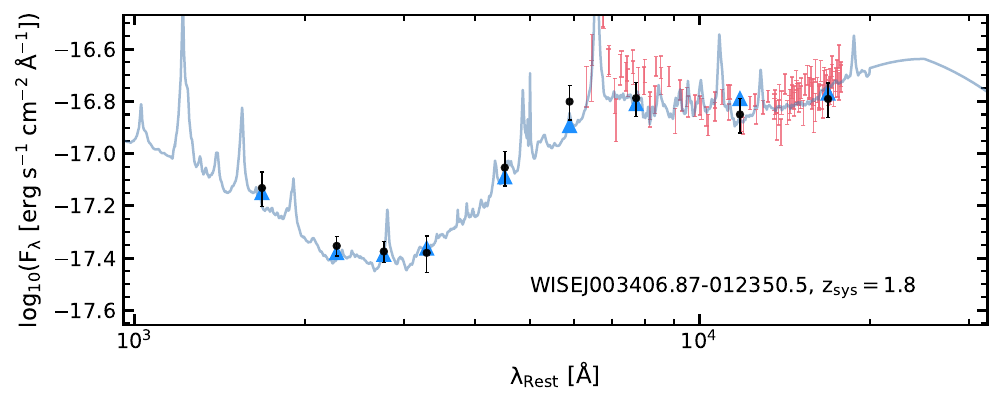} \\
 \includegraphics[scale=0.520, trim={0.3cm, 0.72cm, 0.1cm, 0.0cm}, clip]{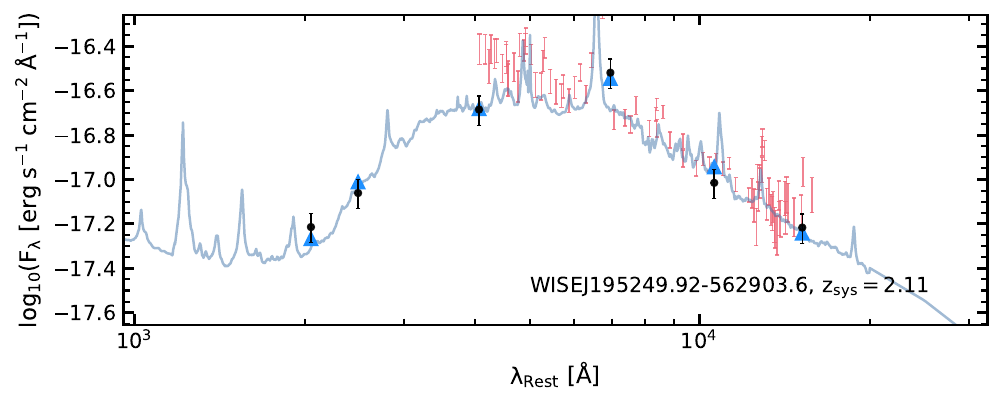} & 
 \includegraphics[scale=0.528, trim={0.3cm, 0.72cm, 0.1cm, 0.0cm}, clip]{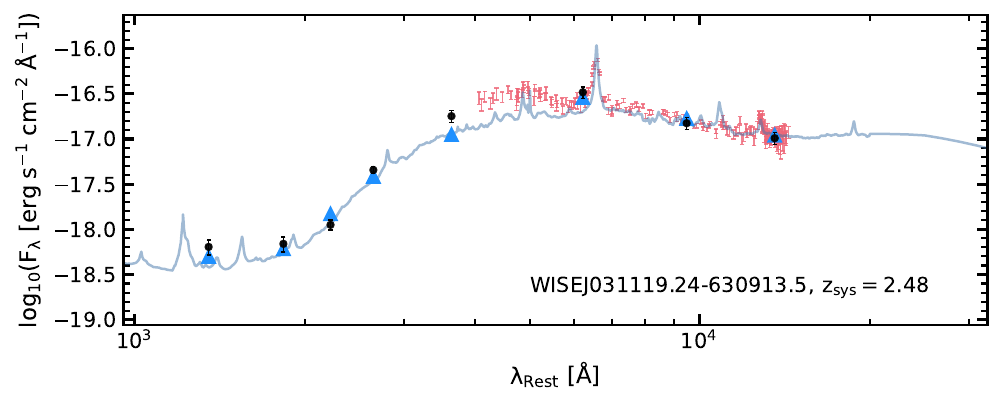} \\
 \includegraphics[scale=0.528, trim={0.3cm, 0.72cm, 0.1cm, 0.0cm}, clip]{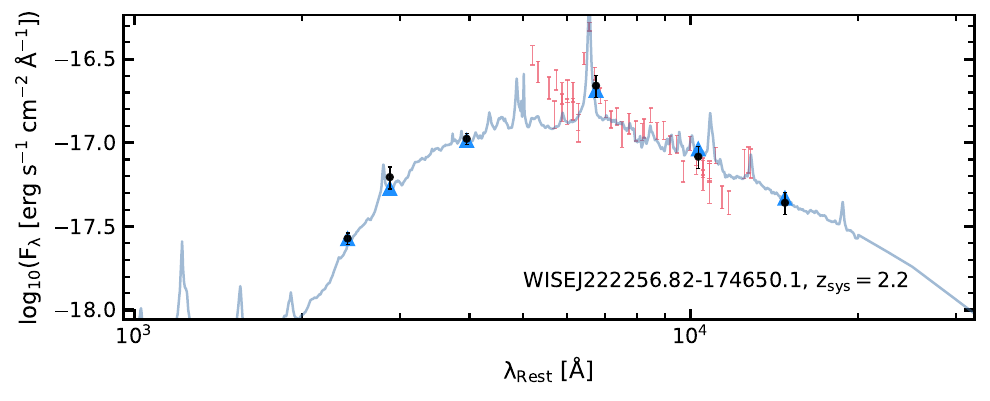} & 
 \includegraphics[scale=0.528, trim={0.3cm, 0.72cm, 0.1cm, 0.0cm}, clip]{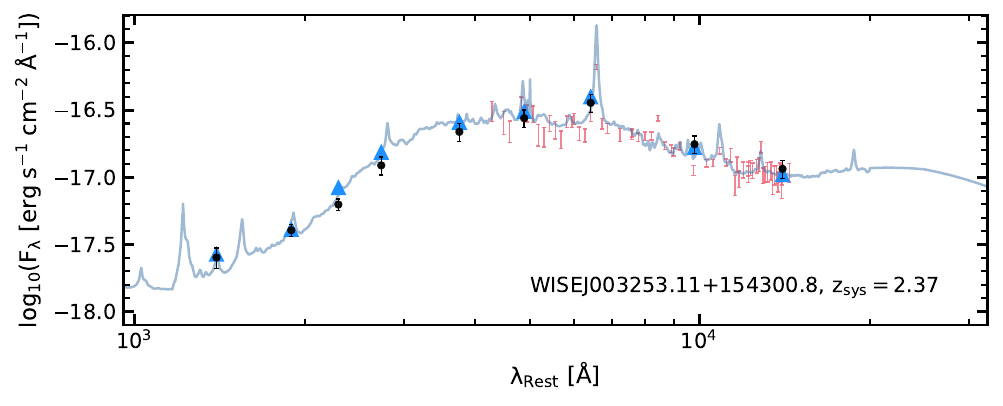} \\
  \includegraphics[scale=0.520, trim={0.3cm, 0.72cm, 0.1cm, 0.0cm}, clip]{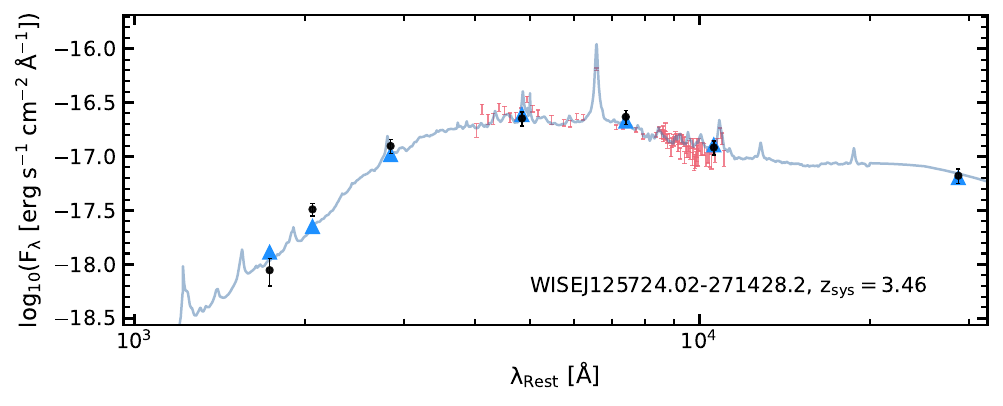} & 
 \includegraphics[scale=0.520, trim={0.3cm, 0.72cm, 0.1cm, 0.0cm}, clip]{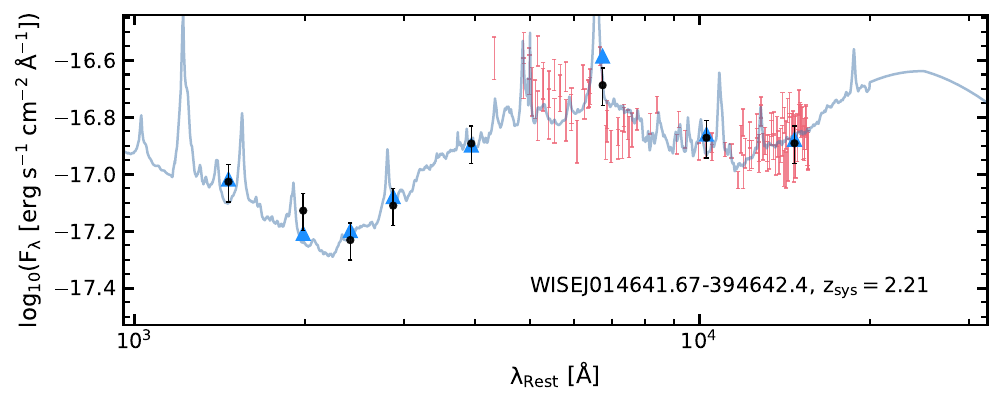} \\
 \includegraphics[scale=0.529, trim={0.3cm, 0.72cm, 0.1cm, 0.0cm}, clip]{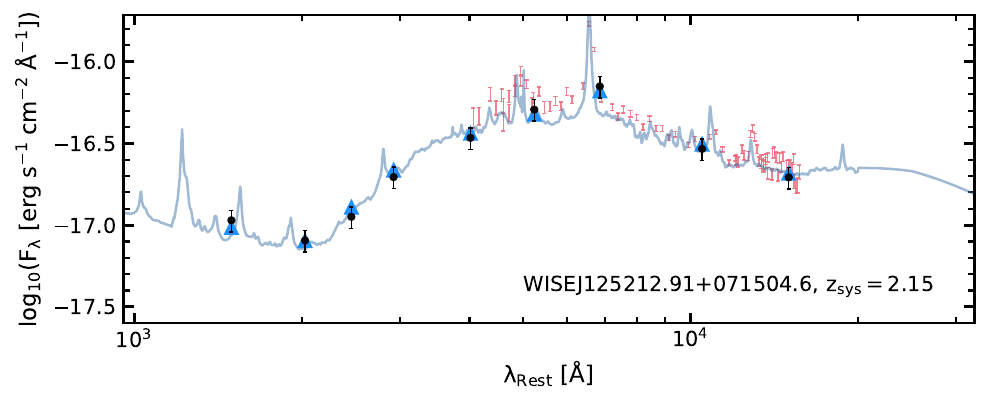} & 
 \includegraphics[scale=0.520, trim={0.3cm, 0.72cm, 0.1cm, 0.0cm}, clip]{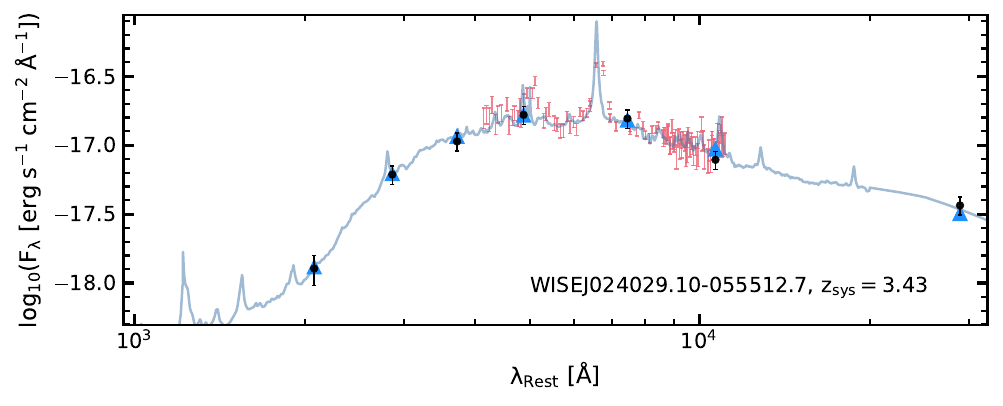} \\ 
 \includegraphics[scale=0.520, trim={0.3cm, 0.22cm, 0.1cm, 0.0cm}, clip]{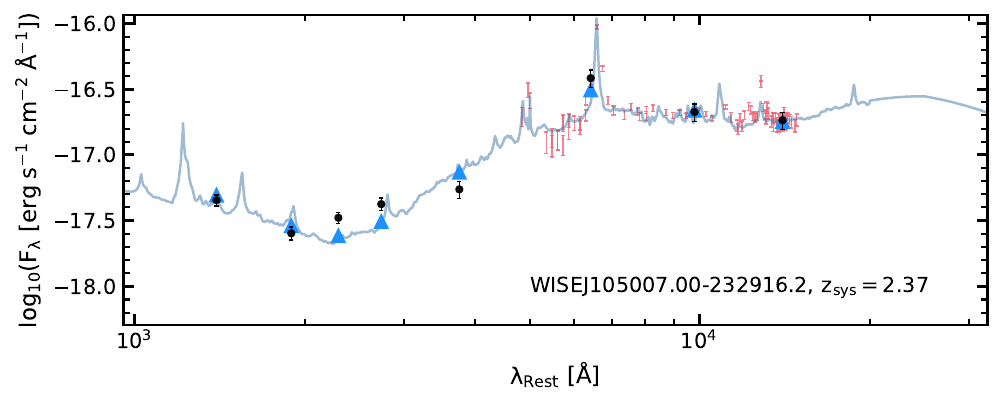} & 
 \includegraphics[scale=0.520, trim={0.3cm, 0.22cm, 0.1cm, 0.0cm}, clip]{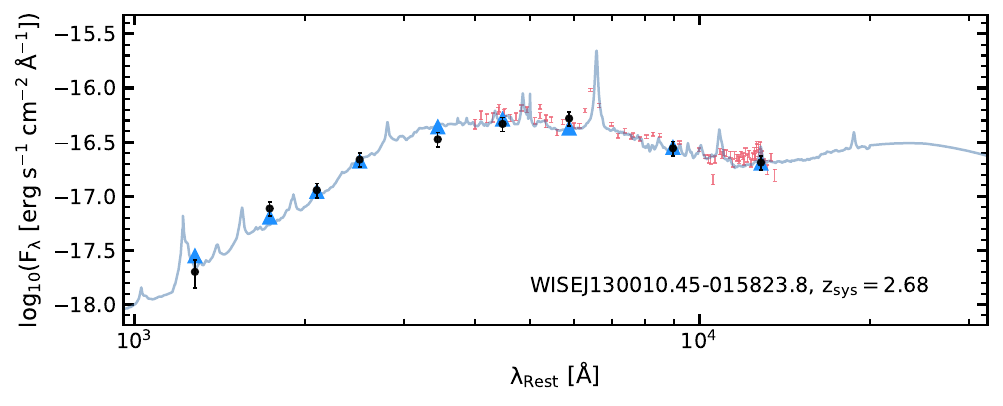} \\ 
 
 \end{tabular}
  \caption{We present the "best-fit" \textsc{qsogen} SEDs for 12 SPHERE\textsc{x}-confirmed HRQs. The broad-band photometric data from DELVE/SDSS, UKIDSS-LAS/UHS/VHS and \textit{WISE} are indicated in black with their associated uncertainties. The best-fit multi-component SED models are shown by the blue lines and triangles and the \spx spectra are overlaid in red.}
 \label{fig:Examples}
\end{figure*}

In this appendix, we present the results from our SED analysis discussed in Section \ref{sec:SED_model}. We present a complete summary of our results in Table \ref{Tab:SED_Results}, which will also be made available as online supplementary material at CDS. In addition, Fig. \ref{fig:Examples} illustrates SED fits for 12 HRQs with a range in redshift, extinction, hot-dust amplitude and scattering fraction.   

\begin{center}
\begin{longtable}{l c c c c c c c} 
    \caption{The best-fit SED parameters and their corresponding MCMC uncertainties for the 76 \spx HRQs fit with \textsc{qsogen}. \label{Tab:SED_Results}} \\
    \hline
      \multicolumn{1}{l} {Object} & \multicolumn{1}{c} {$z_{\rm sys}$} & \multicolumn{1}{c}{log$_{10}\{\lambda$L$_{\lambda} (3000\mathring{A})\}$} & \multicolumn{1}{c}{$\rm E(B-V)$} & \multicolumn{1}{c}{F$_{\rm{UV}}$ [$\%$]} & \multicolumn{1}{c}{$\rm L_{Dust}/L_{Disk}|_{2\mu m}$}& \multicolumn{1}{c}{$\overline{\chi}^2_\nu$} & \multicolumn{1}{c}{UV Excess} \\
     \hline
     \endfirsthead
     \caption{continued.}\\
     \hline
      \multicolumn{1}{l} {Object} & \multicolumn{1}{c} {$z_{\rm sys}$} & \multicolumn{1}{c}{log$_{10}\{\lambda$L$_{\lambda} (3000\mathring{A})\}$} & \multicolumn{1}{c}{$\rm E(B-V)$} & \multicolumn{1}{c}{F$_{\rm{UV}}$ [$\%$]} & \multicolumn{1}{c}{$\rm L_{Dust}/L_{Disk}|_{2\mu m}$}& \multicolumn{1}{c}{$\overline{\chi}^2_\nu$} & \multicolumn{1}{c}{UV Excess} \\
     \hline
    \endhead
    \endfoot
        WISEJ0009+3608     & 2.639$\pm$0.003 & 47.86$\pm$0.04 & 0.97$\pm$0.04 & $-$           & 8.37$\pm$0.09 & 0.6 & N/A   \\
        WISEJ0013+0216     & 2.809$\pm$0.003 & 47.96$\pm$0.07 & 0.83$\pm$0.03 & 0.07$\pm$0.01 & 0.89$\pm$0.01$^b$ & 1.3 & Conf. \\
        WISEJ0020+0815$^a$ & 2.770$\pm$0.004 & 47.64$\pm$0.07 & 0.72$\pm$0.03 & $-$           & 0.09$\pm$0.02$^b$ & 1.9 & Rej.  \\
        WISEJ0032+1543     & 2.368$\pm$0.003 & 47.47$\pm$0.06 & 0.66$\pm$0.02 & 0.05$\pm$0.03 & 0.96$\pm$0.01 & 0.5 & Conf. \\
        WISEJ0034-0123     & 1.801$\pm$0.002 & 47.50$\pm$0.09 & 1.31$\pm$0.07 & 0.12$\pm$0.02 & 1.02$\pm$0.02 & 1.5 & Conf. \\
        WISEJ0043+2832     & 1.931$\pm$0.004 & 47.63$\pm$0.04 & 1.47$\pm$0.05 & 0.9$\pm$0.2   & 0.38$\pm$0.01 & 21.8& Conf. \\
        WISEJ0135+3147     & 2.380$\pm$0.001 & 47.27$\pm$0.03 & 0.66$\pm$0.07 & 0.12$\pm$0.03 & 0.69$\pm$0.01 & 4.1 & Conf. \\
        WISEJ0146+2121     & 2.88$\pm$0.02   & 47.49$\pm$0.03 & 0.68$\pm$0.02 & $-$           & 3.76$\pm$0.04$^b$ & 0.6 & Rej.  \\
        WISEJ0146-3946     & 2.210$\pm$0.003 & 47.41$\pm$0.08 & 0.88$\pm$0.04 & 0.33$\pm$0.06 & 2.13$\pm$0.03 & 0.3 & Conf. \\
        WISEJ0152+2608     & 2.266$\pm$0.009 & 47.08$\pm$0.03 & 0.63$\pm$0.02 & 0.07$\pm$0.02 & 0.21$\pm$0.02 & 2.9 & Conf. \\
        WISEJ0202+1342     & 2.891$\pm$0.003 & 48.07$\pm$0.06 & 0.90$\pm$0.03 & 0.02$\pm$0.01 & 1.05$\pm$0.02$^b$ & 3.3 & Conf. \\
        WISEJ0224-4625     & 2.363$\pm$0.007 & 47.26$\pm$0.04 & 0.71$\pm$0.04 & 0.42$\pm$0.08 & 2.20$\pm$0.04 & 0.4 & Conf. \\
        WISEJ0229-4225     & 2.413$\pm$0.008 & 47.07$\pm$0.06 & 0.53$\pm$0.03 & $-$           & 0.72$\pm$0.02 & 1.0 & Rej.  \\
        WISEJ0236+2637     & 2.500$\pm$0.005 & 47.09$\pm$0.03 & 0.68$\pm$0.02 & $-$           & 0.68$\pm$0.07 & 1.9 & Rej.  \\
        WISEJ0240-0555     & 3.43$\pm$0.04   & 47.88$\pm$0.07 & 0.79$\pm$0.04 & $-$           & 0.12$\pm$0.02 & 0.3 & Rej.  \\
        WISEJ0311-6309     & 2.479$\pm$0.003 & 47.68$\pm$0.09 & 0.92$\pm$0.05 & 0.01$\pm$0.01 & 0.64$\pm$0.01 & 3.2 & Conf. \\
        WISEJ0321+4017     & 2.26$\pm$0.01   & 47.33$\pm$0.03 & 0.79$\pm$0.02 & 0.02$\pm$0.01 & 0.95$\pm$0.01 & 0.6 & Conf. \\
        WISEJ0326+1547$^a$ & 2.308$\pm$0.007 & 47.46$\pm$0.04 & 1.04$\pm$0.04 & $-$           & 0.06$\pm$0.01 & 1.6 & N/A   \\
        WISEJ0338-1048     & 2.165$\pm$0.006 & 47.24$\pm$0.01 & 0.71$\pm$0.09 & 1.49$\pm$0.60 & 1.32$\pm$0.03 & 2.0 & Conf. \\
        WISEJ0347-5643     & 2.270$\pm$0.003 & 47.25$\pm$0.04 & 0.81$\pm$0.05 & 0.26$\pm$0.04 & 2.18$\pm$0.04 & 0.7 & Conf. \\
        WISEJ0418-0822     & 2.299$\pm$0.005 & 47.40$\pm$0.08 & 0.74$\pm$0.04 & 0.17$\pm$0.04 & 1.53$\pm$0.03 & 1.8 & Conf. \\
        WISEJ0439-3909     & 2.240$\pm$0.001 & 47.39$\pm$0.03 & 0.82$\pm$0.02 & 0.25$\pm$0.03 & 0.68$\pm$0.02 & 0.6 & Conf. \\
        WISEJ0442+1040     & 2.383$\pm$0.005 & 47.18$\pm$0.04 & 0.68$\pm$0.02 & 0.12$\pm$0.01 & 1.16$\pm$0.01 & 7.3 & Conf. \\
        WISEJ0514-4705     & 2.145$\pm$0.002 & 47.02$\pm$0.02 & 0.60$\pm$0.08 & 0.38$\pm$0.05 & 4.15$\pm$0.06 & 0.6 & Conf. \\
        WISEJ0612+5629     & 2.590$\pm$0.001 & 47.68$\pm$0.03 & 0.91$\pm$0.03 & $-$           & 3.30$\pm$0.04 & 1.1 & N/A   \\
        WISEJ0746+4722     & 2.34$\pm$0.01   & 47.26$\pm$0.04 & 0.94$\pm$0.03 & 0.31$\pm$0.03 & 1.63$\pm$0.02 & 1.0 & Conf. \\
        WISEJ0800+0517     & 2.351$\pm$0.006 & 47.33$\pm$0.03 & 0.86$\pm$0.02 & $-$           & 0.93$\pm$0.01 & 0.5 & Rej.  \\
        WISEJ0819+1901     & 2.00$\pm$0.02   & 47.23$\pm$0.03 & 0.90$\pm$0.02 & 0.18$\pm$0.02 & 2.52$\pm$0.02 & 0.5 & Conf. \\
        WISEJ0819+3537     & 1.902$\pm$0.004 & 47.54$\pm$0.03 & 1.05$\pm$0.02 & $-$           & 0.64$\pm$0.01 & 0.7 & Rej.  \\
        WISEJ0836+4230     & 2.460$\pm$0.003 & 47.24$\pm$0.04 & 0.70$\pm$0.03 & $-$           & 0.41$\pm$0.01 & 0.5 & N/A   \\
        WISEJ0929+2001     & 3.901$\pm$0.002 & 47.73$\pm$0.03 & 0.55$\pm$0.02 & 0.39$\pm$0.06 & 1.85$\pm$0.02 & 1.3 & Conf. \\
        WISEJ1050-2329     & 2.374$\pm$0.005 & 47.90$\pm$0.06 & 1.19$\pm$0.08 & 0.06$\pm$0.01 & 1.13$\pm$0.02 & 4.5 & Conf. \\
        WISEJ1056+2535     & 2.320$\pm$0.001 & 47.03$\pm$0.03 & 0.63$\pm$0.02 & $-$           & 2.22$\pm$0.02 & 2.6 & Rej.  \\
        WISEJ1140+0634     & 2.164$\pm$0.006 & 47.26$\pm$0.08 & 0.78$\pm$0.04 & $-$           & 1.02$\pm$0.03 & 0.9 & Rej.  \\
        WISEJ1214+4910     & 2.540$\pm$0.001 & 47.40$\pm$0.03 & 0.73$\pm$0.02 & 0.24$\pm$0.03 & 0.84$\pm$0.01 & 1.3 & Conf. \\
        WISEJ1252+0715     & 2.150$\pm$0.001 & 47.74$\pm$0.07 & 0.82$\pm$0.04 & 0.14$\pm$0.02 & 0.66$\pm$0.03 & 1.0 & Conf. \\
        WISEJ1257-2714     & 3.460$\pm$0.002 & 47.99$\pm$0.06 & 0.74$\pm$0.02 & $-$           & 0.63$\pm$0.01 & 1.6 & Rej.  \\
        WISEJ1300-0158     & 2.68$\pm$0.01   & 47.81$\pm$0.07 & 0.62$\pm$0.04 & $-$           & 1.85$\pm$0.03 & 1.8 & Rej.  \\
        WISEJ1306-2210     & 2.330$\pm$0.006 & 47.31$\pm$0.09 & 0.81$\pm$0.05 & 0.17$\pm$0.04 & 1.98$\pm$0.05 & 1.5 & Conf. \\
        WISEJ1323-1750$^a$ & 2.375$\pm$0.009 & 47.15$\pm$0.09 & 0.64$\pm$0.08 & 0.08$\pm$0.02 & 2.05$\pm$0.04 & 1.7 & Conf. \\
        WISEJ1325-4921$^a$ & 2.875$\pm$0.005 & 48.00$\pm$0.04 & 0.91$\pm$0.04 & 0.11$\pm$0.02 & 0.99$\pm$0.02$^b$ & 0.5 & Conf. \\
        WISEJ1336+0228     & 2.282$\pm$0.005 & 47.23$\pm$0.07 & 0.65$\pm$0.04 & 0.28$\pm$0.06 & 0.99$\pm$0.01 & 1.1 & Conf. \\
        WISEJ1341+1005     & 2.39$\pm$0.01   & 47.41$\pm$0.08 & 0.83$\pm$0.05 & 0.32$\pm$0.07 & 2.68$\pm$0.04 & 3.6 & Conf. \\
        WISEJ1351+4026$^a$ & 2.59$\pm$0.01   & 47.19$\pm$0.04 & 0.66$\pm$0.03 & $-$           & 0.57$\pm$0.01 & 0.7 & N/A   \\
        WISEJ1359+3157$^a$ & 1.720$\pm$0.002 & 47.97$\pm$0.08 & 1.42$\pm$0.08 & 0.03$\pm$0.01 & 0.53$\pm$0.02 & 4.8 & Conf. \\
        WISEJ1424+0315     & 2.683$\pm$0.004 & 47.65$\pm$0.08 & 0.85$\pm$0.05 & 0.47$\pm$0.09 & 0.24$\pm$0.01 & 2.0 & Conf. \\
        WISEJ1453+1740     & 2.309$\pm$0.002 & 47.39$\pm$0.03 & 0.75$\pm$0.02 & 0.12$\pm$0.02 & 3.11$\pm$0.03 & 0.4 & Conf. \\
        WISEJ1454+1049     & 3.14$\pm$0.02   & 47.76$\pm$0.07 & 0.67$\pm$0.03 & 0.19$\pm$0.03 & 1.06$\pm$0.03$^b$ & 0.5 & Conf. \\
        WISEJ1455+0324     & 2.627$\pm$0.005 & 47.50$\pm$0.08 & 0.77$\pm$0.04 & 0.07$\pm$0.02 & 0.73$\pm$0.03 & 0.8 & Conf. \\
        WISEJ1511+0031     & 2.690$\pm$0.001 & 47.67$\pm$0.09 & 0.81$\pm$0.06 & 0.37$\pm$0.09 & 1.86$\pm$0.04 & 1.4 & Conf. \\
        WISEJ1558+2057     & 3.000$\pm$0.001 & 47.73$\pm$0.04 & 0.75$\pm$0.03 & 0.63$\pm$0.07 & 0.27$\pm$0.01$^b$ & 2.1 & Conf. \\
        WISEJ1648+3447     & 2.379$\pm$0.004 & 46.96$\pm$0.04 & 0.64$\pm$0.03 & 0.68$\pm$0.07 & 0.95$\pm$0.01 & 1.5 & Conf. \\
        WISEJ1805-6813     & 3.570$\pm$0.001 & 48.17$\pm$0.03 & 0.84$\pm$0.03 & $-$           & 0.08$\pm$0.02 & 0.7 & Rej.  \\
        WISEJ1806+1909     & 3.21$\pm$0.01   & 47.70$\pm$0.04 & 0.68$\pm$0.02 & 0.53$\pm$0.01 & 1.20$\pm$0.09 & 1.4 & Conf. \\
        WISEJ1815-6633     & 2.040$\pm$0.001 & 48.01$\pm$0.04 & 1.04$\pm$0.08 & 0.03$\pm$0.01 & 1.98$\pm$0.03 & 4.6 & Conf. \\
        WISEJ1815+1612     & 2.596$\pm$0.005 & 47.48$\pm$0.04 & 0.89$\pm$0.03 & $-$           & 1.08$\pm$0.01 & 1.1 & N/A   \\
        WISEJ1843+4345     & 2.235$\pm$0.009 & 47.57$\pm$0.04 & 0.89$\pm$0.03 & $-$           & 0.26$\pm$0.01 & 0.2 & N/A   \\
        WISEJ1939+4712     & 2.529$\pm$0.004 & 47.18$\pm$0.03 & 0.65$\pm$0.03 & $-$           & 2.43$\pm$0.03 & 1.9 & N/A   \\
        WISEJ1949+0436     & 2.073$\pm$0.005 & 47.46$\pm$0.04 & 1.18$\pm$0.04 & $-$           & 1.64$\pm$0.02 & 0.4 & N/A   \\
        WISEJ1952-5629     & 2.110$\pm$0.007 & 47.23$\pm$0.04 & 0.68$\pm$0.05 & 0.20$\pm$0.07 & 0.10$\pm$0.01 & 0.6 & Conf. \\
        WISEJ2018+5938     & 2.291$\pm$0.003 & 47.10$\pm$0.03 & 0.73$\pm$0.02 & 0.14$\pm$0.03 & 2.87$\pm$0.03 & 1.6 & Conf. \\
        WISEJ2020+5931     & 2.408$\pm$0.004 & 47.18$\pm$0.04 & 0.66$\pm$0.03 & $-$           & 2.12$\pm$0.02 & 0.5 & N/A   \\
        WISEJ2021-7808     & 2.274$\pm$0.009 & 47.56$\pm$0.06 & 0.93$\pm$0.05 & 0.06$\pm$0.02 & 1.81$\pm$0.03 & 3.1 & Conf. \\
        WISEJ2059+1225     & 2.568$\pm$0.004 & 47.34$\pm$0.04 & 0.63$\pm$0.02 & 0.29$\pm$0.03 & 1.70$\pm$0.02 & 1.5 & Conf. \\
        WISEJ2124+1753     & 2.214$\pm$0.009 & 46.93$\pm$0.03 & 0.57$\pm$0.02 & 0.87$\pm$0.09 & 0.99$\pm$0.01 & 2.6 & Conf. \\
        WISEJ2137-2819     & 2.131$\pm$0.003 & 47.37$\pm$0.09 & 0.83$\pm$0.06 & 0.31$\pm$0.08 & 2.68$\pm$0.07 & 0.4 & Conf. \\
        WISEJ2152+2605     & 2.245$\pm$0.005 & 47.01$\pm$0.03 & 0.67$\pm$0.02 & 0.41$\pm$0.04 & 2.87$\pm$0.03 & 1.8 & Conf. \\
        WISEJ2154+0640$^a$ & 1.929$\pm$0.005 & 47.15$\pm$0.03 & 0.83$\pm$0.02 & 0.06$\pm$0.02 & 1.38$\pm$0.01 & 2.7 & Conf. \\
        WISEJ2213-0534     & 2.332$\pm$0.008 & 47.33$\pm$0.04 & 0.86$\pm$0.02 & $-$           & 1.91$\pm$0.03 & 0.3 & Rej.  \\
        WISEJ2213+3437     & 2.516$\pm$0.008 & 47.33$\pm$0.04 & 0.83$\pm$0.03 & $-$           & 0.58$\pm$0.01 & 0.7 & N/A   \\
        WISEJ2222-1746     & 2.205$\pm$0.005 & 47.26$\pm$0.07 & 0.82$\pm$0.03 & $-$           & 0.01$\pm$0.01 & 0.5 & Rej.  \\
        WISEJ2316+0938     & 2.19$\pm$0.01   & 47.25$\pm$0.08 & 0.70$\pm$0.04 & 0.11$\pm$0.02 & 1.54$\pm$0.03 & 0.8 & Conf. \\
        WISEJ2320+2919$^a$ & 2.22$\pm$0.01   & 46.99$\pm$0.04 & 0.65$\pm$0.02 & 0.24$\pm$0.03 & 1.85$\pm$0.02 & 8.4 & Conf. \\
        WISEJ2342+4149     & 2.121$\pm$0.003 & 47.32$\pm$0.03 & 0.85$\pm$0.02 & $-$           & 1.93$\pm$0.02 & 1.4 & Rej.  \\
        WISEJ2349+0636     & 2.235$\pm$0.009 & 47.07$\pm$0.05 & 0.62$\pm$0.02 & $-$           & 1.72$\pm$0.05 & 3.3 & Rej.  \\
        WISEJ2359+0640$^a$ & 2.45$\pm$0.01   & 47.48$\pm$0.08 & 0.71$\pm$0.04 & 0.28$\pm$0.05 & 2.42$\pm$0.08 & 1.6 & Conf. \\
       \hline

\end{longtable}
\tablefoot{The J2000 coordinates of each HRQ have been truncated to aid readership. We quote $\overline{\chi}^2_\nu$ statistics from our best two-component SED model where the presence of a UV excess is favoured or from our best single-component model, otherwise. Luminosities are given in units of $\rm erg\;s^{-1}$. \\ \tablefoottext{a}{\spx spectra were not used to constrain the rest-frame near-infrared SEDs for these objects.}  \\ \tablefoottext{b}{The \emph{W2} photometry cannot provide robust constraints on $\rm L_{Dust}/L_{Disk}|_{2\mu m}$ at these redshifts}}

\end{center}

\clearpage

\section{Hot-dust amplitude completeness in the \spx HRQ sample} \label{app:Hot-dust_Complete}

\begin{SCfigure*}[0.7][ht!]
\centering
 \includegraphics[scale=0.7, trim={0cm, 0cm, 0cm, 0.2cm}, clip]{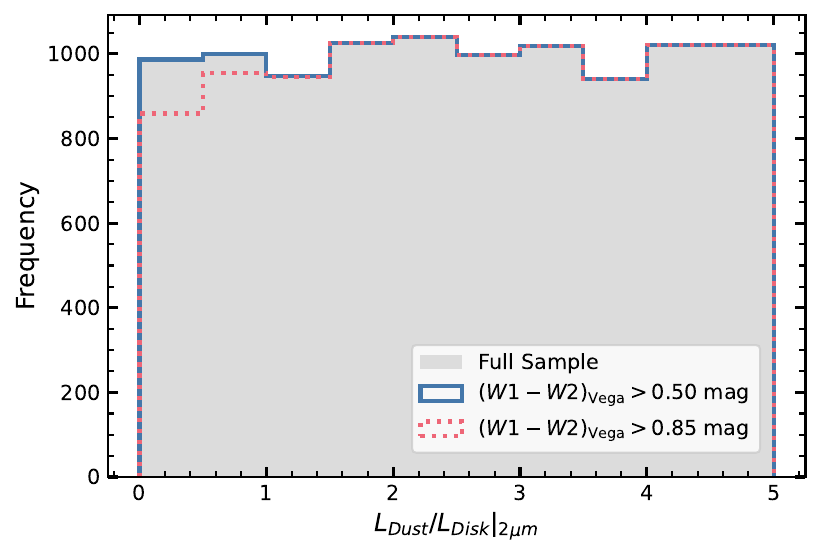} 
  \caption{Histograms illustrating the distribution of hot-dust amplitudes for 10000 simulated \textsc{qsogen} SEDs. The parent sample is presented in grey, the simulated HRQ SEDs which meet the $(W1-W2)_{\rm{Vega}}>0.5$ selection used in \emph{this work} are presented in blue, and the simulated SEDs which meet the original HRQ selection \citep[e.g.][]{2012MNRAS.427.2275B,2015MNRAS.447.3368B,2019MNRAS.487.2594T} are presented in red. The original HRQ selection results in an incomplete sample at $\rm L_{Dust}/L_{Disk}|_{2\mu m}<1$ - as suggested by \citet{S26a}.} 
 \label{fig:dust_completenes}
\end{SCfigure*}

To test whether the distribution of hot-dust amplitudes in the \spx HRQ sample is complete, we generate 10000 \textsc{qsogen} models whose redshift, line-of-sight dust-extinction and hot-dust amplitude are randomly drawn from uniform distributions with the following limits; $\rm z_{sys}=2.0-3.5$, $\rm E(B-V)=0.4-2.0$ and $\rm L_{Dust}/L_{Disk}|_{2\mu m}=0.0-5.0$. The $(W1-W2)_{\rm{Vega}}$ colours are then calculated for each simulated \textsc{qsogen} SED. \autoref{fig:dust_completenes} shows that the \spx selection captures the full diversity of hot dust properties - with all 10000 simulated SEDs meeting the $(W1-W2)_{\rm{Vega}}>0.5$ colour-selection. Conversely, the original HRQ sample \citep[e.g.][]{2012MNRAS.427.2275B,2015MNRAS.447.3368B,2019MNRAS.487.2594T} is only complete at $\rm L_{Dust}/L_{Disk}|_{2\mu m}>1$ - consistent with \citet{S26a}; their Appendix C. We estimate that the completeness of the original HRQ sample is $\simeq87$ per cent at $\rm L_{Dust}/L_{Disk}|_{2\mu m}<0.5$.  

\sidecaptionvpos{figure}{t}

\end{appendix}
\end{document}